\documentclass[prx,onecolumn,showpacs,superscriptaddress,groupedaddress,aps]{revtex4-2}
\usepackage[utf8]{inputenc}
\usepackage{physics}
\usepackage{multirow}
\usepackage{cancel}
\usepackage{amssymb}
\usepackage{amsmath}
\usepackage{amsfonts}
\usepackage{mathtools}
\allowdisplaybreaks
\usepackage{graphicx,bm}
\usepackage{ulem} 

\newcommand{\bphi} {\mbox {\boldmath $\phi$}}

\newcommand{\bxi} {\mbox {\boldmath $\xi$}}

\newcommand{\btheta} {\mbox{\boldmath $\theta$}}

\newcommand{\bp}{{\bf p}}

\newcommand{\bA}{{\bf A}}

\newcommand{\cD}{{\cal D}}

\newcommand{\cF}{{\cal F}}

\newcommand{\cH}{{\cal H}}

\newcommand{\cO}{{\cal O}}

\newcommand{\cR}{{\cal R}}
\newcommand{\cS}{{\cal S}}

\newcommand{\cU}{{\cal U}}

\newcommand{\N}{\mathbb{N}}

\newcommand{\Z}{\mathbb{Z}}

\newcommand{\round}[1]{\ensuremath{\left\lfloor#1\right\rceil}}

\begin{document}

\title{Unveiling the Quantum Toroidal Dipole}

\author{Alexandru-Lucian Nastasia}
\affiliation{Institutul National de Cercetare-Dezvoltare pentru Fizica si Inginerie Nucleara Horia Hulubei
}
\affiliation{University of Bucharest, Faculty of Physics
}
\email{alexandru.nastasia@theory.nipne.ro}
\author{Mircea Dolineanu}
\affiliation{Institutul National de Cercetare-Dezvoltare pentru Fizica si Inginerie Nucleara Horia Hulubei
}
\affiliation{University of Bucharest, Doctoral School of Physics
}
\email{mircea.dolineanu@theory.nipne.ro}
\author{Dragoș-Victor Anghel}
\affiliation{Institutul National de Cercetare-Dezvoltare pentru Fizica si Inginerie Nucleara Horia Hulubei
}
\affiliation{BLTP, JINR, Dubna, Moscow region, 141980, Russia
}
\email{dragos@theory.nipne.ro}

\pacs{Put your PACS codes here}

\begin{abstract}
	The electromagnetic response of matter is governed by three fundamental multipole families: electric, magnetic, and toroidal. While the electric and magnetic are cornerstones of physics, the toroidal dipole (TD) has eluded direct, quantitative measurement for over 60 years. Its far-field signature is masked by the electric dipole, and its behavior in the quantum regime remains largely unexplored. We address this long-standing problem by presenting a complete quantum-mechanical formalism for the TD in a nanostructure and proposing the first spectroscopic method for its direct measurement. We analyze a particle confined to a toroidal manifold subjected to an external current. We demonstrate that the resulting Aharonov-Bohm-like energy shifts in the system's spectrum are directly proportional to the expectation value of the TD operator. The transition energies exhibit a linear dependence on this current, with a quantized slope that directly reveals the change in the TD quantum number between eigenstates. This provides a clear experimental blueprint to unveil, measure, and characterize this elusive third multipole moment and its quantum nature, opening new avenues in quantum metamaterials, nanoscience, and the study of fundamental symmetries.
\end{abstract}

\maketitle


\section{Introduction} \label{sec_intro}

The interaction between matter and electromagnetic fields is fundamentally described by the multipole expansion of charge-current distributions. For over a century, physics and technology have been built on the foundation of two multipole families: the electric and the magnetic. However, a third, independent multipole family—the toroidal—is required for a complete description of electromagnetism. First conceptualized by Zel'dovich in 1957 as the anapole, a parity-violating static moment in nuclear weak interactions \cite{SovPhysJETP.6.1184.1958.Zeldovich}, the concept was generalized by Dubovik and colleagues into a full, dynamic multipole family, distinct in its symmetries from its electric and magnetic counterparts \cite{Sov.24.1965.Dubovik, SovJ.5.318.1974.Dubovik}.

Despite their fundamental nature, toroidal moments have remained one of the most elusive quantities in physics. Their primary challenge is that their far-field radiation pattern is identical to that of an electric dipole, effectively masking their distinct contribution in conventional scattering and emission experiments. This has spurred decades of effort to find indirect evidence of their existence across all scales. In nuclear physics, the toroidal moment has been identified as a key component of low-energy $E1$ modes and pygmy dipole resonances \cite{PhysRevLett.120.182501.2018.Nesterenko, PhysRevLett.133.232502.2024.Neumann}, while the related static anapole moment was famously measured in atomic cesium \cite{Science.275.1759.1997.Wood}. In solid-state physics and metamaterials, the field has seen a resurgence in exploring toroidal order in multiferroics and, most notably, in nanostructured metamaterials designed to exhibit strong toroidal responses \cite{NatureMat.15.263.2016.Papasimakis, PhysRevLett.105.073901.2010.Kaelberer}.

This progress, however, faces two profound limitations. First, existing evidence for toroidal dipoles is almost entirely \textit{indirect}. It relies on measuring far-field scattering or absorption and fitting these data to simulations that \textit{assume} a toroidal current distribution. A direct, quantitative measurement of the toroidal dipole moment itself has never been achieved. Second, the vast majority of studies in metamaterials and nanosystems treat the toroidal dipole \textit{classically}. As these systems shrink to the nanoscale, a full quantum-mechanical formalism is essential. The toroidal dipole, as a quantum-mechanical quantity, has not been rigorously explored, nor has its potential quantization been observed.

This paper bridges this fundamental gap. We provide a complete quantum-mechanical formalism for the toroidal dipole in a nanoscale system. Building on our previous work which established the toroidal dipole operator as a self-adjoint (Hermitian) operator and thus a true quantum observable \cite{JPA30.3515.1997.Anghel, PhysicaA.598.127377.2022.Dolineanu, PhysicaScripta.98.015223.2023.Anghel}, we now propose the first-ever method for its \textit{direct} and \textit{quantitative} measurement. We analyze a canonical quantum system—a particle on a toroidal manifold—and show that its energy levels shift in response to an external current (via the Aharonov-Bohm effect). We prove that this spectroscopic energy shift is directly proportional to the expectation value of the toroidal dipole operator. This establishes a clear experimental blueprint to measure the toroidal moment by observing transition energies. Furthermore, we show that the change in the toroidal moment between eigenstates is quantized, a prediction that can now be experimentally verified.

The remainder of this paper is organized as follows. In Sec.~\ref{sec_framework}, we present the complete theoretical framework, deriving the quantum-mechanical Hamiltonian and the toroidal dipole operator for a particle confined to a toroidal manifold, and establishing the Aharonov-Bohm periodicity of the system. In Sec.~\ref{sec_results}, we present our main numerical results, solving the eigenvalue problem to find the complete energy spectrum and the toroidal dipole expectation values, and we analyze the direct link between the energy spectrum and the toroidal moment. Finally, in Sec.~\ref{sec_conclusions}, we summarize our findings and discuss their direct experimental implications for a new class of spectroscopic measurements.

\section{Theoretical Framework} \label{sec_framework}

\begin{figure}[t]
	\centering
	\includegraphics[height=4 cm]{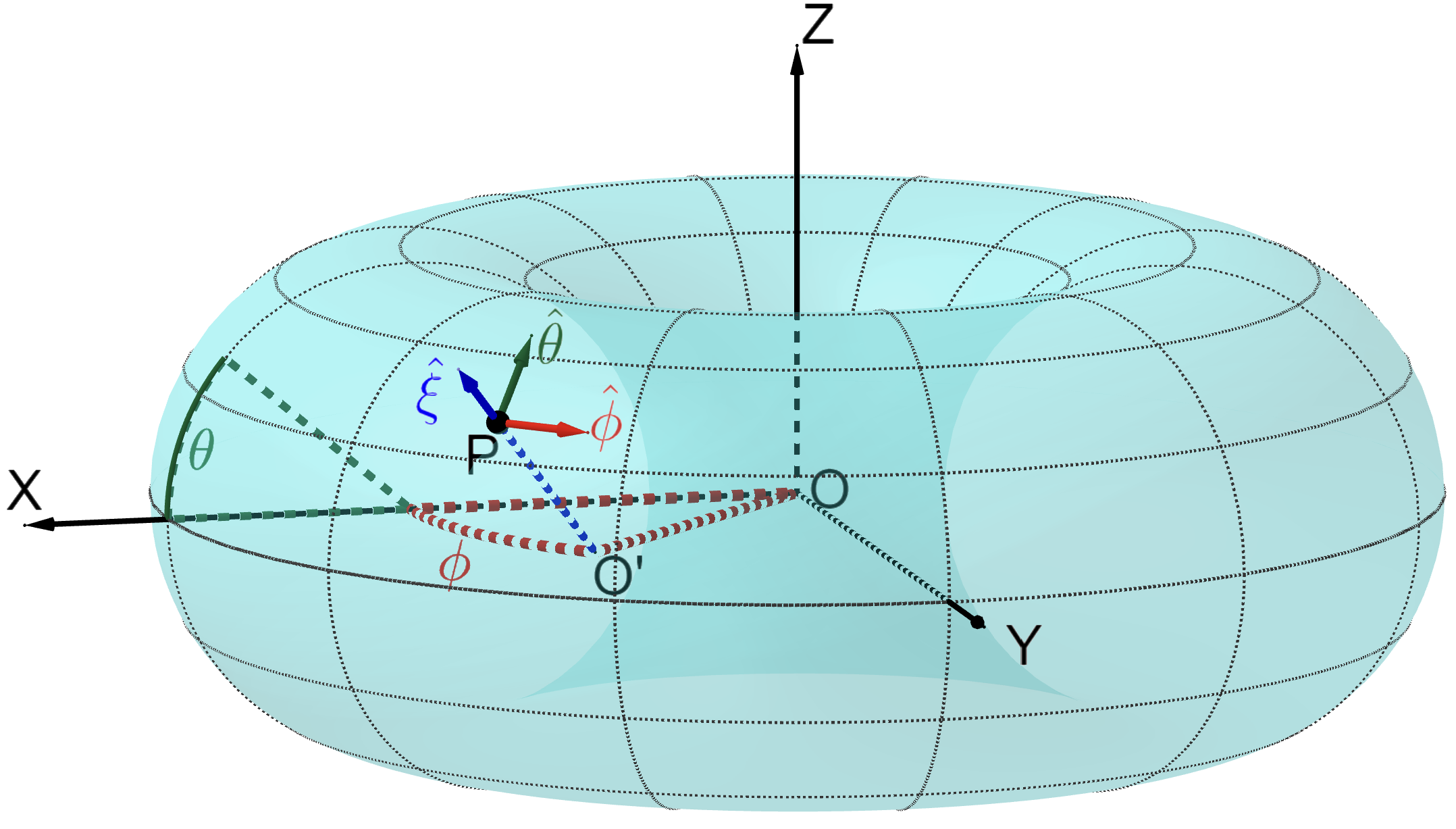}
	\caption{The system consists of a particle of charge $q$ and mass $m_\text{p}$ on a thin layer between between two tori.
		Both these tori have the same center $O$, rotation axis $z$, and major radii $R$, but they have different minor radii, $r$ and $r+\xi_{max}$, such that $r+\xi_{max} < R$ and $\xi_{max}\ll r$.
		The system is described in the orthogonal curvilinear coordinates $\theta, \phi, \xi$, where $\phi \in [0,2\pi)$ is the azimuthal angle, $\theta \in [0,2\pi)$ is the angle between the minor and the major radii, and $\xi \in [0,\xi_{\max}]$ is the coordinate that runs along the minor radius.
		The unit vectors of the local coordinates system are $\hat{\btheta}, \hat{\bphi}, \hat{\bxi}$.
		A straight filiform current $I$ of length $L$ runs along the $z$ axis, symmetrically with respect to the $(xy)$ plane.
		We assume that $L \gg R$ and the circuit is closed in such a way that only the straight portion $L$ significantly influences the particle.
	}
	\label{fig_tor}
\end{figure}

We consider a quantum particle of mass $m_p$ and charge $q$ confined to a thin toroidal membrane or major radius $R$ and minor (interior) radius $r$, as depicted in Fig.~\ref{fig_tor}. The system is described in a set of orthogonal curvilinear coordinates $(\theta, \phi, \xi)$, where $\theta, \phi \in [0, 2\pi)$ are the poloidal and azimuthal angles, respectively, and $\xi \in [0, \xi_{\max}]$ is the radial coordinate along the minor radius. The particle's wavefunctions $\psi(\theta, \phi, \xi)$ belong to the Hilbert space of square-integrable functions and obey the boundary conditions:
\begin{equation}
	\psi(\theta, \phi, \xi) = \psi(\theta+2\pi, \phi, \xi) = \psi(\theta, \phi + 2\pi, \xi)
	\quad \text{and} \quad
	\psi(\theta, \phi, 0) = \psi(\theta, \phi, \xi_{\max}) = 0
	. \label{b_cond}
\end{equation}
To simplify the expressions throughout our analysis, we introduce a set of dimensionless quantities. Geometric parameters are scaled by the major radius $R$, such that $\alpha \equiv r/R \equiv 1/a$ (the inverse aspect ratio), $\alpha_{\xi} \equiv \xi/R$, and $\alpha_{\xi_{\max}} \equiv \xi_{\max}/R$. Operators are also made dimensionless, denoted by a tilde, by factoring out characteristic physical scales.

\subsection{The Hamiltonian}

With zero electrostatic potential, the system's Hamiltonian is given by the kinetic energy term, which includes coupling to a magnetic vector potential $\vb{A}$:
\begin{equation}
	\hat{\cH} = \frac{1}{2m_p} \left(\hat{\vb{p}} - q\vb{A}\right)^2.
	\label{H_tot}
\end{equation}
The canonical momentum operator $\hat{\vb{p}} = -i\hbar\nabla$ has the following Hermitian components in toroidal coordinates:
\begin{subequations} \label{p_tfx}
	\begin{align}
		\hat{p}_\theta^\text{free}
		& = \frac{-i\hbar}{R} \left[\frac{1}{\alpha + \alpha_{\xi}} \frac{\partial}{\partial\theta} - \frac{1}{2} \frac{\sin\theta}{1 + (\alpha + \alpha_{\xi})\cos\theta}\right]
		\equiv \frac{\hbar}{R} \, \tilde{p}_\theta^\text{free}
		, \label{p_t} \\
		\hat{p}_\phi^\text{free}
		&= \frac{-i\hbar}{R} \frac{1}{1 + (\alpha + \alpha_{\xi})\cos\theta} \frac{\partial}{\partial\phi}
		\equiv \frac{\hbar}{R} \, \tilde{p}_\phi^\text{free}
		, \label{p_f} \\
		\hat{p}_\xi^\text{free}
		& =  \frac{-i\hbar}{R} \left\{\frac{\partial}{\partial\alpha_\xi} + \frac{1+ 2(\alpha + \alpha_{\xi}) \cos\theta}{2 (\alpha + \alpha_{\xi}) [1+(\alpha + \alpha_{\xi}) \cos\theta]}\right\}
		\equiv \frac{\hbar}{R} \, \tilde{p}_\xi^\text{free}
		. \label{p_x}
	\end{align}
\end{subequations}
In the presence of the field, the dimensionless components of the kinetic momentum $\tilde{\bp} \equiv (\hat{\bp} - q\vb{A}) / (\hbar/R)$ are $\tilde{p}_j = \tilde{p}_j^\text{free} - \tilde{A}_{j}$, where $\tilde{\bA} \equiv (qR/\hbar) \bA$ is the dimensionless vector potential.

The total Hamiltonian in Eq.~(\ref{H_tot}) can be separated into a field-free part $\hat{\cH}_\text{free}$ and a field-dependent part $\hat{\cH}_\bA$. The field-free Hamiltonian, studied in~\cite{PhysicaA.598.127377.2022.Dolineanu}, is decomposed based on the coordinates:
\begin{subequations} \label{H_free_contribs}
\begin{align}
	\hat{\cH}^{(\xi)}_\text{free} &\equiv
	\frac{(\hat{p}_\xi^\text{free})^2}{2m_p}
	= - \frac{\hbar^2}{2m_pR^2} \left(
	\frac{\partial^2}{\partial\alpha_{\xi}^2} + \frac{1+2(\alpha+\alpha_{\xi})\cos\theta}{2 (\alpha+\alpha_{\xi}) [1+(\alpha+\alpha_{\xi})\cos\theta]} \frac{\partial}{\partial\alpha_{\xi}} - \frac{1}{4 (\alpha+\alpha_{\xi})^2 [1+(\alpha+\alpha_{\xi})\cos\theta]^2}
	\right)
	\equiv \frac{\hbar^2}{2m_pR^2} \tilde{\cH}^{(\xi)}_\text{free}
	, \label{H_free_xi} \\
	\hat{\cH}^{(\theta)}_\text{free}
	&= - \frac{\hbar^2}{2m_p R^2} \left(
	\frac{1}{(\alpha +\alpha_{\xi})^2} \frac{\partial^2}{\partial \theta^2}
	- \frac{\sin\theta}{\left(\alpha +\alpha_{\xi}\right) \left[1 + (\alpha +\alpha_{\xi})\cos\theta\right]} \frac{\partial}{\partial\theta}
	+ \frac{1}{4 [1+(\alpha +\alpha_{\xi})\cos\theta]^2 (\alpha +\alpha_{\xi})^2} \right)
	\equiv \frac{\hbar^2}{2m_p R^2} \tilde{\cH}^{(\theta)}_\text{free}
	, \label{Hl_torus} \\
	\hat{\cH}^{(\phi)}_\text{free} &\equiv \frac{(\hat{p}_\phi^\text{free})^2}{2m_p}
	= - \frac{\hbar^2}{2m_p R^2} \frac{1}{\left[1 + (\alpha +\alpha_{\xi}) \cos\theta\right]^2} \frac{\partial^2}{\partial \phi^2}
	\equiv \frac{\hbar^2}{2m_p R^2} \tilde{\cH}^{(\phi)}_\text{free}
	. \label{Hphi_torus}
\end{align}
The sum
\begin{equation}
	\tilde{\cH}^{\text{(2D)}}_\text{free} \equiv \tilde{\cH}^{(\theta)}_\text{free} + \tilde{\cH}^{(\phi)}_\text{free}
	\label{H2D_free}
\end{equation}
\end{subequations}
is the 2D component of the free Hamiltonian, that was studied in~\cite{PhysicaA.598.127377.2022.Dolineanu}.
The term from Eq.~(\ref{H_tot}) which describes the coupling to the magnetic field, say,  $\hat{\cH}_\bA$, can be decomposed into parts that are linear ($\tilde{\cH}_{\bA,1}$) and quadratic ($\tilde{\cH}_{\bA,2}$) in the components of the dimensionless vector potential $\tilde{\bA}$. These contributions are further separated into 2D and transverse ($\xi$) components as follows:
\begin{subequations} \label{H_I_contribs}
\begin{eqnarray}
	\hat{\cH}_\bA^{\text{(2D)}}
	&\equiv& \frac{\hbar^2}{2m_pR^2} \tilde{\cH}_\bA^{\text{(2D)}}
	\equiv \frac{\hbar^2}{2m_pR^2} \left[\tilde{\cH}_{\bA,1}^{\text{(2D)}} + \tilde{\cH}_{\bA,2}^{\text{(2D)}}\right]
	, \label{cH2D_desc_A1_A2} \\
	\tilde{\cH}_{\bA,1}^{\text{(2D)}}
	&\equiv&
	2i \left[\frac{\tilde{A}_\theta}{\alpha + \alpha_{\xi}} \pdv{\theta} + \frac{\tilde{A}_\phi}{1 + (\alpha + \alpha_{\xi}) \cos\theta} \pdv{\phi}\right]
	+
	i \frac{\partial\{[1+(\alpha + \alpha_{\xi})\cos\theta] \tilde{A}_\theta\}/\partial\theta + (\alpha + \alpha_{\xi}) \partial\tilde{A}_\phi/\partial\phi}{(\alpha + \alpha_{\xi}) [1 + (\alpha + \alpha_{\xi})\cos\theta]}
	, \label{cH2D_A1} \\
	\tilde{\cH}_{\bA,2}^{\text{(2D)}}
	&\equiv& \tilde{A}_\theta^2 + \tilde{A}_\phi^2
	, \label{cH2D_A2}\\
	\hat{\cH}_\bA^{(\xi)}
	&\equiv& \frac{\hbar^2}{2m_pR^2} \tilde{\cH}_\bA^{(\xi)}
	\equiv \frac{\hbar^2}{2m_pR^2} \left[\tilde{\cH}_{\bA,1}^{(\xi)} + \tilde{\cH}_{\bA,2}^{(\xi)}\right]
	, \label{cHxi_desc_A1_A2} \\
	\tilde{\cH}_{\bA,1}^{(\xi)}
	&\equiv& 2i \left(\tilde{A}_\xi \pdv{\alpha_\xi} + \frac{1}{2 (\alpha+\alpha_{\xi}) [1 + (\alpha+\alpha_{\xi})\cos\theta]} \pdv{\{(\alpha+\alpha_{\xi}) [1 + (\alpha+\alpha_{\xi})\cos\theta] \tilde{A}_\xi\}}{\alpha_\xi} \right)
	, \label{cHxi_A1} \\
	\hat{\cH}_{\bA,2}^{(\xi)}
	&\equiv& \tilde{A}_\xi^2
	, \label{cHxi_A2}
\end{eqnarray}
The total linear and quadratic terms are:
\begin{equation}
	\hat{\cH}_{\bA,1} \equiv \hat{\cH}_{\bA,1}^{\text{(2D)}} + \hat{\cH}_{\bA,1}^{(\xi)}
	, \quad
	\hat{\cH}_{\bA,2} \equiv \hat{\cH}_{\bA,2}^{\text{(2D)}} + \hat{\cH}_{\bA,2}^{(\xi)}
	= \tilde{A}_\theta^2 + \tilde{A}_\phi^2 + \tilde{A}_\xi^2 = \tilde{A}^2
	. \label{cH_A2}
\end{equation}
\end{subequations}
Adding~(\ref{H2D_free}) and (\ref{cH2D_desc_A1_A2}) we obtain the 2D Hamiltonian, acting in the $\theta$ and $\phi$ coordinates,
\begin{subequations} \label{Htot_2D_xi}
\begin{equation}
	\tilde{\cH}^{\text{(2D)}} \equiv \tilde{\cH}_\text{free}^{\text{(2D)}} + \tilde{\cH}_\bA^{\text{(2D)}}
	\label{H2D_tot}
\end{equation}
Similarly, the Hamiltonian corresponding to the $\xi$ direction is
\begin{equation}
	\hat{\cH}^{(\xi)} \equiv \frac{\left[\hat{p}_\xi - q A_\xi\right]^2}{2m_p}
	\equiv \hat{\cH}_\bA^{(\xi)} + \frac{\hat{p}_\xi^2}{2m_p}
	\equiv \frac{\hbar^2}{2m_p R^2} \tilde{\cH}^{(\xi)}
	. \label{Hxi_tot}
\end{equation}
Finally, we reobtain
\begin{equation}
	\hat{\cH} = \hat{\cH}^\text{(2D)} + \hat{\cH}^{(\xi)}
	\equiv \frac{\hbar^2}{2m_p R^2} \tilde{\cH} .
	\label{H_tot_v2}
\end{equation}
\end{subequations}

\subsection{The Toroidal Dipole Operator} \label{subsec_T3}

The second key operator for our analysis is the toroidal dipole (TD) moment. Its projection onto the $z$-axis is defined in Cartesian coordinates as:
\begin{equation} \label{def_Ti_op}
	\hat T_3 \equiv \frac{1}{10 m_p} \sum_{j=1}^{3} \left(2 z x_j - r^2 \delta_{3j} \right) \hat p_j .
\end{equation}
and gets the following form in toroidal coordinates
\begin{eqnarray}
	\hat{T}_3 = -\frac{i\hbar}{10m_p}[z\rho \hat{\bm{\rho}}-(2\rho^2 +z^2) \hat{\bm{z}}] \left[\hat{\bm{\theta}}\left(\frac{1}{r}\frac{\partial}{\partial\theta}-\frac{iq}{\hbar}A_\theta\right) + \hat{\bm{\xi}} \left(\frac{\partial}{\partial \xi}-\frac{iq}{\hbar}A_\xi\right)\right] ,
	\label{eq_T3_theta_q}
\end{eqnarray}
where $\hat{\bm{z}}$ and $\hat{\bm{\rho}}$ are the unit vectors along the cylindrical coordinates.
We split $\hat{T}_3$ into the hermitic components~\cite{PhysicaA.598.127377.2022.Dolineanu}
\begin{subequations} \label{gen_T3_theta_xi}
	\begin{eqnarray}
		\hat{T}_3^{(\theta)}
		&=& \frac{\hbar R}{10m_p}\left\{
		\tilde{T}_{3,0}^{(\theta)} + \tilde{T}_{3,\bA}^{(\theta)}
		- \tilde{T}_\theta\right\}
		\equiv \frac{\hbar R}{10m_p} \tilde{T}_3^{(\theta)}
		, \label{T3_theta} \\
		\hat{T}_3^{(\xi)}
		&=& \frac{\hbar R}{10m_p}\left\{
		\tilde{T}_{3,0}^{(\xi)} + \tilde{T}_{3,\bA}^{(\xi)}
		- \tilde{T}_\xi\right\}
		\equiv \frac{\hbar R}{10m_p} \tilde{T}_3^{(\xi)}
		, \label{T3_xi}
	\end{eqnarray}
	%
	where we introduced the dimensionless operators $\tilde{T}_3^{(\theta)}$, $\tilde{T}_3^{(\xi)}$,
	%
	\begin{eqnarray}
		&& \tilde{T}_\theta
		= i \frac{\sin\theta \left\{ 9 \left(\alpha +\alpha_{\xi}\right)^{2} \cos^{2}\theta + 2 \left(\alpha +\alpha_{\xi}\right) \left[ 2 \left(\alpha +\alpha_{\xi}\right)^{2}+5\right] \cos \! \left(\theta \right)+3 \left(\alpha +\alpha_{\xi}\right)^{2}+2\right\}}{2 \left(1+\left(\alpha +\alpha_{\xi}\right) \cos \! \left(\theta \right)\right) \left(\alpha +\alpha_{\xi}\right)}
		= - \tilde{T}_\xi
		, \label{tT_theta} \\
		&& \tilde{T}_{3,0}^{(\theta)} \equiv
		i \left\{
		3 \cos^{2}\theta + \frac{2 \left[\left(\alpha +\alpha_{\xi}\right)^{2}+1\right] \cos\theta}{\alpha + \alpha_{\xi}} + 1
		\right\} \frac{\partial}{\partial\theta}
		, \nonumber \\
		&& \tilde{T}_{3,\bA}^{(\theta)} \equiv
		\left\{
		3 \cos^{2}\theta + \frac{2 \left[\left(\alpha +\alpha_{\xi}\right)^{2}+1\right] \cos\theta}{(\alpha+\alpha_{\xi})} + 1
		\right\}
		(\alpha+\alpha_{\xi}) \tilde{A}_\theta
		, \label{defs_T_theta} \\
		&& \tilde{T}_{3,0}^{(\xi)} \equiv i \sin\theta \left[ 3 \left(\alpha+\alpha_{\xi}\right) \cos\theta + (\alpha+\alpha_{\xi})^{2}+2\right] \frac{\partial}{\partial \alpha_{\xi}} ,
		\quad
		\tilde{T}_{3,\bA}^{(\xi)} \equiv \sin\theta \left[ 3 \left(\alpha+\alpha_{\xi}\right) \cos\theta + (\alpha+\alpha_{\xi})^{2}+2\right] \tilde{A}_\xi .
		\label{defs_T_xi}
	\end{eqnarray}
\end{subequations}

In the limit $\alpha_{\xi} \to 0$, from Eqs.~(\ref{gen_T3_theta_xi}) we obtain
\begin{subequations} \label{gen_T3_theta_xi_0}
	\begin{eqnarray}
		\tilde{T}_3^{(\theta)}
		&=& \left\{ \left[
		3 \alpha\cos^{2}\theta + 2 \left(\alpha^{2}+1\right) \cos\theta + \alpha
		\right]
		\left(\frac{i}{\alpha} \frac{\partial}{\partial\theta} + \tilde{A}_\theta\right)- \tilde{T}_\theta\right\}
		, \label{T3_theta_0} \\
		\tilde{T}_3^{(\xi)}
		&=& \left\{ \sin\theta \left( 3\alpha\cos\theta + \alpha^{2}+2\right) \left( i \frac{\partial}{\partial \alpha_{\xi}} + \tilde{A}_\xi\right)
		- \tilde{T}_\xi\right\}
		, \label{T3_xi_0} \\
		\tilde{T}_\theta
		&=& i \frac{\sin\theta \left[ 9\alpha^{2}\cos^{2}\theta + 2\alpha\left( 2\alpha^{2} + 5\right) \cos\theta + 3\alpha^{2}+2\right]}{2 \left(1+ \alpha\cos\theta\right) \alpha}
		= - \tilde{T}_\xi
		. \label{tT_theta_0}
	\end{eqnarray}
\end{subequations}
%

\subsection{Aharonov-Bohm periodicity} \label{subsec_periodicity_2D}

If the torus has very thin walls (i.e., $\xi_{\max}/r \to 0$), then $\hat{\cH}^{\text{(2D)}}$ alone leads to a dynamics which is periodic in the intensity of the vector field.
We can see this by introducing the unitary transformation
\begin{eqnarray}
	\cU (\theta, \phi, \xi) &\equiv& \exp[-\frac{iq}{\hbar}\int_0^\theta (r+\xi) A_\theta(\theta',\phi,\xi)  \, \dd\theta']
	= \exp[-i\int_0^\theta (\alpha+\alpha_\xi) \tilde{A}_\theta(\theta',\phi,\xi)  \, \dd\theta']
	\nonumber \\
	&\approx& \exp[-i\int_0^\theta \alpha \tilde{A}_\theta(\theta',\phi,0)  \, \dd\theta']
	\equiv \cU (\theta, \phi, 0) ,
	\label{def_cU}
\end{eqnarray}
which we apply to the Hilbert space of wavefunctions and to the operators that act on these wavefunctions.
Using Eqs.~(\ref{H2D_free}) and~(\ref{cH2D_desc_A1_A2}), together with the cylindrical symmetry of the problem ($\tilde{A}_\phi = 0$ and $\partial\bA/\partial\phi = 0$), we obtain
\begin{subequations} \label{H_transf}
\begin{eqnarray}
	\cU \tilde{H}^{\text{(2D)}} \cU^{-1}
	&=& \exp[-i \alpha \int_0^\theta \tilde{A}_\theta(\theta',\phi) \, \dd\theta']
	\left(
	- \frac{1}{\alpha^2} \pdv[2]{\theta}
	+ \frac{\sin{\theta}}{\alpha \left(1+ \alpha\cos{\theta}\right)} \pdv{\theta}
	+ \frac{2 i \tilde{A}_\theta}{\alpha} \pdv{\theta}
	\right)
	\exp[i \alpha \int_0^\theta \tilde{A}_\theta(\theta',\phi) \, \dd\theta']
	\nonumber \\
	&&
	- \frac{1}{4 \alpha^2 \left(1 + \alpha\cos{\theta}\right)^2}
	+ \frac{i}{\alpha (1 + \alpha\cos\theta)} \pdv{[\tilde{A}_\theta (1+\alpha\cos\theta)]}{\theta}
	+ \tilde{A}_\theta^2
	\label{H_transf_0}
\end{eqnarray}
with
\begin{eqnarray}
	- \cU \frac{1}{\alpha^2} \pdv[2]{\theta} \cU^{-1}
	&=&
	A_\theta^2(\theta)
	- \frac{i}{\alpha} \frac{\partial A_\theta(\theta)}{\partial\theta}
	- \frac{2i A_\theta(\theta)}{\alpha} \frac{\partial}{\partial\theta}
	- \frac{1}{\alpha^2} \pdv[2]{\theta}\label{H_transf_1} \\
	\cU \frac{\sin{\theta}}{\alpha \left(1+ \alpha\cos{\theta}\right)} \pdv{\theta} \cU^{-1}
	&=&
	i \frac{\sin{\theta} A_\theta(\theta)}{\left(1+ \alpha\cos{\theta}\right)}
	+ \frac{\sin{\theta}}{\alpha \left(1+ \alpha\cos{\theta}\right)}
	\pdv{\theta}
	\label{H_transf_2} \\
	\cU \frac{2 i \tilde{A}_\theta}{\alpha} \pdv{\theta} \cU^{-1}
	&=& - 2 \tilde{A}_\theta^2
	+ \frac{2 i \tilde{A}_\theta}{\alpha} \pdv{\theta}
	\label{H_transf_3} \\
	\frac{i}{\alpha (1 + \alpha\cos\theta)} \pdv{[\tilde{A}_\theta (1+\alpha\cos\theta)]}{\theta}
	&=& \frac{i}{\alpha} \pdv{\tilde{A}_\theta}{\theta}
	- \frac{i \sin\theta}{(1 + \alpha\cos\theta)} \tilde{A}_\theta
	\label{H_transf_4}
\end{eqnarray}
\end{subequations}
Adding all terms~(\ref{H_transf}), we get
\begin{eqnarray}
	\cU \tilde{H}^{\text{(2D)}} \cU^{-1}
	&=& - \frac{1}{\alpha^2} \pdv[2]{\theta}
	+ \frac{\sin{\theta}}{\alpha \left(1+ \alpha\cos{\theta}\right)} \pdv{\theta}
	- \frac{1}{4 \alpha^2 \left(1 + \alpha\cos{\theta}\right)^2}
	\equiv \tilde{\cH}_\text{free}^{\text{(2D)}}
	, \label{H_transf_U}
\end{eqnarray}
which is the free Hamiltonian.


Similarly, we obtain
\begin{equation}
	\cU \tilde{T}_3^{(\theta)} \cU^{-1}
	= \left. \tilde{T}_3^{(\theta)} \right|_{\bA = 0}
	\quad \text{and} \quad
	\cU \tilde{p}_\theta \cU^{-1}
	=
	\tilde{p}_\theta^\text{free} \equiv \left. \tilde{p}_\theta \right|_{\bA = 0} .
	\label{T3_p_theta_U}
\end{equation}

Any wavefunction $\psi(\theta, \phi, \xi) \approx \psi(\theta, \phi, 0)$ satisfies the periodic boundary condition~(\ref{b_cond}) and, under the unitary transformation $\cU$, becomes
\begin{subequations} \label{psi_cU_tot}
\begin{equation}
	\psi^{(\cU)}(\theta, \phi, 0) \equiv \cU(\theta, \phi, 0)\psi(\theta, \phi, 0) .
	\label{psi_cU}
\end{equation}
The transformed wavefunctions $\psi^{(\cU)}(\theta, \phi, 0)$ satisfy the modified boundary conditions
\begin{equation}
	\psi^{(\cU)}(2\pi, \phi, 0) \equiv \cU(2\pi, \phi, 0)\psi(0, \phi, 0)
	= \exp(-i\theta_\bA) \psi(0, \phi, 0) ,
	\quad \text{where} \quad
	\theta_\bA \equiv \int_0^{2\pi} \alpha \tilde{A}_\theta(\theta',\phi,0)  \, \dd\theta'
	\equiv \frac{q}{\hbar} \Phi_B
	. \label{psi_cU_bc}
\end{equation}
\end{subequations}
where $\Phi_B$ is the magnetic flux inside the torus--independent of $\phi$.
We notice that for
\begin{equation}
	\theta_\bA  = 2\pi n \quad \text{and} \quad \theta_\bA  = 2\pi \left(n+\frac{1}{2}\right) \quad
	\text{(where $n\in \Z$),}
	\label{cond_per}
\end{equation}
$\psi^{(\cU)}(\theta, \phi, 0)$ satisfies periodic [$\psi^{(\cU)}(2\pi, \phi, 0) = \psi^{(\cU)}(0, \phi, 0)$] and anti-periodic [$\psi^{(\cU)}(2\pi, \phi, 0) = - \psi^{(\cU)}(0, \phi, 0)$] boundary conditions, respectively.

\subsection{The electromagnetic field} \label{subsec_ElM}

The vector field produced by the filiform current $I$, of length $L$, flowing along the $z$ axis in Fig.~\ref{fig_tor} is
\begin{subequations} \label{A}
	\begin{equation} \label{A_dervs}
		A_\phi = 0, \quad
		A_\theta = \frac{\mu I}{2\pi} \ln{\left(\frac{2L}{R + (r+\xi)\cos\theta}\right)} \cos\theta , \quad
		A_\xi = \frac{\mu I}{2\pi} \ln{\left(\frac{2L}{R + (r+\xi)\cos\theta}\right)} \sin\theta ,
	\end{equation}
	or
	\begin{equation} \label{tA_dervs}
		\tilde{A}_\phi = 0, \quad
		\tilde{A}_\theta = \frac{I}{2I_0} \ln{\left(\frac{2L}{R [1 + (\alpha+\alpha_{\xi})\cos\theta]}\right)} \cos\theta, \quad
		\tilde{A}_\xi = \frac{I}{2I_0} \ln{\left(\frac{2L}{R [1 + (\alpha+\alpha_{\xi})\cos\theta]}\right)} \sin\theta ,
	\end{equation}
\end{subequations}
where $I_0 \equiv \pi\hbar/(\mu_0qR)$ will be our first unit of current.
Plugging Eqs.~(\ref{A}) into~(\ref{cond_per}), we obtain
\begin{equation}
	\theta_\bA \equiv \frac{q}{\hbar} \Phi_B(r)
	= \frac{- I}{2 I_0 a} \int_0^{2\pi} \ln(a + \cos\theta) \cos\theta \, \dd\theta
	= - 2\pi \frac{I}{I_0} \frac{\left(a-\sqrt{a^2-1}\right)}{2a} ,
	\label{cond_per_I}
\end{equation}
which gives us the period and our main unit of current,
\begin{equation} \label{I_per}
	I_\text{per} \equiv I_0 \frac{2a}{\left(a-\sqrt{a^2-1}\right)} .
\end{equation}
%


\subsection{The basis vectors for the Hilbert space} \label{subsec_new}

For the functions on the torus of finite thickness shell, we introduce the basis
\begin{subequations} \label{basis}
\begin{equation}
	\cF_{n,m,k}^{(\Lambda)} (\theta, \phi, \xi)
	= \frac{\sin(\pi k \xi/\xi_\text{max}) e^{i(n\theta + m\phi)}}{\pi \sqrt{2 \xi_\text{max} (r+\xi) [R+(r+\xi)\cos\theta]}}
	\equiv \frac{\sin(\pi k \alpha_\xi/\alpha_{\xi_\text{max}}) e^{i(n\theta + m\phi)}}{\pi R^{3/2} \sqrt{2 \alpha_{\xi_\text{max}} (\alpha + \alpha_\xi) [1+(\alpha + \alpha_\xi)\cos\theta]}}
	\equiv \frac{\tilde{\cF}_{n,m,k}^{(\Lambda)} (\theta, \phi, \alpha_\xi)}{R^{3/2}}
	\label{basis_def}
\end{equation}
where $n, m \in \Z$ (integers) and $k\in \N^*$ (positive integer).
The functions~(\ref{basis_def}) satisfy the normalization conditions
\begin{eqnarray}
	&& \langle \cF_{n,m,k}^{(\Lambda)} | \cF_{n',m',k'}^{(\Lambda)} \rangle
	\equiv \int\limits_{0}^{2\pi} \dd\theta \int\limits_{0}^{2\pi} \dd\phi \int\limits_{0}^{\xi_\text{max}} \dd\xi \, (r+\xi)
	[R+(r+\xi)\cos\theta] \left({\cF_{n,m,k}^{(\Lambda)}} (\theta, \phi, \xi)\right)^* \cF_{n',m',k'}^{(\Lambda)} (\theta, \phi, \xi)
	= \delta_{nn'} \delta_{mm'} \delta_{kk'}
	\nonumber \\
	&& = \langle \tilde{\cF}_{n,m,k}^{(\Lambda)} | \tilde{\cF}_{n',m',k'}^{(\Lambda)} \rangle
	\equiv \int\limits_{0}^{2\pi} \dd\theta \int\limits_{0}^{2\pi} \dd\phi \int\limits_{0}^{\alpha_{\xi_\text{max}}} \dd\alpha_{\xi} \, (\alpha+\alpha_{\xi}) [1+(\alpha+\alpha_{\xi})\cos\theta]
	\left({\tilde{\cF}_{n,m,k}^{(\Lambda)}} (\theta, \phi, \alpha_{\xi})\right)^* \tilde{\cF}_{n',m',k'}^{(\Lambda)} (\theta, \phi, \alpha_{\xi})
	\label{norm_FF}
\end{eqnarray}
and are eigenfunctions of $\hat{L}_3$ (the projection of the angular momentum on the $z$ axis) and of  $\hat{p}_\theta^\text{free}$ when $\xi_{\max} \to 0$:
\begin{equation}
	\hat{L}_3 \cF_{n,m,k}^{(\Lambda)} = m\hbar \cF_{n,m,k}^{(\Lambda)} ,
	\quad
	\left. \hat{p}_\theta^\text{free} \cF_{n,m,k}^{(\Lambda)}\right|_{\xi_{\max}\to0} = \frac{n\hbar}{r} \cF_{n,m,k}^{(\Lambda)}
	.
	\label{eigf_pt}
\end{equation}
\end{subequations}
%


\section{Results} \label{sec_results}

We calculate the matrix elements of the Hamiltonian and toroidal dipole operator $\tilde{T}_3$, in the basis~(\ref{basis}).
Using these, we numerically solve the eigenvalues and eigenvectors problem of the Hamiltonian and then calculate the expectation values of $\tilde{T}_3$ on the Hamiltonian eigenvectors.

\subsection{The matrix elements of the Hamiltonian} \label{subsec_H_matrix}

The matrix elements of $\tilde{\cH}^{\text{(2D)}}_\text{free}$ were calculated in~\cite{PhysicaA.598.127377.2022.Dolineanu}.
In the limit $\alpha_{\xi_{\max}} \to 0$, keeping only the lowest order term in the Taylor expansion with respect to $\alpha_{\xi_{\max}}$, we have
\begin{equation}
	\langle \tilde{\cF}^{(\Lambda)}_{n_2-n,m_1, k_1} | \tilde{\cH}^\text{(2D)}_\text{free} | \tilde{\cF}^{(\Lambda)}_{n_2,m_2, k_2} \rangle
	= a^2 \left[\left(n_2^{2}-\frac{1}{4}\right) \delta_{n,0}
	+\frac{\left({|n|} \sqrt{a^{2}-1}+a \right) \left(\sqrt{a^{2}-1}-a \right)^{{|n |}}}{\left(a^{2}-1\right)^{\frac{3}{2}}} \left(m_1^{2}-\frac{1}{4}\right)
	\right] \delta_{m_1m_2} \delta_{k_1k_2}
	. \label{H_elms_matr}
\end{equation}
Explicitly, the diagonal and off-diagonal elements are
\begin{subequations} \label{H_diag_nondiag}
\begin{eqnarray}
	\langle \tilde{\cF}^{(\Lambda)}_{n_2,m,k} | \tilde{\cH}^\text{(2D)}_\text{free} | \tilde{\cF}^{(\Lambda)}_{n_2,m,k} \rangle
	&=& a^2 \left[\left(n_2^{2}-\frac{1}{4}\right)
	+\frac{a}{\left(a^{2}-1\right)^{\frac{3}{2}}} \left(m^{2}-\frac{1}{4}\right)
	\right]
	= \langle \tilde{\cF}^{(\Lambda)}_{-n_2,m,k} | \tilde{\cH}^\text{(2D)}_\text{free} | \tilde{\cF}^{(\Lambda)}_{-n_2,m,k} \rangle
	, \label{H_diag} \\
	\langle \tilde{\cF}^{(\Lambda)}_{n_2-n,m, k} | \hat{\cH}^\text{(2D)}_\text{free} | \tilde{\cF}^{(\Lambda)}_{n_2,m, k} \rangle
	&=& \frac{a^2 \left({|n|} \sqrt{a^{2}-1}+a \right) \left(\sqrt{a^{2}-1}-a \right)^{{|n|}}}{\left(a^{2}-1\right)^{\frac{3}{2}}} \left(m^{2}-\frac{1}{4}\right)
	\qquad (n \ne 0)
	, \label{H_nondiag}
\end{eqnarray}
\end{subequations}
respectively.

The matrix elements of the $\xi$ component of the free Hamiltonian are
\begin{equation}
	\left\langle \tilde{\cF}_{n_2-n,m,k}^{(\Lambda)} \left| \tilde{\cH}^{(\xi)}_\text{free} \right| \tilde{\cF}_{n_2,m',k'}^{(\Lambda)} \right\rangle
	= \delta_{mm'} \delta_{kk'} \delta_{n,0} \frac{\pi^2 {k}^{2}}{\alpha_{\xi_{\max}}^{2}}
	. \label{F_d2dxi_F}
\end{equation}
%

To calculate the matrix elements of $\tilde{\cH}_\bA$, we plug~(\ref{tA_dervs}) into~(\ref{H_I_contribs}) and obtain
%
\begin{subequations} \label{H_elmg}
\begin{eqnarray}
	\tilde{\cH}^\text{(2D)}_{\bA,1} &\equiv& \frac{i I}{I_0} \left( \frac{1}{\alpha+\alpha_{\xi}} \ln\left[\frac{1}{1+(\alpha+\alpha_{\xi})\cos\theta} \frac{2L}{R}\right] \left\{
	\cos\theta \pdv{\theta}
	- \frac{\sin\theta [1 + 2 (\alpha+\alpha_{\xi})\cos\theta]}{2 [1 + (\alpha+\alpha_{\xi})\cos\theta]}
	\right\}
	+ \frac{1}{4} \frac{\sin(2\theta) }{1 + (\alpha+\alpha_{\xi})\cos\theta}
	\right)
	\nonumber \\
	\tilde{\cH}^\text{(2D)}_{\bA,2} &\equiv& \frac{I^2}{4I_0^2} \log^2\left[\frac{1}{1 + (\alpha+\alpha_{\xi})\cos\theta} \frac{2L}{R}\right] \cos^2\theta
	\nonumber \\
	\tilde{\cH}^{(\xi)}_{\bA,1} &\equiv& \frac{i I}{I_0} \left(
	\ln(\frac{1}{1 + (\alpha + \alpha_{\xi})\cos\theta}\frac{2L}{R}) \sin\theta \left\{ \frac{\partial}{\partial\alpha_{\xi}} + \frac{1 + 2(\alpha+\alpha_{\xi})\cos\theta}{2[1 + (\alpha+\alpha_{\xi})\cos\theta] (\alpha+\alpha_{\xi})} \right\}
	- \frac{1}{4} \frac{\sin(2\theta)}{1 + (\alpha + \alpha_{\xi})\cos\theta} \right)
	\nonumber \\
	\tilde{\cH}^{(\xi)}_{\bA,2} &\equiv& \frac{I^2}{4 I_0^2} \ln^2\left\{\frac{2L}{R [1 + (\alpha + \alpha_{\xi})\cos\theta]}\right\} \sin^2\theta
	. \nonumber
\end{eqnarray}
\end{subequations}
Now we can observe that the matrix elements of $\tilde{\cH}^\text{(2D)}_{\bA}$ and $\tilde{\cH}^{\xi}_{\bA}$ can only be of the order $\alpha_{\xi_{\max}}^{-1}$ or smaller (concrete calculations will show that they actually are of the order $\alpha_{\xi_{\max}}^0$ or smaller), so
\begin{equation}
	\lim_{\alpha_{\xi_{\max}}\to0}\frac{\left\langle \tilde{\cF}_{n_2-n,m,k}^{(\Lambda)} \left| \tilde{\cH}^\text{(2D)}_\bA \right| \tilde{\cF}_{n_2,m,k}^{(\Lambda)} \right\rangle}{\left\langle \tilde{\cF}_{n_2,m,1}^{(\Lambda)} \left| \tilde{\cH}^{(\xi)}_\text{free} \right| \tilde{\cF}_{n_2,m,1}^{(\Lambda)} \right\rangle}
	=
	\lim_{\alpha_{\xi_{\max}}\to0}\frac{\left\langle \tilde{\cF}_{n_2-n,m,k}^{(\Lambda)} \left| \tilde{\cH}^{(\xi)}_\bA \right| \tilde{\cF}_{n_2,m,k'}^{(\Lambda)} \right\rangle}{\left\langle \tilde{\cF}_{n_2,m,1}^{(\Lambda)} \left| \tilde{\cH}^{(\xi)}_\text{free} \right| \tilde{\cF}_{n_2,m,1}^{(\Lambda)} \right\rangle}
	= 0 ,
	\label{m_elems_ratios}
\end{equation}
for any $n_2$, $n$, $m$.
Equations~(\ref{H_elms_matr}) and (\ref{m_elems_ratios}) imply that in the limit $\alpha_{\xi_{\max}} \to 0$ all the matrix elements of $\tilde{\cH} - \tilde{\cH}^{(\xi)}_\text{free}$ are negligible in comparison to $\langle \tilde{\cF}_{n_2,m,k}^{(\Lambda)} | \tilde{\cH}^{(\xi)}_\text{free} | \tilde{\cF}_{n_2,m,k}^{(\Lambda)} \rangle$, for any (finite) $n_2$, $m$, and $k$.
Therefore, $\tilde{\cH}^{(\xi)}_\text{free}$ contributes only the additive constant $\pi^2/\alpha_{\xi_{\max}}^{2}$ (divergent in the limit $\alpha_{\xi_{\max}} \to 0$) to the total energy of the system and fixes $k = k' = 1$--that is, the lowest energy level in the direction $\xi$.
%
Using this observation, we may calculate the matrix elements of $\tilde{\cH}^\text{(2D)}_{\bA,1}$ and $\tilde{\cH}^\text{(2D)}_{\bA,2}$.
After integrating over $\alpha_{\xi}$, $\phi$, $\theta$, and taking again only the lowest order in $\alpha_{\xi_{\max}}$, we obtain
\begin{eqnarray}
	&& \left\langle \tilde{\cF}_{n_2-n,m,1}^{(\Lambda)} \left| \tilde{\cH}^\text{(2D)}_{\bA,1} \right| \tilde{\cF}_{n_2,m',1}^{(\Lambda)} \right\rangle
	= \delta_{mm'} \frac{Ia}{I_0}
	\left[
	n_2 \left(\frac{I_{n+1}^{(\ln)}(a) + I_{n-1}^{(\ln)}(a)}{2} - \log(\frac{2aL}{R}) \frac{\delta_{n+1} + \delta_{n-1}}{2} \right)
	\right. \nonumber \\
	&& \left. + \frac{I_{n+1}^{(\ln)}(a) - I_{n-1}^{(\ln)}(a)}{4}
	- \log(\frac{2aL}{R}) \frac{\delta_{n+1} - \delta_{n-1}}{4}
	+ \frac{I_{n+2}(a) - I_{n-2}(a)}{8}
	\right]
	\label{tH2DI1_matrix1}
\end{eqnarray}
where
\begin{equation}
	\begin{cases}
	I_n(a) & \equiv \frac{1}{2\pi} \int_{0}^{2\pi} e^{in\theta}/({a+\cos\theta}) \, \dd\theta = \frac{\left(-a + \sqrt{a^2-1}\right)^{|n|}}{\sqrt{a^2-1}} \\
	%
	I_n^{(\ln)}(a) & \equiv \frac{1}{2\pi} \int_{0}^{2\pi} e^{in\theta} \ln(a + \cos\theta)\,\dd\theta
	= \ln(\frac{a+\sqrt{a^2-1}}{2})_{n=0} - \left.\frac{\left(-a + \sqrt{a^2-1}\right)^{|n|}}{|n|}\right|_{n\ne0}
	\end{cases}
	\label{In_Inln}
\end{equation}
(see the proof of the first Eq.~\ref{In_Inln} in Appendix~\ref{app_sec_AnSol}).
Pluggind the expressions~(\ref{In_Inln}) into~(\ref{tH2DI1_matrix1}) we get
\begin{eqnarray}
	&& \left\langle \tilde{\cF}_{n_2-n,m,1}^{(\Lambda)} \left| \tilde{\cH}^\text{(2D)}_{\bA,1} \right| \tilde{\cF}_{n_2,m',1}^{(\Lambda)} \right\rangle
	= \delta_{mm'} \frac{Ia}{I_0} \frac{2n_2 - n}{2}
	\left\{ \vphantom{\left.\frac{ \left(a|n| + \sqrt{a^2-1}\right)}{(n^2-1)}\right|_{|n|>1}}
	\left(a - \sqrt{a^2-1}\right)_{n=0}
	\right. \nonumber \\
	&& \left. - \frac{1}{2} \left[
	\log(\frac{4aL}{R \left(a+\sqrt{a^2-1}\right)}) + \frac{\left(-a + \sqrt{a^2-1}\right)^{2}}{2}
	\right]_{|n|=1}
	+ \left(-a + \sqrt{a^2-1}\right)^{|n|} \left. \frac{ \left(a|n| + \sqrt{a^2-1}\right)}{(n^2-1)}
	\right|_{|n|>1}
	\right\} .
	\label{tH2DI1_matrix2}
\end{eqnarray}

From~(\ref{H_elmg}) we obtain
\begin{eqnarray}
	&& \left\langle \tilde{\cF}_{n_2-n,m,1}^{(\Lambda)} \left| \tilde{\cH}_{\bA,1}^{(\xi)} \right| \tilde{\cF}_{n_2,m',1}^{(\Lambda)} \right\rangle
	= \frac{\delta_{mm'}}{2 } \left\{\frac{1}{\alpha_{\xi_{\max}}} \int_{0}^{1} \dd x\, \sin(2\pi x) \int_{0}^{2\pi} \dd\theta \, e^{i n \theta} \sin\theta
	\ln\left[\frac{1}{1+\left(\alpha + x\alpha_{\xi_{\max}} \right)\cos\theta} \frac{2L}{R}\right]
	\right. \nonumber \\
	&& \left. - \frac{1}{2\pi} \int_{0}^{1} \dd x\, \sin^2(\pi x) \int_{0}^{2\pi} \dd\theta
	\frac{e^{i n \theta} \sin(2\theta)}{1+\left(\alpha + x \alpha_{\xi_{\max}}\right) \cos\theta}
	\right\}
	\label{H2_matr_elem}
\end{eqnarray}
Expanding the integrands in Taylor series and keeping only the terms proportional to $\alpha_{\xi_{\max}}^{-1}$ and $\alpha_{\xi_{\max}}^{0}$ (which remain different from zero in the limit $\alpha_{\xi_{\max}}\to0$), upon integration over $x$ we obtain
\begin{equation}
	\left\langle \tilde{\cF}_{n_2-n,m,1}^{(\Lambda)} \left| \tilde{\cH}_{\bA,1}^{(\xi)} \right| \tilde{\cF}_{n_2,m',1}^{(\Lambda)} \right\rangle = 0.
	\label{H_xi_A1_0}
\end{equation}

\begin{figure}[t]
	\centering
	\includegraphics[width=0.4\linewidth]{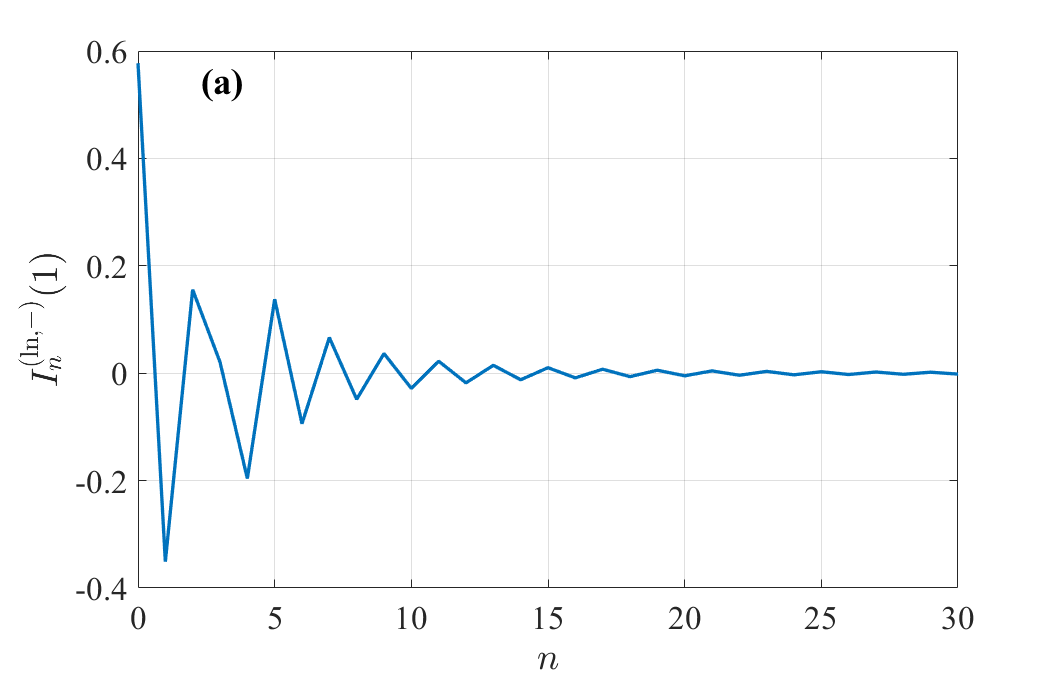}
	\includegraphics[width=0.4\linewidth]{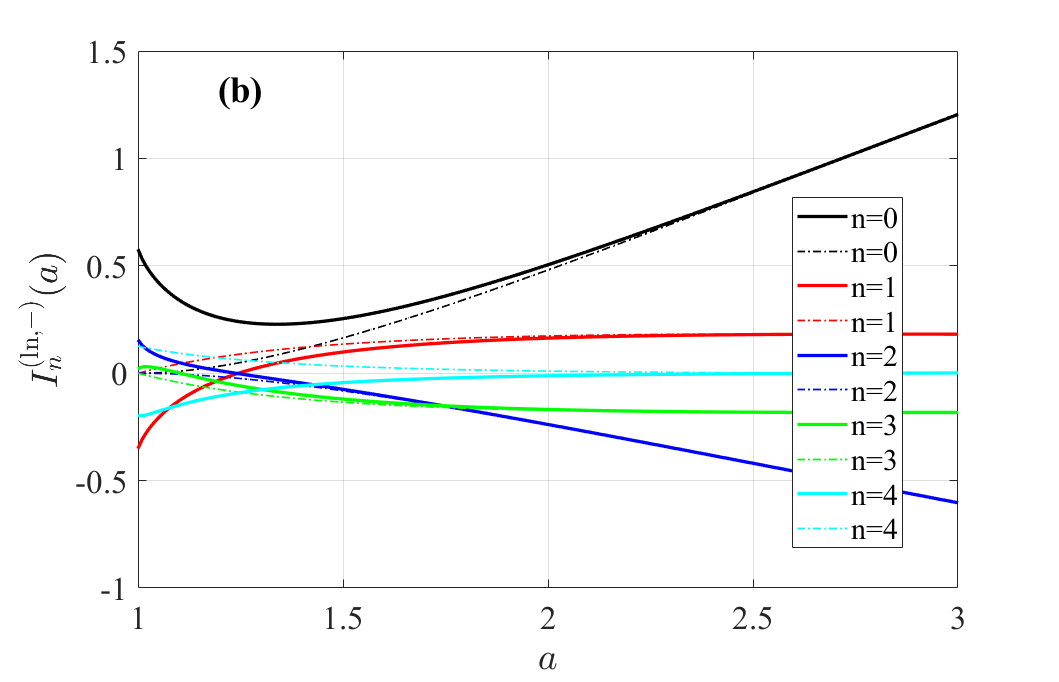}
	\caption{(a) The function $\tilde{I}_n^{(\ln^2,-)}(1)$ vs. $n$ and (b) $\tilde{I}_n^{(\ln^2,-)}(a)$ vs $a$, for the values $n = -4, \ldots, 4$, as indicated in the legend. We observe that the asymptotic values are very rapidly approached as $a$ increases. The asymptotic expressions for $n = -4, \ldots, 4$ are given in Eq.~(\ref{In_ln2_p}) and for larger values of $|n|$ are given in~(\ref{asympt_exp_tIn_ln2}).}
	\label{fig_inln2ma1}
\end{figure}

For the two remaining terms we have:
\begin{subequations} \label{H_A2_2Dxi}
\begin{eqnarray}
	&& \left\langle \tilde{\cF}_{n_2-n,m,k}^{\Lambda} \left| \tilde{\cH}^\text{(2D)}_{\bA,2} \right| \tilde{\cF}_{n_2,m',k'}^{\Lambda} \right\rangle
	= \delta_{mm'} \delta_{kk'}
	\frac{I^2}{16 I_0^2} \left\{\log(\frac{2aL}{R})^2 (\delta_{n+2} + 2 \delta_n + \delta_{n-2})
	\right. \nonumber \\
	&& \left. - 2 \log(\frac{2aL}{R}) \left[I_{n+2}^{(\ln)}(a) + 2I_n^{(\ln)}(a) + I_{n-2}^{(\ln)}(a)\right]
	+ \left[I_{n+2}^{(\ln^2)}(a) + 2I_n^{(\ln^2)}(a) + I_{n-2}^{(\ln^2)}(a)\right] \right\} ,
	\label{tH2DI_matrix1} \\
	&& \left\langle \tilde{\cF}_{n_2-n,m,k}^{\Lambda} \left| \tilde{\cH}^{(\xi)}_{\bA,2} \right| \tilde{\cF}_{n_2,m',k'}^{\Lambda} \right\rangle
	= \delta_{mm'} \delta_{kk'}
	\frac{I^2}{16 I_0^2} \left\{\log(\frac{2aL}{R})^2 (-\delta_{n+2} + 2 \delta_n - \delta_{n-2})
	\right. \nonumber \\
	&& \left. - 2 \log(\frac{2aL}{R}) \left[-I_{n+2}^{(\ln)}(a) + 2I_n^{(\ln)}(a) - I_{n-2}^{(\ln)}(a)\right]+ \left[-I_{n+2}^{(\ln^2)}(a) + 2I_n^{(\ln^2)}(a) - I_{n-2}^{(\ln^2)}(a)\right]
	\right\} ,
	\label{tHxiI_matrix1}
\end{eqnarray}
\end{subequations}
where (see Appendix~\ref{app_sec_asympt_Iln2})
\begin{equation}
	I_n^{(\ln^2)}(a) \equiv \frac{1}{2\pi} \int_{0}^{2\pi} e^{in\theta} \ln^2(a + \cos\theta) \dd \theta ,
	\quad \text{with} \quad
	I_n^{(\ln^2)}(a) \stackrel{a\gg1}{\sim} \ln^2a \delta_{n,0} + \frac{\ln a}{a} \delta_{|n|,1} + \frac{\left(1 -\ln a\right)}{4a^{2}} \left(\delta_{|n|,2} + 2\delta_{n,0}\right).
	\label{In_ln2}
\end{equation}
Plugging~(\ref{In_Inln}) and (\ref{In_ln2}) into~(\ref{H_A2_2Dxi}) we obtain
\begin{subequations} \label{H_A2_2Dxi2}
\begin{eqnarray}
	&& \left\langle \tilde{\cF}_{n_2-n,m,1}^{\Lambda} \left| \tilde{\cH}^\text{(2D)}_{\bA,2} \right| \tilde{\cF}_{n_2,m,1}^{\Lambda} \right\rangle
	=
	\frac{I^2}{8 I_0^2} \left[ \log(\frac{2aL}{R}) \left(
	\left\{\log\left[\frac{8aL}{R \left(a+\sqrt{a^2-1}\right)^2}\right] + \left(-a + \sqrt{a^2-1}\right)^{2} \right\}_{n=0}
	\right. \right. \nonumber \\
	&&
	+ \left. \frac{\left(-a + \sqrt{a^2-1}\right) \left[\left(-a + \sqrt{a^2-1}\right)^{2} +9\right]}{3} \right|_{|n|=1}
	+ \frac{1}{2} \Biggl\{\ln\left[\frac{8aL}{R \left(a+\sqrt{a^2-1}\right)^2}\right]
	+ \frac{ \left[\left(-a + \sqrt{a^2-1}\right)^{2} + 4\right]}{2}
	\nonumber \\
	&& \left. \left. \times \left(-a + \sqrt{a^2-1}\right)^{2}
	\Biggr\}_{|n|=2}
	+ \left. \frac{4 \left(-a+\sqrt{a^2-1}\right)^{|n|} \left[a^{2} n^{2} + 2a|n| \sqrt{a^{2}-1} -2\right]}{|n|\left(n^2-4\right)} \right|_{|n|>2}\right)
	+ {\tilde{I}_n^{(\ln^2, +)}}(a) \right]
	, \label{H_2D_A2} \\
	&& \left\langle \tilde{\cF}_{n_2-n,m,1}^{\Lambda} \left| \tilde{\cH}^{(\xi)}_{\bA,2} \right| \tilde{\cF}_{n_2,m,1}^{\Lambda} \right\rangle
	= \frac{I^2}{8 I_0^2} \left[ - \log(\frac{2aL}{R}) \left(
	- \left\{ \log\left[\frac{8aL}{R \left(a+\sqrt{a^2-1}\right)^2}\right] - \left(-a + \sqrt{a^2-1}\right)^{2} \right\}_{n=0}
	\right. \right. \label{H_xi_A2} \\
	&&
	 + \left.\frac{2\left(-a +\sqrt{a^{2}-1}\right)^2 \left(a + 2\sqrt{a^{2}-1}\right)}{3} \right|_{|n|=1}
	+ \frac{1}{2} \Biggl\{ \log\left[\frac{8aL}{R \left(a+\sqrt{a^2-1}\right)^2}\right]
	+ \frac{\left(3a + 5\sqrt{a^{2}-1}\right)}{2}
	\nonumber \\
	&& \left. \left.
	\times \left(-a +\sqrt{a^{2}-1}\right)^{3} \Biggr\}_{|n|=2}
	+ \left.\frac{4\left(-a + \sqrt{a^2-1}\right)^{|n|} \left[ \left(a^{2}-1\right) n^{2} + 2a|n|\sqrt{a^{2}-1} + 2 \right]}{|n| (n^2-4)}\right|_{|n|>2}
	\right)
	+ {\tilde{I}_n^{(\ln^2, -)}}(a) \right] ,
	\nonumber
\end{eqnarray}
where
%
%
\begin{equation}
\begin{cases}
	{\tilde{I}_n^{(\ln^2, \pm)}}(a) \equiv \frac{1}{2\pi} \int_{0}^{2\pi} e^{in\theta} [1\pm\cos(2\theta)] \ln^2(a + \cos\theta) \dd \theta
	, \quad \text{with $\tilde{I}_n^{(\ln^2, \pm)}(1) =$ finite for any integer $n$ and} \vphantom{\frac{1}{\frac{a}{b}}}\\
	{\tilde{I}_n^{(\ln^2,\pm)}}(a) \stackrel{a\gg1}{\sim} \ln^2a \left(\delta_{n,0} \pm \frac{\delta_{|n|,2}}{2}\right)
	+ \frac{\ln a}{a} \left[\left(1 \pm \frac{1}{2}\right) \delta_{|n|,1} \pm \frac{\delta_{|n|,3}}{2} \right]
	+ \frac{\left(1 -\ln a \right)}{4a^{2}} \left[ (2 \pm 1) \delta_{n,0} + (1\pm1) \delta_{|n|,2} \pm \frac{\delta_{|n|,4}}{2} \right] .
\end{cases}
\label{In_ln2_p}
\end{equation}
\end{subequations}
Adding Eqs.~(\ref{H_A2_2Dxi2}) and using the definitions~(\ref{cH_A2}) we obtain
\begin{eqnarray}
	\left\langle \tilde{\cF}_{n_2-n,m,1}^{(\Lambda)} \left| \tilde{\cH}_{\bA,2}^\text{(2D)} + \tilde{\cH}_{\bA,2}^{(\xi)} \right| \tilde{\cF}_{n_2,m,1}^{(\Lambda)} \right\rangle
	&=&
	\frac{I^2}{4 I_0^2} \left\{
	\log(\frac{2aL}{R}) \log\left[\frac{8aL}{R \left(a+\sqrt{a^2-1}\right)^2}\right]_{n=0}
	\right. \nonumber \\
	&& \left.
	+ 2 \log(\frac{2aL}{R}) \left.\frac{\left(-a + \sqrt{a^2-1}\right)^{|n|}}{|n|}\right|_{n\ne0}
	+ I_n^{(\ln^2)}(a) \right\}
	. \label{tHA2_matrix1}
\end{eqnarray}
Figure~\ref{fig_inln2ma1} illustrates the behavior of the functions $\tilde{I}_n^{(\ln^2,-)}(a)$.
Subplot (a) represents $\tilde{I}_n^{(\ln^2, -)}(1)$ versus $n$ for $n=1, \ldots, 30$, showing a rapid decrease in magnitude as $n$ increases.
Subplot (b) represents $\tilde{I}_n^{(\ln^2,\pm)}(a)$ versus $a$ for $n = 0, \ldots, 4$ (solid lines), demonstrating a quick convergence to their asymptotic approximations from Eq.~(\ref{In_ln2_p}) (dash-dot lines).
These empirical results are explained by the analytical behavior for large $a$.
According to Eq.~(\ref{In_ln2_p}), the approximations for $\tilde{I}_0^{(\ln^2,\pm)}(a)$ and $\tilde{I}_2^{(\ln^2,\pm)}(a)$ are dominated by terms proportional to $\ln^2a$, while for $n \neq 0, 2$, the leading term scales as $\ln a/a^n$.
This dependence on $1/a^n$ accounts for a sharp decay of $\left|\tilde{I}_n^{(\ln^2, -)}(a)\right|$ with increasing values of both $n$ and $a$.

Therefore, considering the freeze-out conditions~(\ref{F_d2dxi_F}) and (\ref{H_xi_A1_0}), we conclude that in the limit $\alpha_{\xi_{\max}}\to0$, the system is described by the effective Hamiltonian
\begin{equation}
	\tilde{\cH}_{\text{eff}} \equiv \tilde{\cH}^{\text{(2D)}} + \tilde{\cH}^{(\xi)}_{\bA,2} \equiv \tilde{\cH}^{\text{(2D)}} + \tilde{A}_\xi^2
	. \label{def_Heff}
\end{equation}
Applying the transformation $\cU$ of Section~\ref{subsec_periodicity_2D} to Eq.~(\ref{def_Heff}), we obtain
\begin{equation}
	\tilde{\cH}^{(\cU)}_{\text{eff}}  \equiv \cU \tilde{\cH}_{\text{eff}}  \cU^{-1} = \tilde{\cH}_\text{free}^\text{(2D)} + \tilde{\cH}_{\bA,2}^{(\xi)} .
	\label{Heff_cU}
\end{equation}
Therefore, in general, the eigenvalues and eigenfunctions problem of $\tilde{\cH}$ is reduced to that of $\tilde{\cH}^{(\cU)}$, but on a space of functions with modified boundary conditions $\{\psi^{(\cU)}\}$~(\ref{psi_cU_bc}).

\subsection{The matrix elements of the toroidal dipole} \label{subsec_T3_m}

Using again Taylor expansions in $\alpha_{\xi_{\max}}$, from the expressions of Section~\ref{subsec_T3} we obtain in the lowest order:
\begin{subequations} \label{T3_matrices}
\begin{eqnarray}
	&& \left\langle \cF_{n_2-n,m,1}^{(\Lambda)} \left| \tilde{T}_{3,0}^{(\theta)} \right| \cF_{n_2,m',1}^{(\Lambda)} \right\rangle
	= \delta_{mm'} \left\{
	- \frac{3}{4} \left(n_{2} + \frac{n}{4}\right)_{|n|=2}
	- \frac{1}{a} \left[n_2\left(a^{2}+1\right) - \frac{n (a^2-2)}{4} \right]_{|n|=1}
	- \frac{5n_2}{2}\delta_{n}
	\right. \nonumber \\
	&& \left.
	+ \text{sgn}(n) \frac{(a^2-1) \left(-a+\sqrt{a^2-1}\right)^{|n|}}{2}
	\right\} ,
	\label{calc_T3t_0} \\
	&& \left\langle \cF_{n_2-n,m,1}^{(\Lambda)} \left| \tilde{T}_{3,0}^{(\xi)} \right| \cF_{n_2,m',1}^{(\Lambda)} \right\rangle
	= \delta_{mm'}
	\left\{
	\left.\frac{9n}{16}\right|_{|n|=2} + \left.\frac{n\left(a^2 + 4\right)}{4a}\right|_{|n|=1}
	- \frac{\text{sgn}(n) \left(a^2 -1\right) \left(-a+\sqrt{a^2-1}\right)^{|n|}}{2}
	\right\} ,
	\label{calc_T3x_0} \\
	&& \left\langle \cF_{n_2-n,m,1}^{(\Lambda)} \left| \tilde{T}_{3,\bA}^{(\theta)} \right| \cF_{n_2,m',1}^{(\Lambda)} \right\rangle
	= \frac{\delta_{mm'}}{2a^2} \frac{I}{I_0}
	\left\{ \ln(\frac{2aL}{R}) \left[
	\frac{3a}{4} \frac{\delta_{n+3} + \delta_{n-3}}{2} + \left(a^{2}+1\right)  \frac{\delta_{n+2} + \delta_{n-2}}{2} + \frac{13a}{4} \frac{\delta_{n+1} + \delta_{n-1}}{2}
	\right. \right. \label{calc_T3t_A} \\
	&& \left. \left. + \left(a^{2}+1\right) \delta_n
	\right]
	- \frac{3a}{4} \frac{I_{n+3}^{(\ln)}(a) + I_{n-3}^{(\ln)}(a)}{2} - \left(a^{2}+1\right)  \frac{I_{n+2}^{(\ln)}(a) + I_{n-2}^{(\ln)}(a)}{2} - \frac{13a}{4} \frac{I_{n+1}^{(\ln)}(a) + I_{n-1}^{(\ln)}(a)}{2} - \left(a^{2}+1\right) I_{n}^{(\ln)}(a)
	\right\} , \nonumber \\
	&& \left\langle \cF_{n_2-n,m,1}^{(\Lambda)} \left| \tilde{T}_{3,\bA}^{(\xi)} \right| \cF_{n_2,m',1}^{(\Lambda)} \right\rangle
	= \frac{\delta_{mm'}}{18a^2} \frac{I}{I_0}
	\left(
	\ln(\frac{2aL}{R}) \left[4(1 + 2a^2) \delta_n + 3a (\delta_{n+1} + \delta_{n-1}) - 2(1 + 2a^2) (\delta_{n+2} + \delta_{n-2})
	\right. \right. \nonumber \\
	&& \left. - 3a \left(\delta_{n+3} + \delta_{n-3}\right)\right]
	- \left\{ 4(1 + 2a^2) I^{(\ln)}_{n}(a) + 3a \left[ I^{(\ln)}_{n+1}(a) + I^{(\ln)}_{n-1}(a) \right] - 2(1 + 2a^2) \left[ I^{(\ln)}_{n+2}(a) + I^{(\ln)}_{n-2}(a) \right]
	\right. \nonumber \\
	&& \left. \left. - 3a \left[ I^{(\ln)}_{n+3}(a) + I^{(\ln)}_{n-3}(a) \right] \right\}
	\vphantom{\frac{1}{1}}\right)
	\label{calc_T3xiA_In}
\end{eqnarray}
\end{subequations}
Using Eqs.~(\ref{In_Inln}) we get
\begin{subequations} \label{calc_T3tx_A2}
\begin{eqnarray}
	&& \left\langle \cF_{n_2-n,m,1}^{(\Lambda)} \left| \tilde{T}_{3,\bA}^{(\theta)} \right| \cF_{n_2,m',1}^{(\Lambda)} \right\rangle
	= \frac{\delta_{mm'}}{2a^2} \frac{I}{I_0}
	\scalebox{1.2}{$\Biggl($}
	\Biggl\{
	\left(a^{2}+1\right) \ln\left[\frac{4aL}{R \left(a+\sqrt{a^2-1}\right)}\right]
	- 2 a \left(a-\sqrt{a^{2}-1}\right)-\frac{1}{2}
	\Biggr\}_{n=0}
	\nonumber \\
	&& + \Biggl\{
	\frac{13a}{8} \ln\left[\frac{4aL}{R \left(a+\sqrt{a^2-1}\right)}\right]
	+ \left(-\frac{1}{12} a^{4}+\frac{3}{8} a^{2}+\frac{4}{3}\right) \sqrt{a^{2}-1}+\frac{a^{5}}{12}-\frac{5 a^{3}}{12}-\frac{61 a}{32}
	\Biggr\}_{|n| = 1}
	\nonumber \\
	&& + \Biggl\{
	\frac{\left(a^{2}+1\right)}{2} \ln\left[\frac{4aL}{R \left(a+\sqrt{a^2-1}\right)}\right]
	+ \frac{\left(-a +\sqrt{a^{2}-1}\right)^{3}}{8} \left[\left(\frac{139 a^{2}}{5}-3\right) \sqrt{a^{2}-1}+\frac{a}{5} \left(141 a^{2}-\frac{259}{3}\right)\right]
	\Biggr\}_{|n| = 2}
	\nonumber \\
	&& + \Biggl\{
	\frac{3a}{8} \ln\left[\frac{4aL}{R \left(a+\sqrt{a^2-1}\right)}\right]
	+ \frac{\left(-a + \sqrt{a^2-1}\right)^{3}}{5} \left[\frac{\left(168 a^{3}+53 a \right) \sqrt{a^{2}-1}}{32}+\frac{19 a^{4}}{4}-\frac{47 a^{2}}{96}-\frac{4}{3}\right]
	\Biggr\}_{|n|=3} \nonumber \\
	&& - \frac{\left(-a +\sqrt{a^{2}-1}\right)^{|n|}}{|n| \left(n^{2}-1\right) \left(n^{2}-4\right) \left(n^{2}-9\right)}
	\Biggl\{a \sqrt{a^{2}-1}\, \left[\left(5 a^{2}-3\right) n^{4}-\left(5 a^{2}-9\right) n^{2}+72\right] |n|
	\nonumber \\
	&& +a^{2} \left(a^{2}-1\right) n^{6}+\left(5 a^{4}-7 a^{2}+4\right) n^{4}-2 \left(3 a^{4}-25 a^{2}+20\right) n^{2}+36 a^{2}+36 \Biggr\}_{|n|>3}
	\scalebox{1.2}{$\Biggr)$} ,
	\label{calc_T3t_A2} \\
	&& \left\langle \cF_{n_2-n,m,1}^{(\Lambda)} \left| \tilde{T}_{3,\bA}^{(\xi)} \right| \cF_{n_2,m',1}^{(\Lambda)} \right\rangle
	= \frac{\delta_{mm'}}{2a^2} \frac{I}{I_0}
	\scalebox{1.2}{$\Biggl($}
	\Biggl\{
	\frac{(1 + 2a^2)}{2} \ln\left[\frac{4aL}{R \left(a+\sqrt{a^2-1}\right)}\right]
	+ \frac{3 a \left(-a +\sqrt{a^{2}-1}\right)}{2}+\frac{1}{4}
	\Biggr\}_{n=0}
	\nonumber \\
	&& + \Biggl\{
	\frac{3a}{8} \ln\left[\frac{4aL}{R \left(a+\sqrt{a^2-1}\right)}\right]
	+ \frac{\left(2 a^{4}-a^{2}+8\right) \left(-a +\sqrt{a^{2}-1}\right)}{24}+\frac{a^{3}}{24}-\frac{25 a}{96}
	\Biggr\}_{|n|=1}
	\nonumber \\
	&& + \Biggl\{
	- \frac{(1 + 2a^2)}{4} \ln\left[\frac{4aL}{R \left(a+\sqrt{a^2-1}\right)}\right]
	-\frac{a \left(4 a^{4}-8 a^{2}+19\right) \left(-a +\sqrt{a^{2}-1}\right)}{20}-\frac{a^{4}}{10}+\frac{7 a^{2}}{40}-\frac{5}{16}
	\Biggr\}_{|n|=2}
	\nonumber \\
	&& + \Biggl\{
	- \frac{ 3a}{8} \ln\left[\frac{4aL}{R \left(a+\sqrt{a^2-1}\right)}\right]
	- \left(-a +\sqrt{a^{2}-1}\right)^{3} \left[\frac{a\left(168 a^{2}+69 \right) \sqrt{a^{2}-1}}{160} + \frac{19 a^{4}}{20} + \frac{13 a^{2}}{96} - \frac{7}{15}\right]
	\Biggr\}_{|n|=3}
	\nonumber \\
	&& + \frac{\left(-a + \sqrt{a^2-1}\right)^{|n|}}{|n| \left(n^{2}-1\right) \left(n^{2}-2\right) \left(n^{2}-3\right)} \Biggl\{
	\left(a^{2}-1\right)^{2} n^{6}
	+ 5a \left(a^{2}-1\right) \sqrt{a^{2}-1}\, |n|^{5}
	+\left(a^{2}-1\right) \left(5 a^{2}+12\right) n^{4}
	\nonumber \\
	&& - a \left(5 a^{2}-41\right) \sqrt{a^{2}-1}\, |n|^{3}
	- \left(6 a^{4}-13 a^{2}-29\right) n^{2}-54 \sqrt{a^{2}-1}\, a |n|
	-18 (2a^{2}-1)
	\Biggr\}_{|n|>3}
	\scalebox{1.2}{$\Biggl)$}
	. \label{calc_T3xiA_fin}
\end{eqnarray}
\end{subequations}
It is interesting to note that both~(\ref{calc_T3t_0}) and (\ref{calc_T3x_0}) do not form hermitian matrices and this happens because $\tilde{T}_{3,0}^{(\theta)}$ and $\tilde{T}_{3,0}^{(\xi)}$ are not hermitian operators.
Moreover, while $\tilde{T}_{3,0}^{(\xi)}$ has matrix elements which are antisymmetric in $n$, whereas the sum $\tilde{T}_{3,0}^{(\theta)} + \tilde{T}_{3,0}^{(\xi)}$ is hermitian, we have the relations:
\begin{subequations} \label{sym_T3}
\begin{eqnarray}
	&& \langle \cF_{n_2-n,m,1}^{(\Lambda)} | \tilde{T}_{3,0}^{(\xi)} | \cF_{n_2,m',1}^{(\Lambda)} \rangle
	= - \langle \cF_{n_2,m,1}^{(\Lambda)} | \tilde{T}_{3,0}^{(\xi)} | \cF_{n_2-n,m',1}^{(\Lambda)} \rangle
	\label{T3x_asym} \\
	&& \left\langle \cF_{n_2-n,m,1}^{(\Lambda)} \left| \tilde{T}_{3,0}^{(\xi)} + \tilde{T}_{3,0}^{(\theta)} \right| \cF_{n_2,m',1}^{(\Lambda)} \right\rangle
	= - \delta_{mm'} \left[
	\left.\frac{3 \left(2n_{2} -n\right)}{8}\right|_{|n|=2}
	+ \left. \frac{\left(a^{2}+1\right) (2n_2 - n)}{2a}\right|_{|n|=1}
	+ \left. \frac{5n_2}{2} \right|_{n=0}
	\right] \nonumber \\
	&& = \left\langle \cF_{n_2-n,m,1}^{(\Lambda)} \left| \tilde{T}_3^{(\theta)} (I = 0) \right| \cF_{n_2,m',1}^{(\Lambda)} \right\rangle
	= \left\langle \cF_{n_2-n,m,1}^{(\Lambda)} \left| \tilde{T}_{3,0}^{(\theta)} \right| \cF_{n_2,m',1}^{(\Lambda)} \right\rangle
	+ \left\langle \cF_{n_2,m,1}^{(\Lambda)} \left| \tilde{T}_{3,0}^{(\theta)} \right| \cF_{n_2-n,m',1}^{(\Lambda)} \right\rangle
	, \label{T30}
\end{eqnarray}
\end{subequations}
where $\tilde{T}_3^{(\theta)} (I = 0) \equiv \tilde{T}_{3,0}^{(\theta)} - \tilde{T}_\theta$~(\ref{gen_T3_theta_xi}) is the toroidal dipole at $\bA=0$~\cite{PhysicaA.598.127377.2022.Dolineanu}.

From Eqs.~(\ref{T3_matrices}) and (\ref{calc_T3tx_A2}) we have
\begin{eqnarray}
	&& \left\langle \cF_{n_2-n,m,1}^{(\Lambda)} \left| \tilde{T}_{3,\bA}^{(\theta)} + \tilde{T}_{3,\bA}^{(\xi)} \right| \cF_{n_2,m',1}^{(\Lambda)} \right\rangle
	=
	\delta_{mm'} \frac{I}{2a^2I_0}
	\left\{\ln(\frac{2aL}{R})\left[\frac{\delta_{n+2} + \delta_{n-2}}{4} + 2a(\delta_{n+1} + \delta_{n-1}) + \frac{3+4a^2}{2}\delta_{n}\right]
	\right. \nonumber \\
	&& \left. - \left[\frac{I_{n+2}^{(\ln)}(a) + I_{n-2}^{(\ln)}(a)}{4} + 2a(I_{n+1}^{(\ln)}(a) + I_{n-1}^{(\ln)}(a)) + \frac{3+4a^2}{2}I_{n}^{(\ln)}(a)\right]\right\}
	\nonumber \\ 
	&&
	=
	\delta_{mm'} \frac{I}{2a^2I_0} \left(
	\frac{1}{4} \left\{ \ln\left[\frac{4aL}{R \left(a+\sqrt{a^2-1}\right)}\right]
	+ \frac{\left(-a +\sqrt{a^{2}-1}\right)^{2} \left(-74 a^{2}+33-70 a \sqrt{a^{2}-1}\right)}{12}
	\right\}_{|n|=2}
	\right. \nonumber \\
	&&
	+ \left\{
	2a \ln\left[\frac{4aL}{R \left(a+\sqrt{a^2-1}\right)}\right]
	+ \frac{\left(-a +\sqrt{a^{2}-1}\right) \left(7 a^{2}+10+5 a \sqrt{a^{2}-1}\right)}{6}
	\right\}_{|n|=1}
	\nonumber \\
	&& + \left\{ \frac{3+4a^2}{2} \ln\left[\frac{4aL}{R \left(a+\sqrt{a^2-1}\right)}\right]
	+ \frac{\left(-a +\sqrt{a^{2}-1}\right) \left(15 a +\sqrt{a^{2}-1}\right)}{4}
	\right\}_{n=0}
	\nonumber \\
	&& \left. + \left(-a + \sqrt{a^2-1}\right)^{|n|} \left\{
	\frac{- \left(a^{2}-1\right) n^{4} - 2a\sqrt{a^{2}-1} |n|^{3}+\left(5 a^{2}-7\right) n^{2} + 14a \sqrt{a^{2}-1} |n| +8 a^{2}+6}{|n| (n^2-1) (n^{2}-4)}
	\right\}_{|n|>2}
	\right) .
	\label{T3A_fin}
\end{eqnarray}

Using the Eqs.~(\ref{T3_p_theta_U}) and (\ref{sym_T3}) we obtain
\begin{equation}
	\tilde{T}_3^{(\cU)} \equiv \cU \tilde{T}_3 \cU^{-1}
	= \left. \tilde{T}_3^{(\theta)} \right|_{\bA = 0}
	+ \tilde{T}_{3,\bA}^{(\xi)}
	= \left. \tilde{T}_3^{(\theta)} \right|_{\bA = 0} + \frac{2a^2 + 3a\cos\theta + 1}{a^2} \sin\theta \tilde{A}_\xi .
	\label{T2_cU}
\end{equation}

\subsection{Numerical results} \label{subsec_num_res}

Hereafter, we explicitly emphasize the current dependence of operators by using expressions such as $\tilde{\mathcal{H}}^\text{(2D)}(I)$, $\tilde{\mathcal{H}}(I)$, $\tilde{T}_3^\theta(I)$, and $\tilde{T}_3(I)$.  For a fixed wavevector index $k=1$ and each azimuthal quantum number $m$, we numerically diagonalize the Hamiltonians $\tilde{\mathcal{H}}(I)$ and $\tilde{\mathcal{H}}^\text{(2D)}(I)$ within an 801-dimensional space spanned by the basis functions $\left\{\mathcal{F}_{n_2,m,1}^{(\Lambda)}\right\}$, where $n_2$ ranges from -400 to 400.  The energy levels are indexed by $l = 0, 1, 2, \ldots$, where $l=0$ corresponds to the ground state, and increasing values of $l$ represent successively higher energy levels. This basis dimensionality is sufficient to ensure numerical convergence for the examples presented below.

\begin{figure}[t]
	\centering
	\includegraphics[width=55mm]{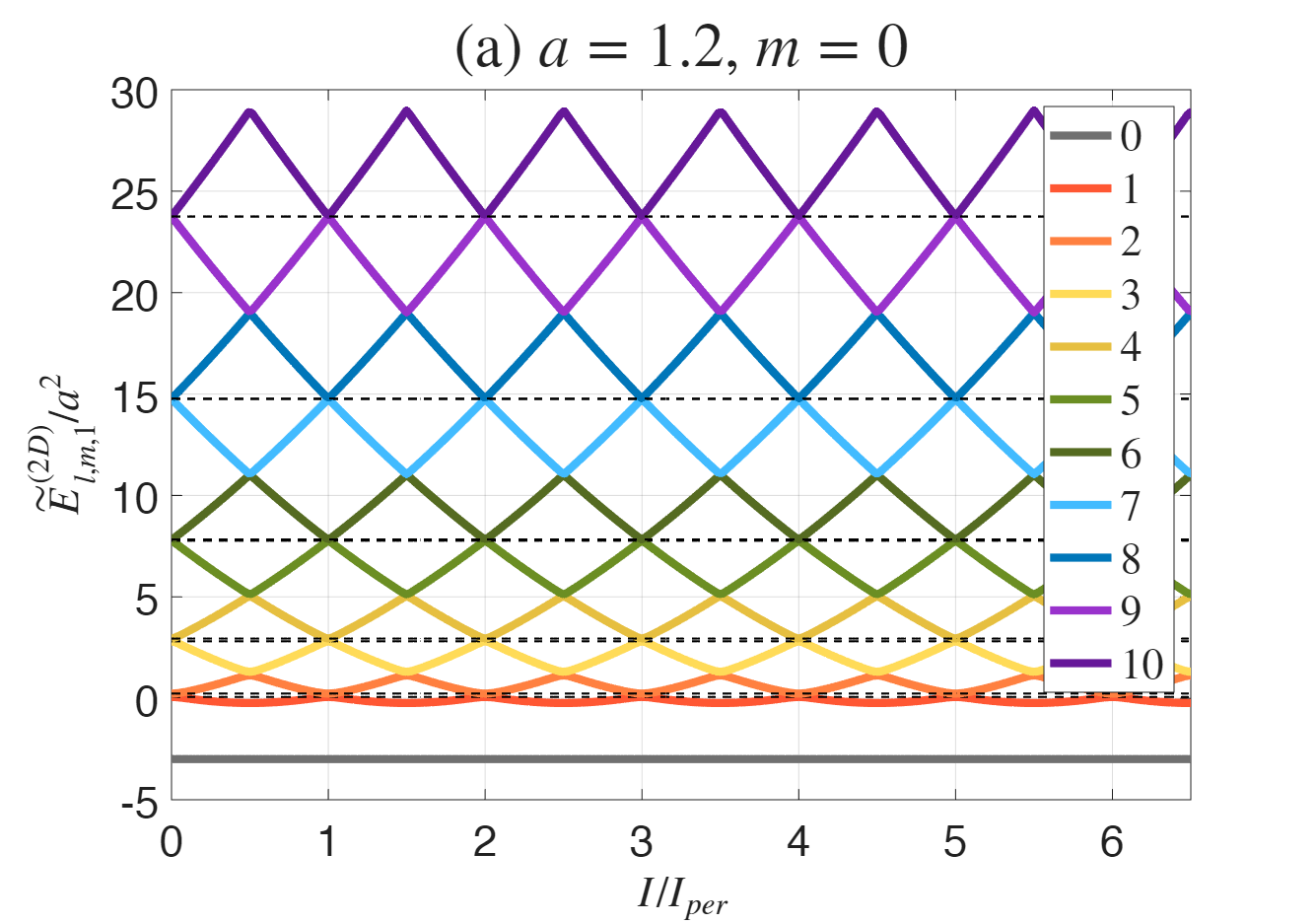}
	\includegraphics[width=55mm]{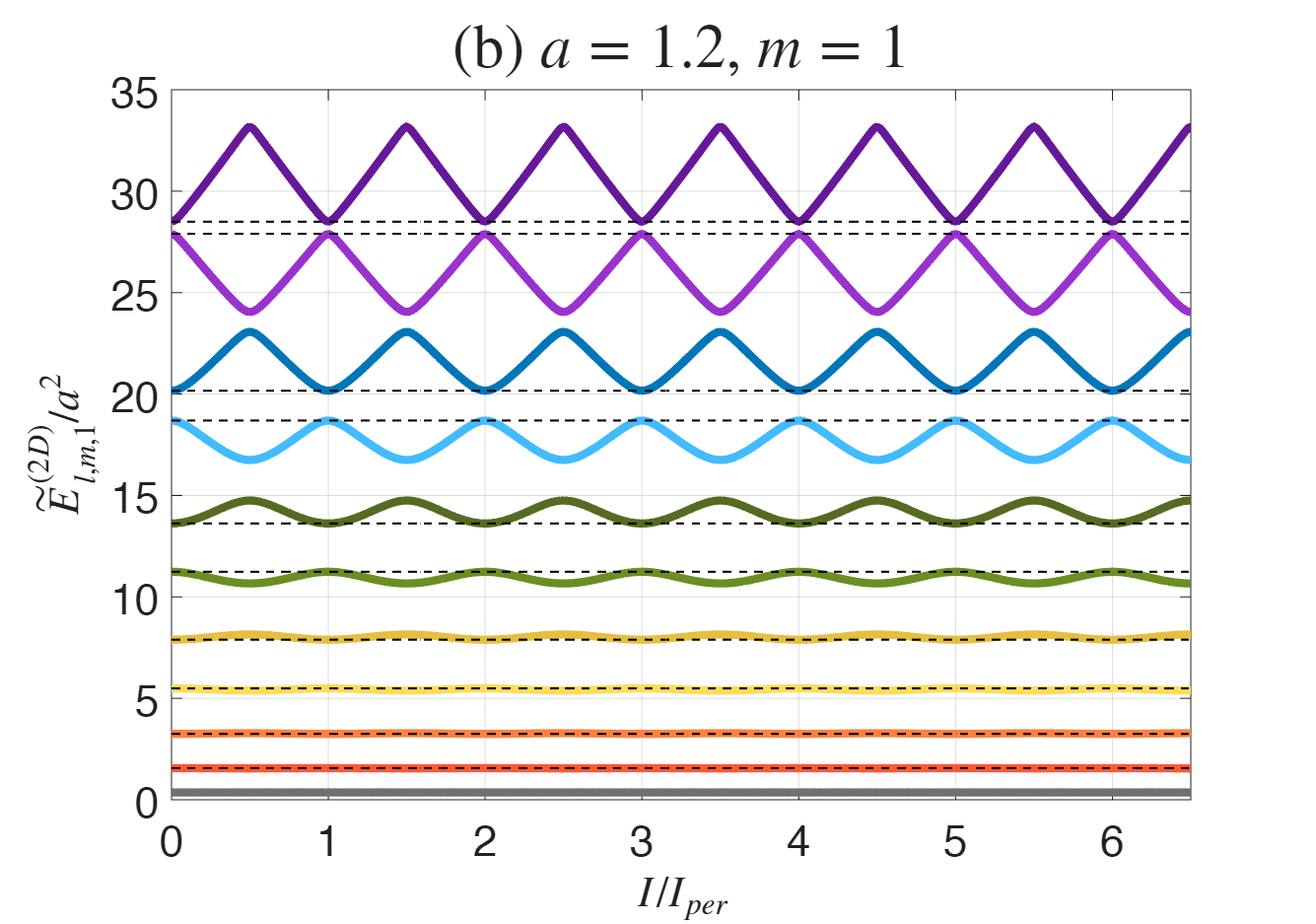}
	\includegraphics[width=55mm]{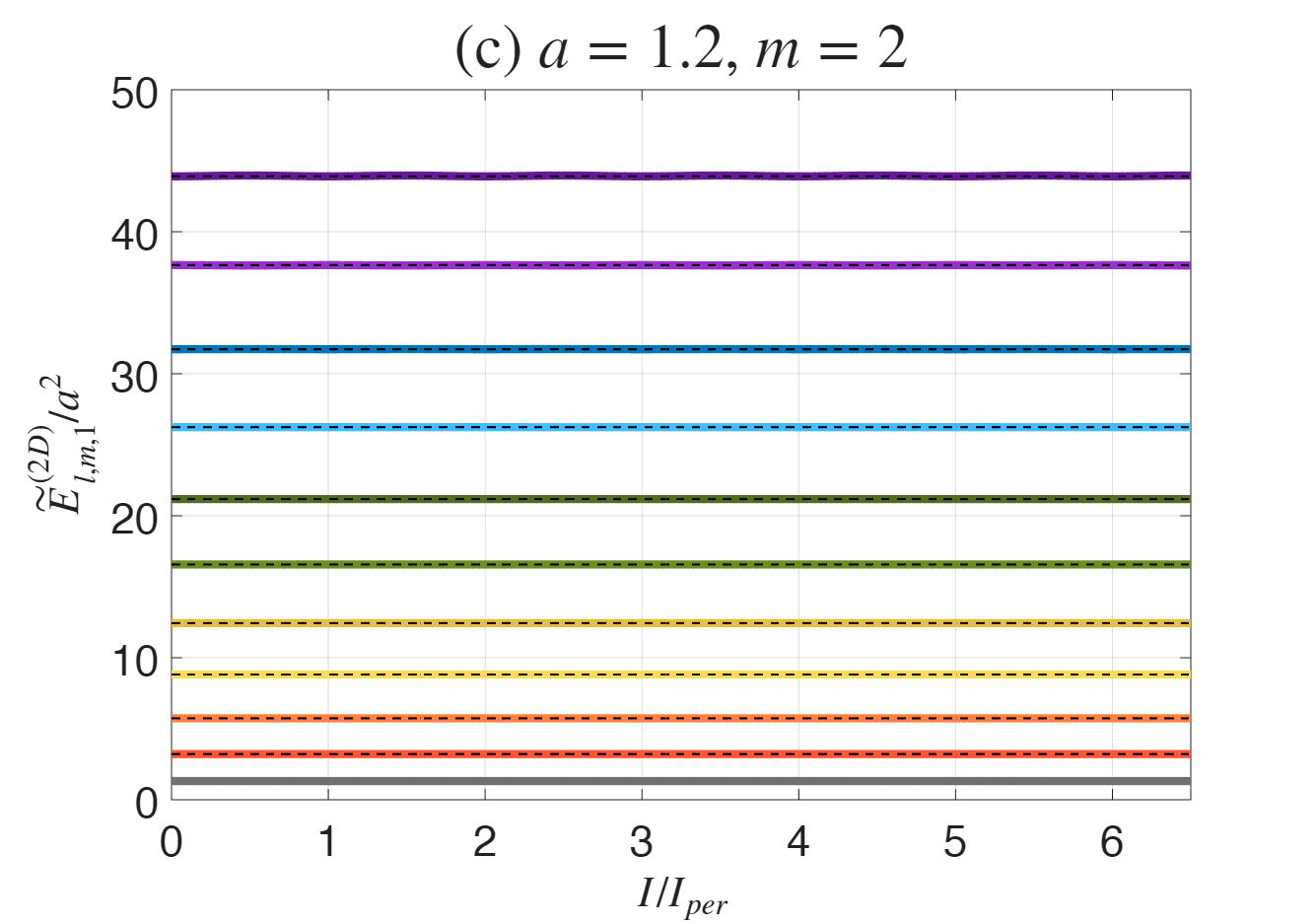}
	\includegraphics[width=55mm]{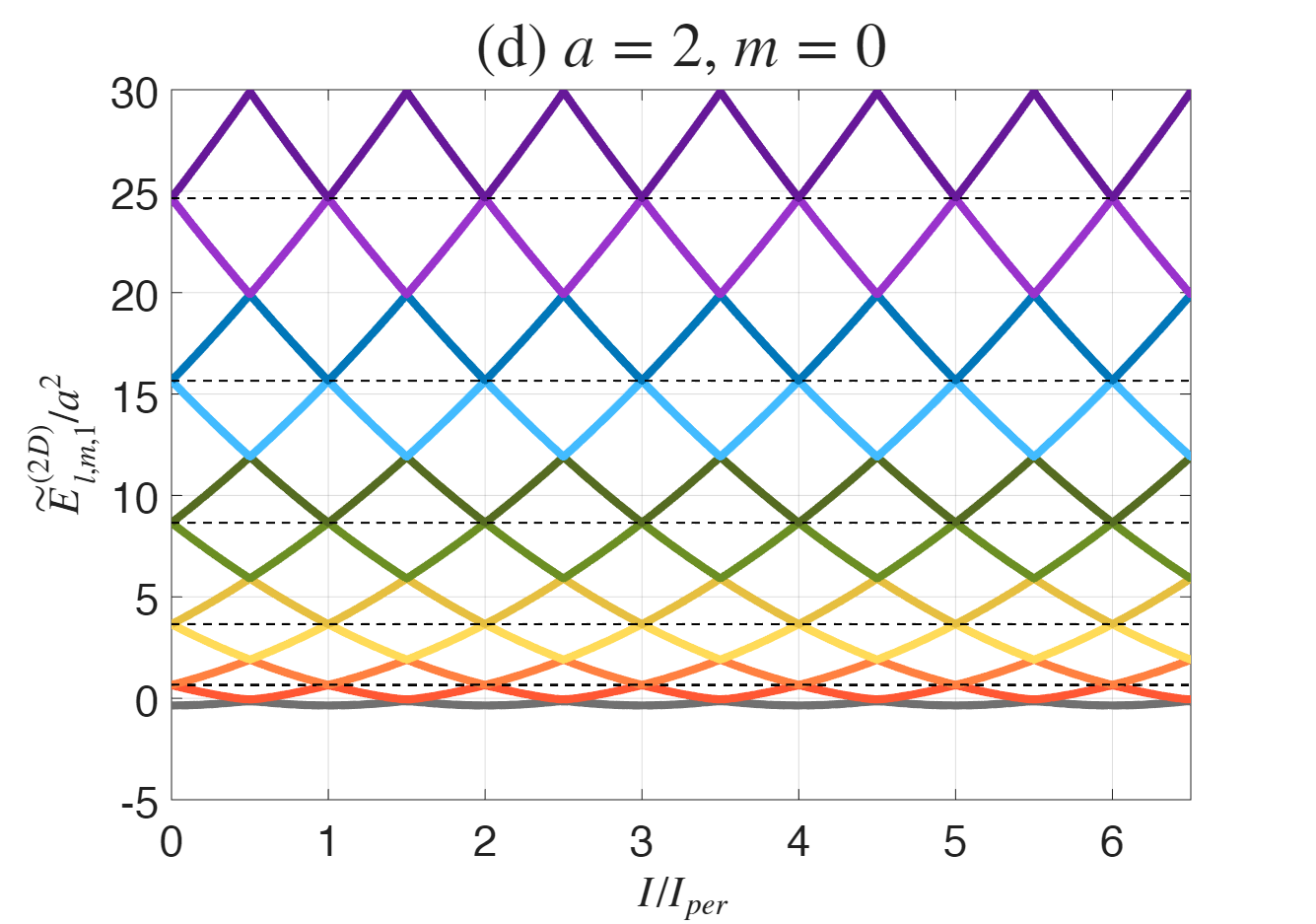}
	\includegraphics[width=55mm]{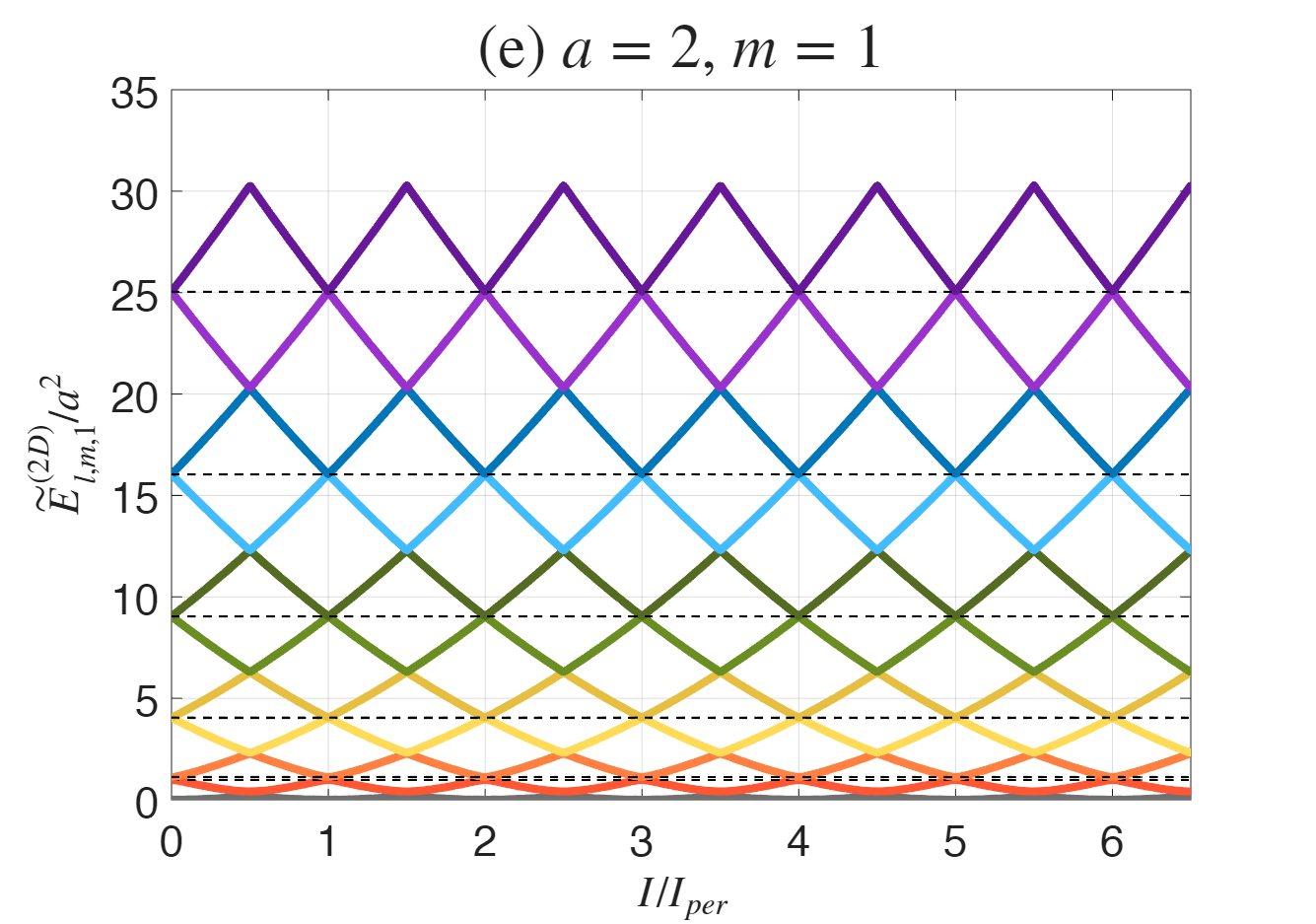}
	\includegraphics[width=55mm]{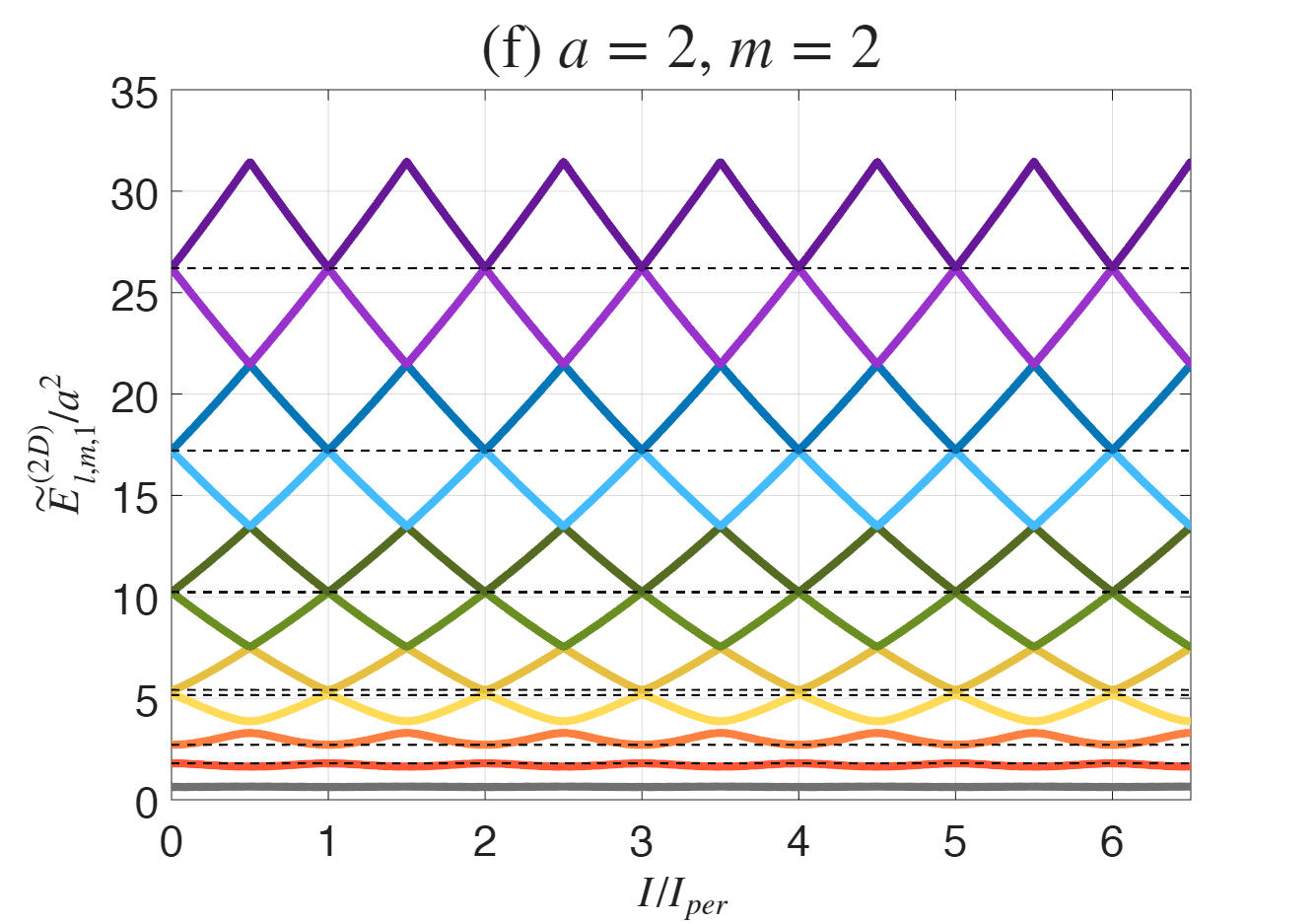}
	\includegraphics[width=55mm]{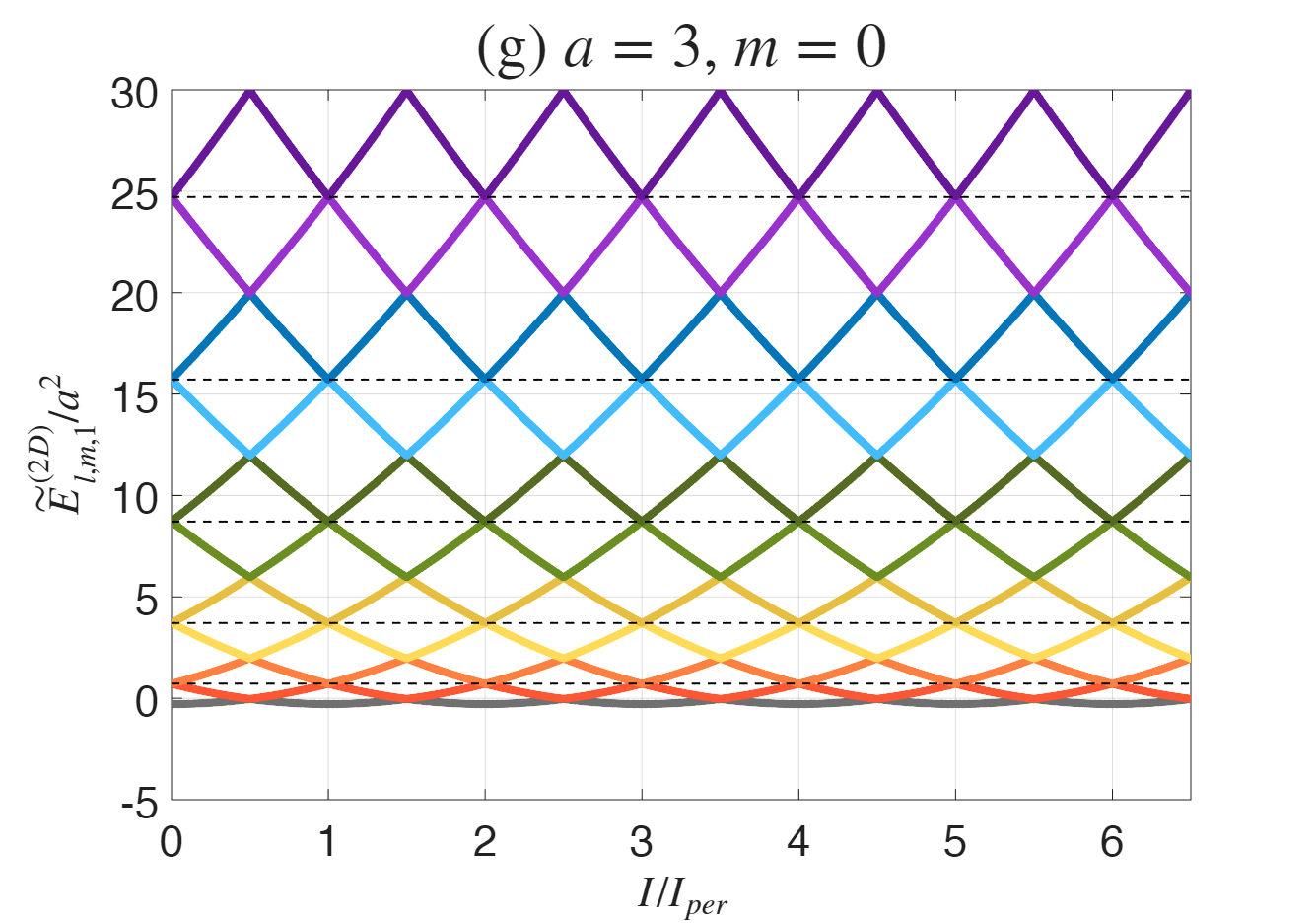}
	\includegraphics[width=55mm]{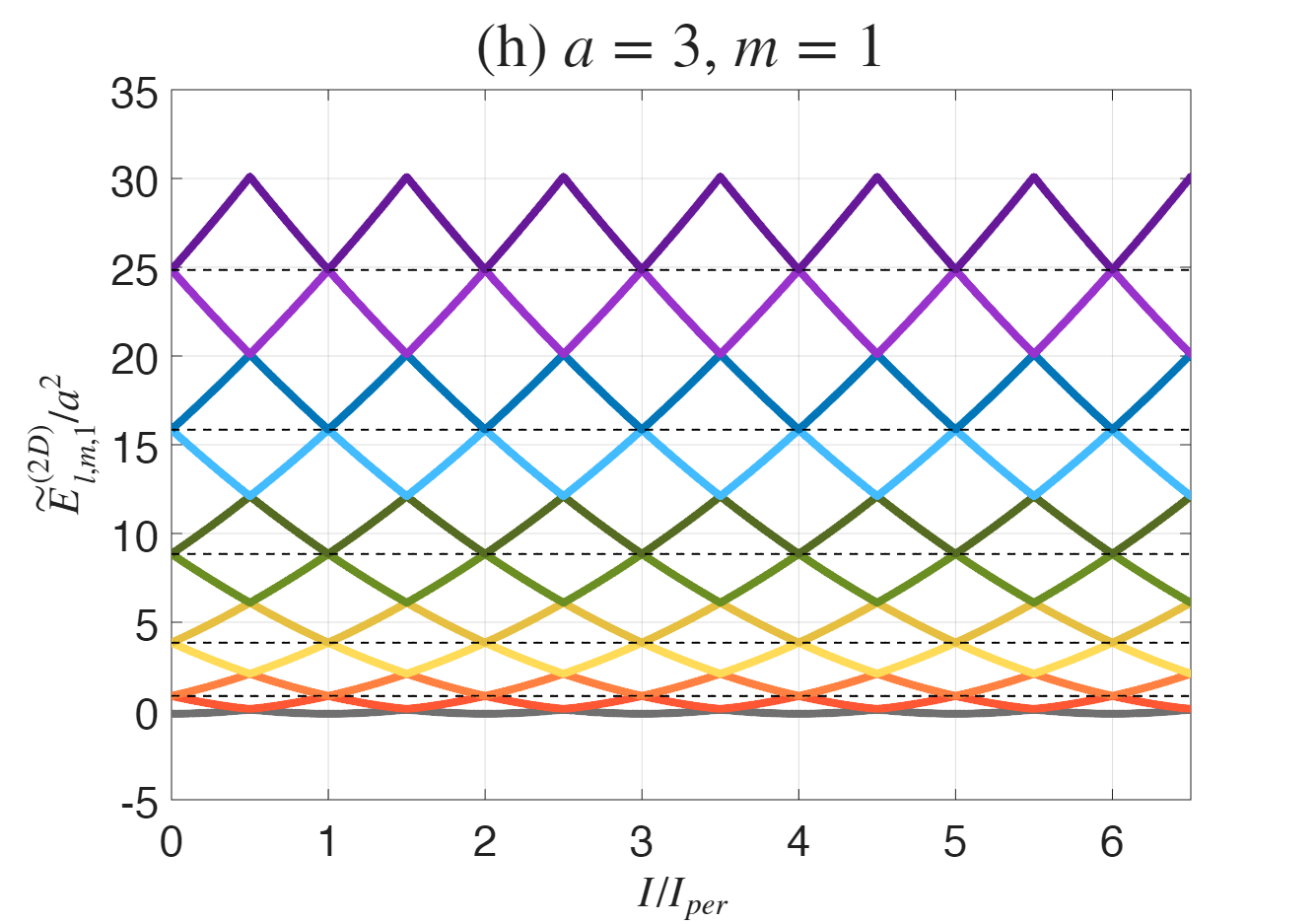}
	\includegraphics[width=55mm]{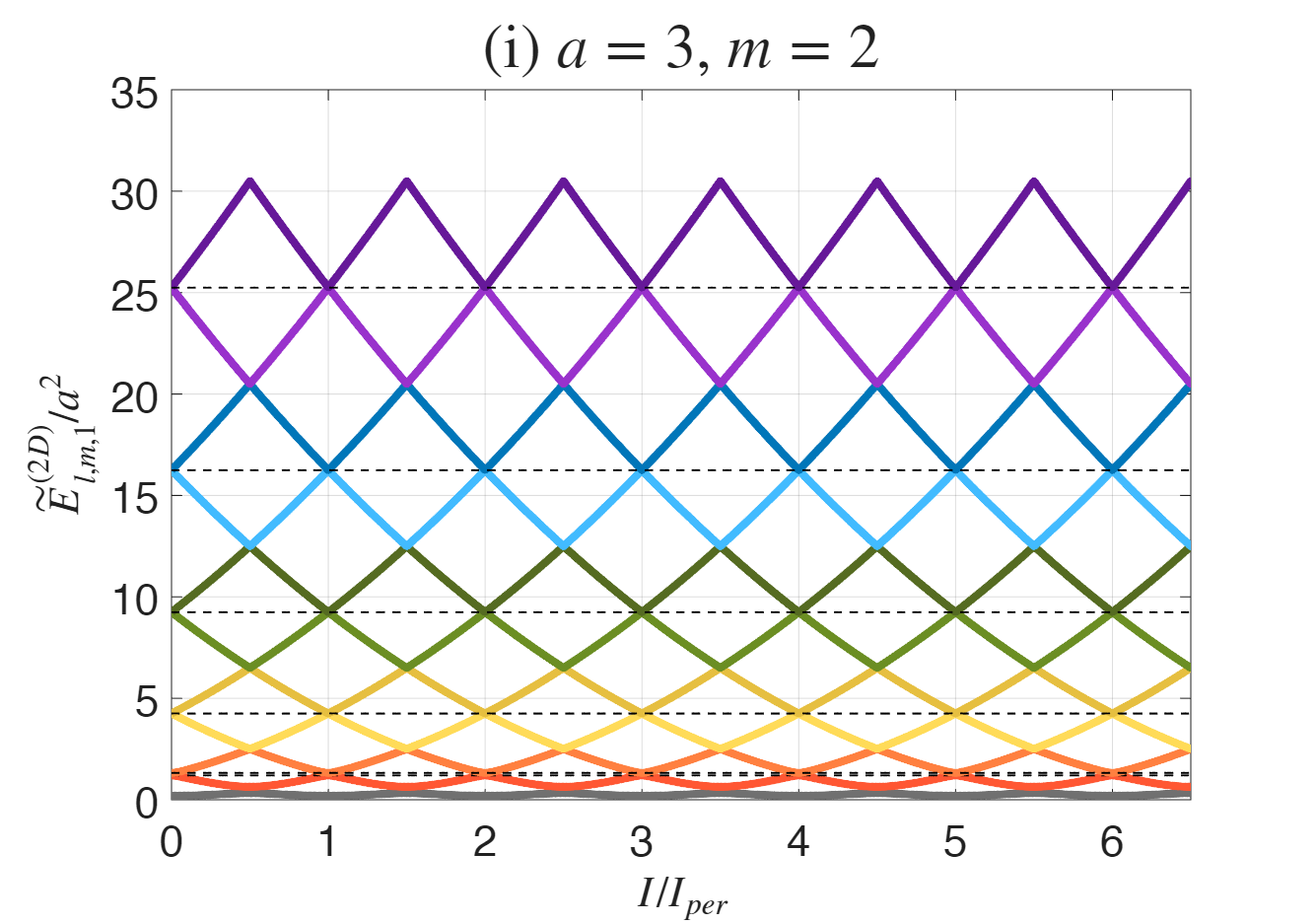}
	\caption{The first eleven eigenvalues of $\tilde{H}^{(2D)}$, $\tilde{E}^{(2D)}_{l,m,1}$ (with $l=0,1,\ldots,10$), scaled by $a^2$, are shown as functions of $I/I_{per}$, with $a = 1.2, 2, 3$ and $m = 0, 1, 2$, as indicated in each plot. The legend, shown in (a), is the same for every plot: 0 represents the ground state, 1 the first excited state, and so on.
	}
	\label{fig_H_2D}
\end{figure}

\begin{figure}[t]
	\centering
	\includegraphics[width=55mm]{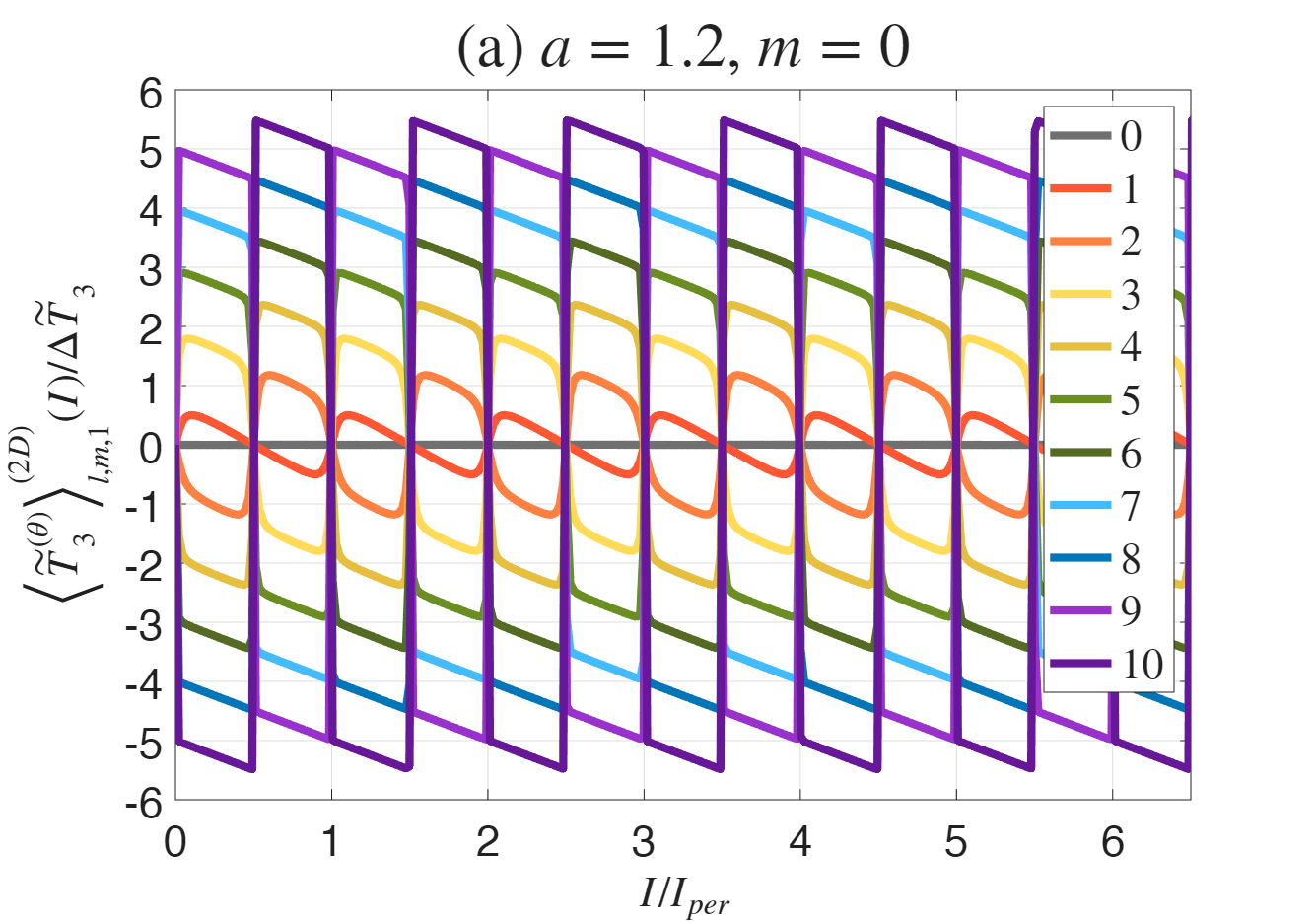}
	\includegraphics[width=55mm]{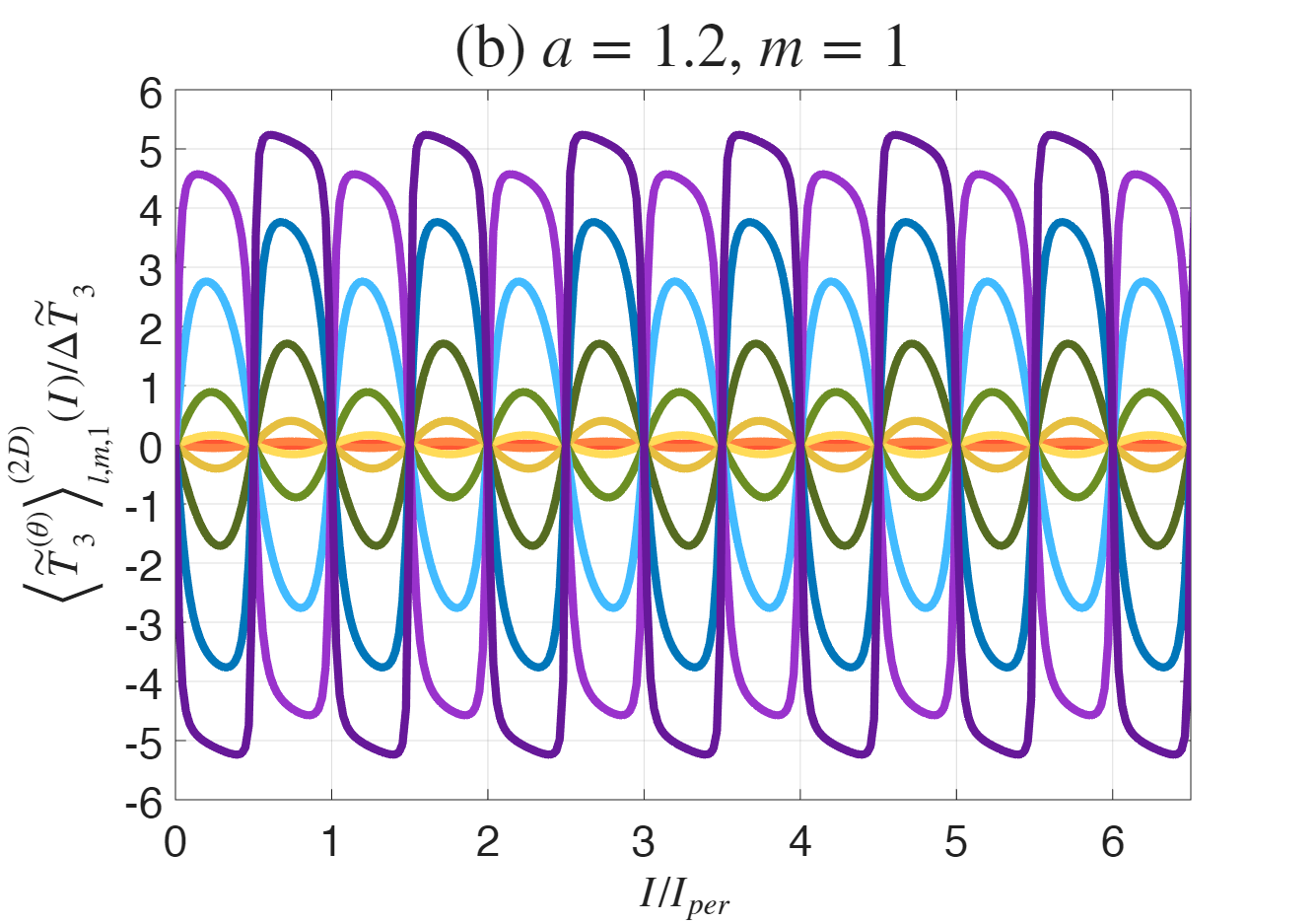}
	\includegraphics[width=55mm]{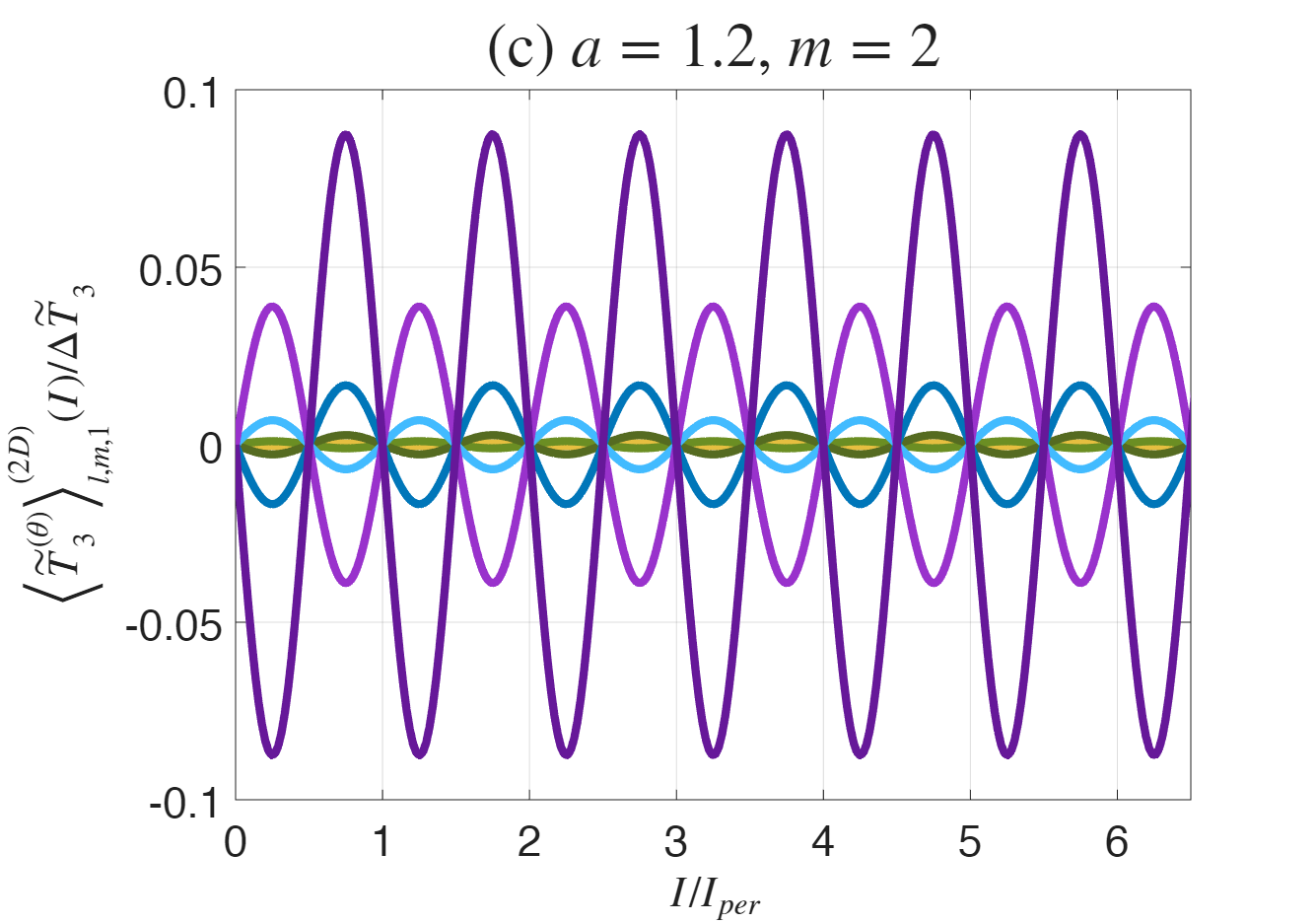}
	\includegraphics[width=55mm]{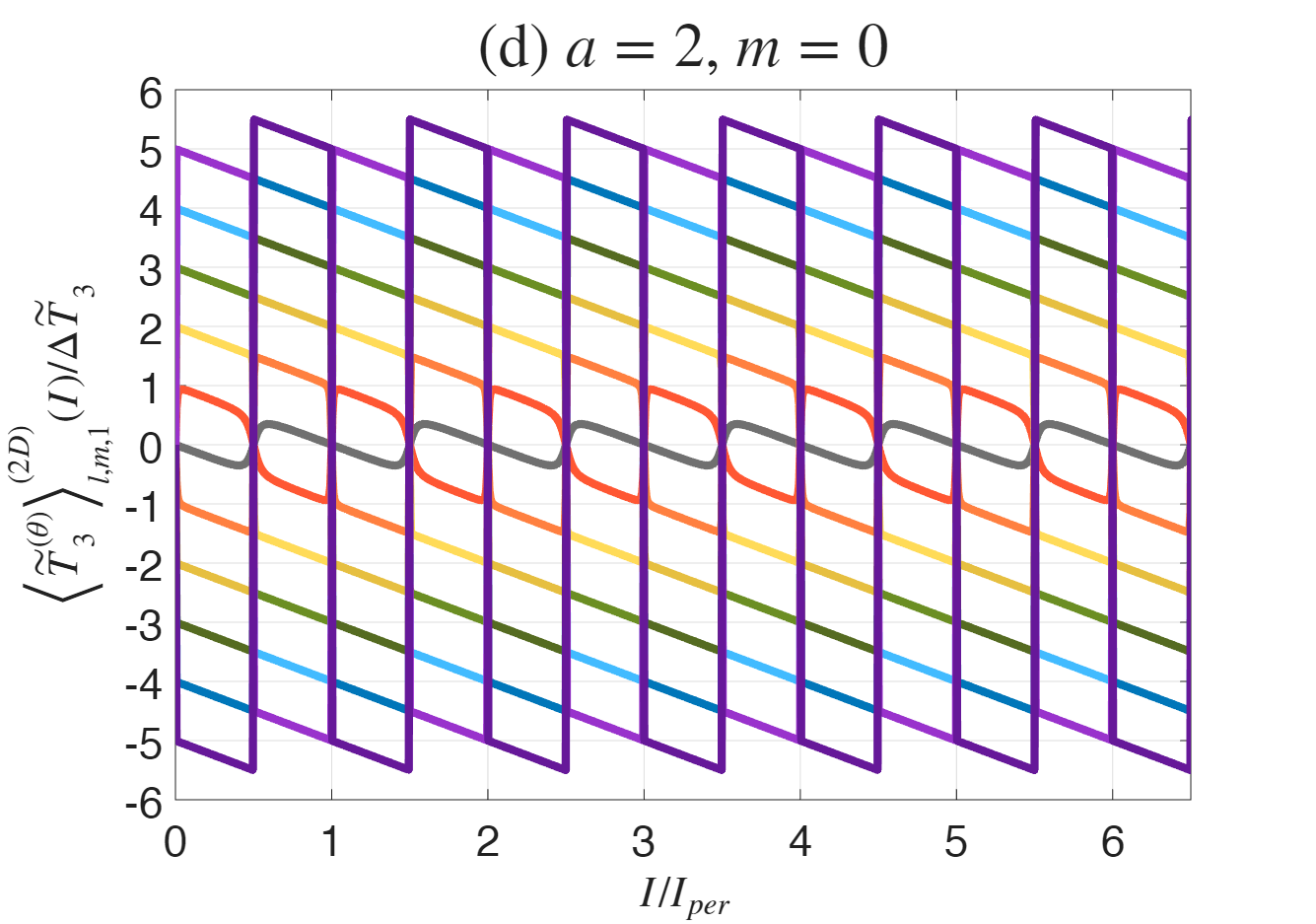}
	\includegraphics[width=55mm]{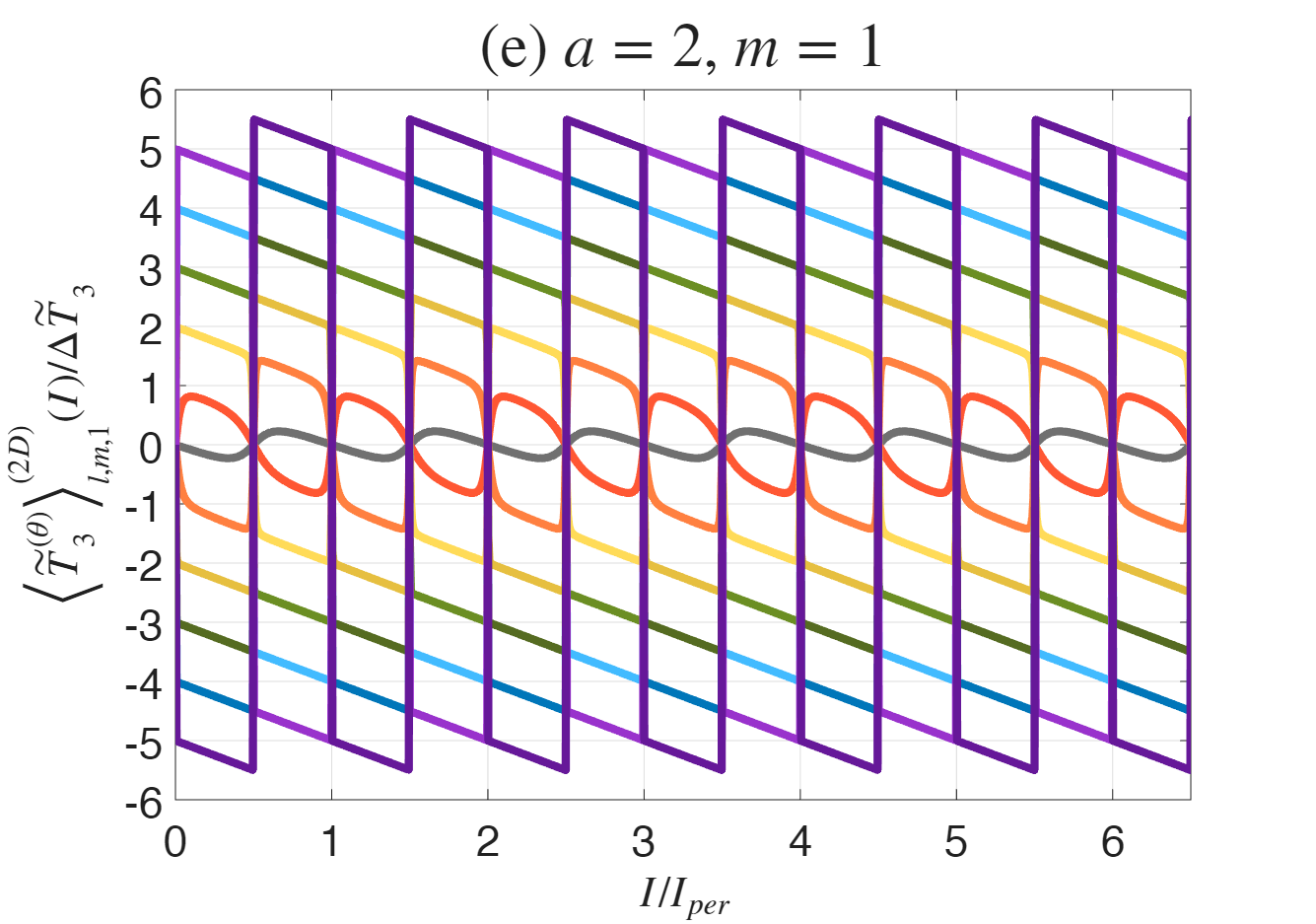}
	\includegraphics[width=55mm]{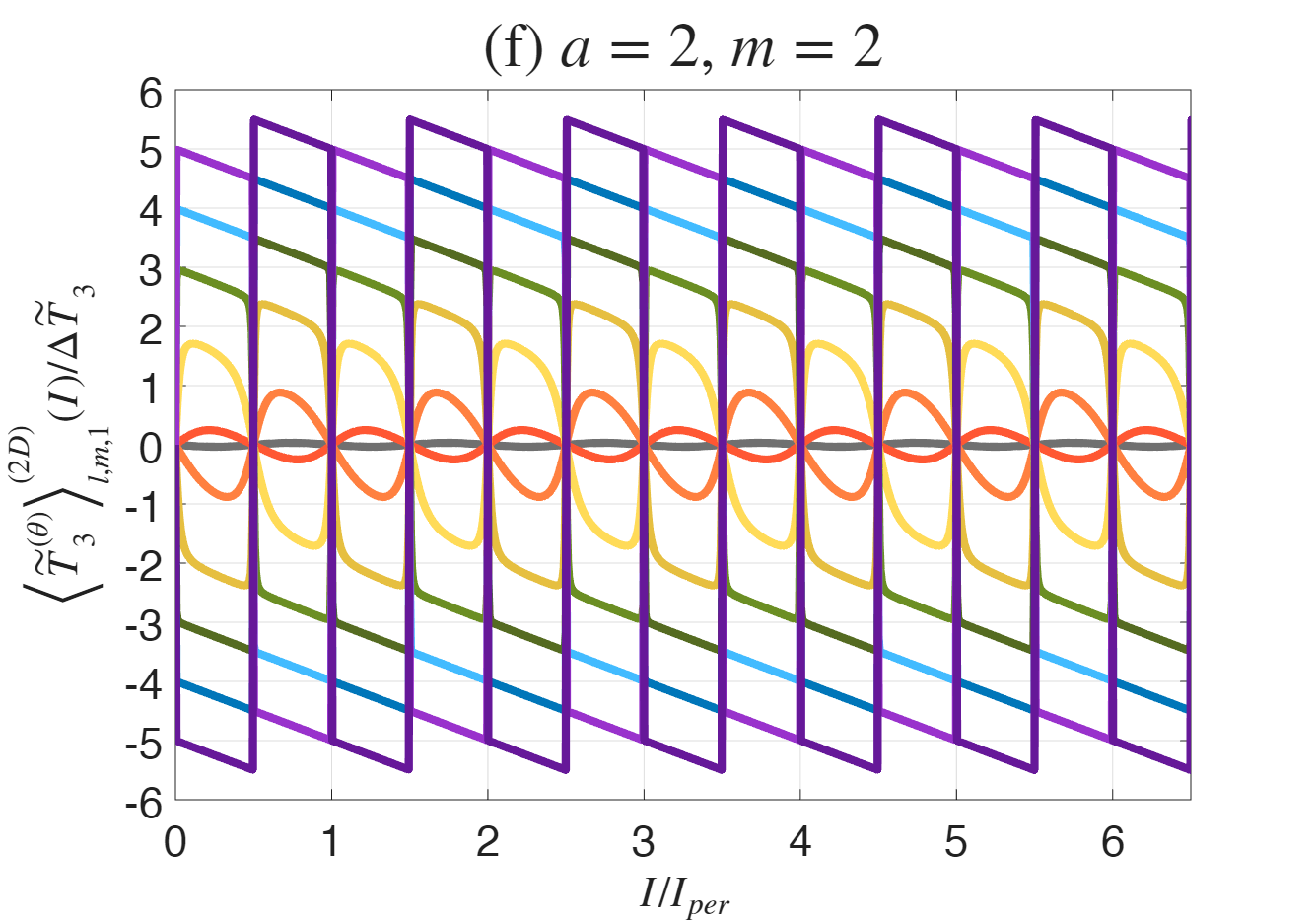}
	\includegraphics[width=55mm]{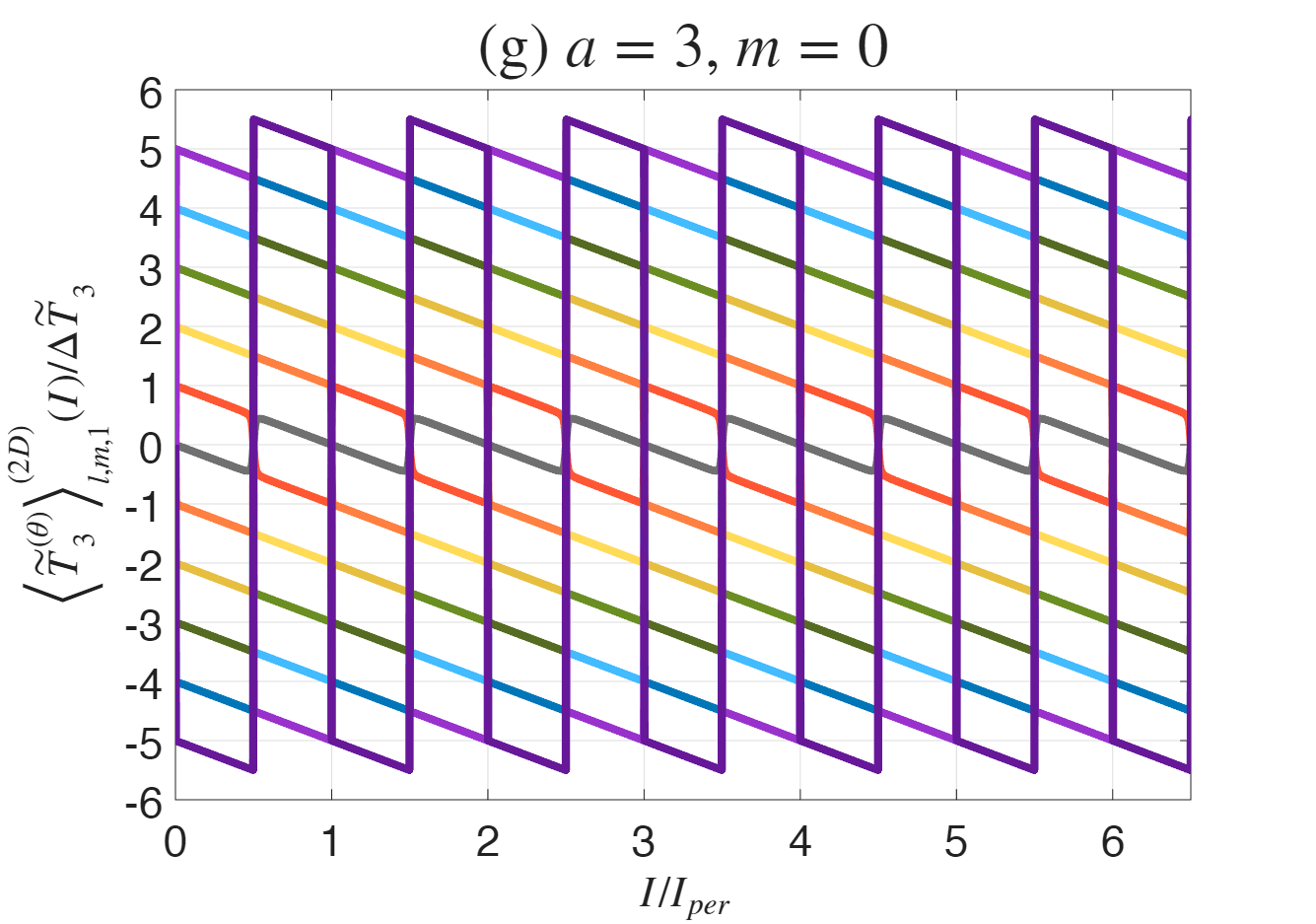}
	\includegraphics[width=55mm]{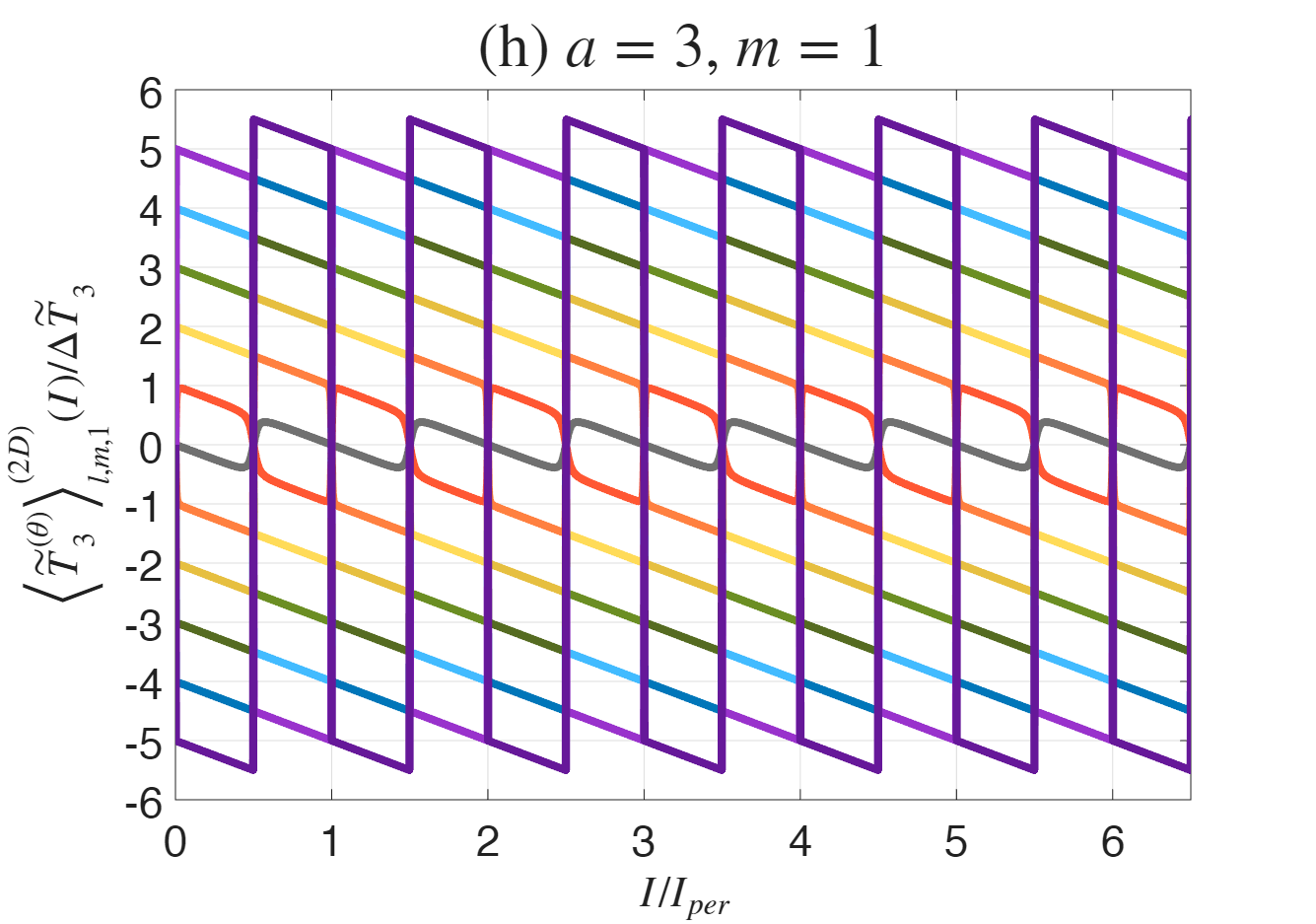}
	\includegraphics[width=55mm]{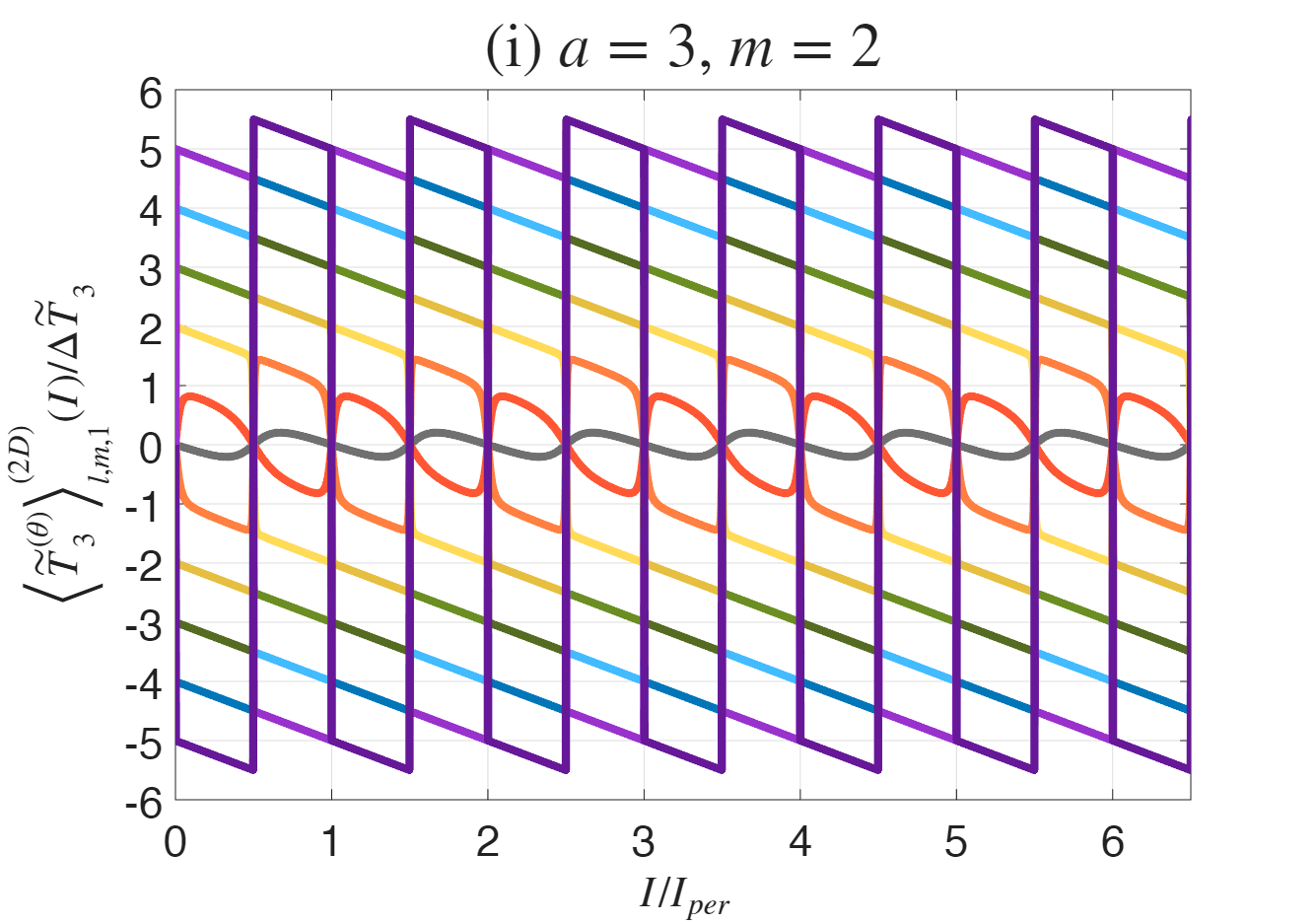}
	\caption{The expectation values $\left\langle \vcenter{\hbox{$\tilde{T}^{(\theta)}_3$}} \right\rangle^{\vcenter{\hbox{\scriptsize (2D)}}}_{\vcenter{\hbox{\scriptsize l,m,1}}}(I)$, scaled by $\Delta\tilde{T}_3 = 2.5$, vs. $I/I_{per}$ for the first eleven eigenstates, $a = 1.2, 2, 3$, and $m = 0, 1, 2$, as indicated in each plot. The legend, shown in (a), is the same in every plot: 0 represents the ground state, 1 the first excited state, and so on.
	}
	\label{fig_T3_2D}
\end{figure}

First, we illustrate the periodicity of the 2D dynamics, as proven in Section~\ref{subsec_periodicity_2D}.  Let $\tilde{E}_{l,m,1}^{(2D)}$ and $| \tilde{\psi}_{l,m,1}^\text{(2D)}(I) \rangle$ be the eigenvalues and eigenfunctions of the 2D Hamiltonian $\tilde{\mathcal{H}}^\text{(2D)}(I)$ for different values of $a$ and $m$ (with the ground state corresponding to $l=0$):
\begin{subequations} \label{2D_probl}
\begin{equation}
	\tilde{\mathcal{H}}^\text{(2D)}(I) |\tilde{\psi}_{l,m,1}^\text{(2D)}(I) \rangle = \tilde{E}_{l,m,1}^{(2D)} |\tilde{\psi}_{l,m,1}^\text{(2D)}(I) \rangle,
	\label{2D_probl_H}
\end{equation}
where $\tilde{E}_{l,m,1}^{(2D)} \le \tilde{E}_{l',m,1}^{(2D)}$ for any $0 \le l < l'$.  Figure~\ref{fig_H_2D} displays $\tilde{E}_{l,m,1}^{(2D)}/a^2$, whereas Fig.~\ref{fig_T3_2D} shows the expectation values of $\tilde{T}_3^{(\theta)}$, normalized to $\Delta\tilde{T}_3=2.5$~(\ref{T3_I_approx}), on the eigenstates $| \tilde{\psi}_{l,m,1}^\text{(2D)}(I) \rangle$:
\begin{equation}
	\frac{\langle\tilde{T}_3^{(\theta)}\rangle_{l,m,1}^\text{(2D)}(I)}{\Delta\tilde{T}_3} \equiv \frac{\langle \tilde{\psi}_{l,m,1}^\text{(2D)}(I) | \tilde{T}_3(I) | \tilde{\psi}_{l,m,1}^\text{(2D)}(I) \rangle}{\Delta\tilde{T}_3}
	. \label{2D_probl_T3}
\end{equation}
\end{subequations}

\begin{figure}[t]
	\centering
	\includegraphics[width=0.48\linewidth]{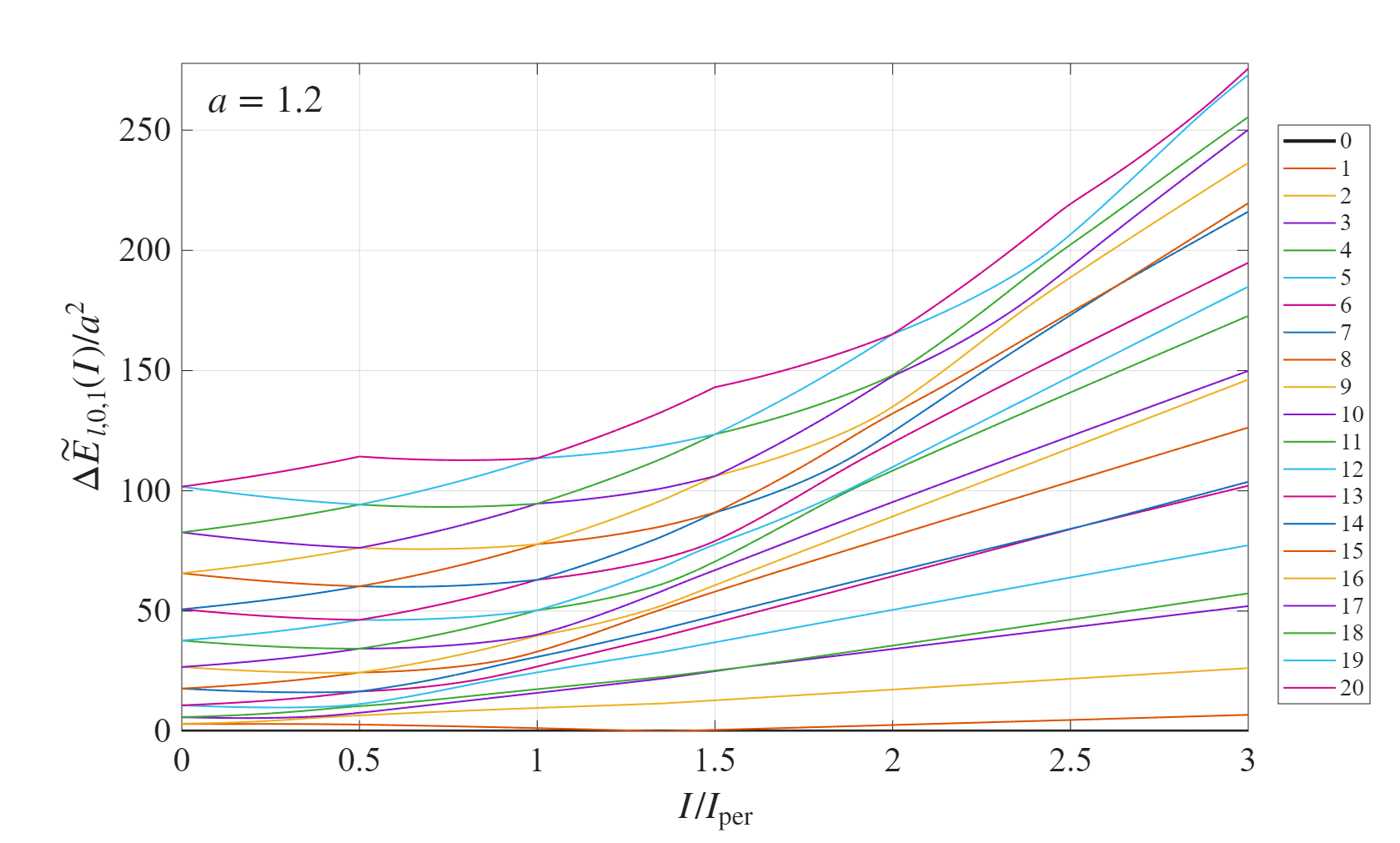}
	\includegraphics[width=0.48\linewidth]{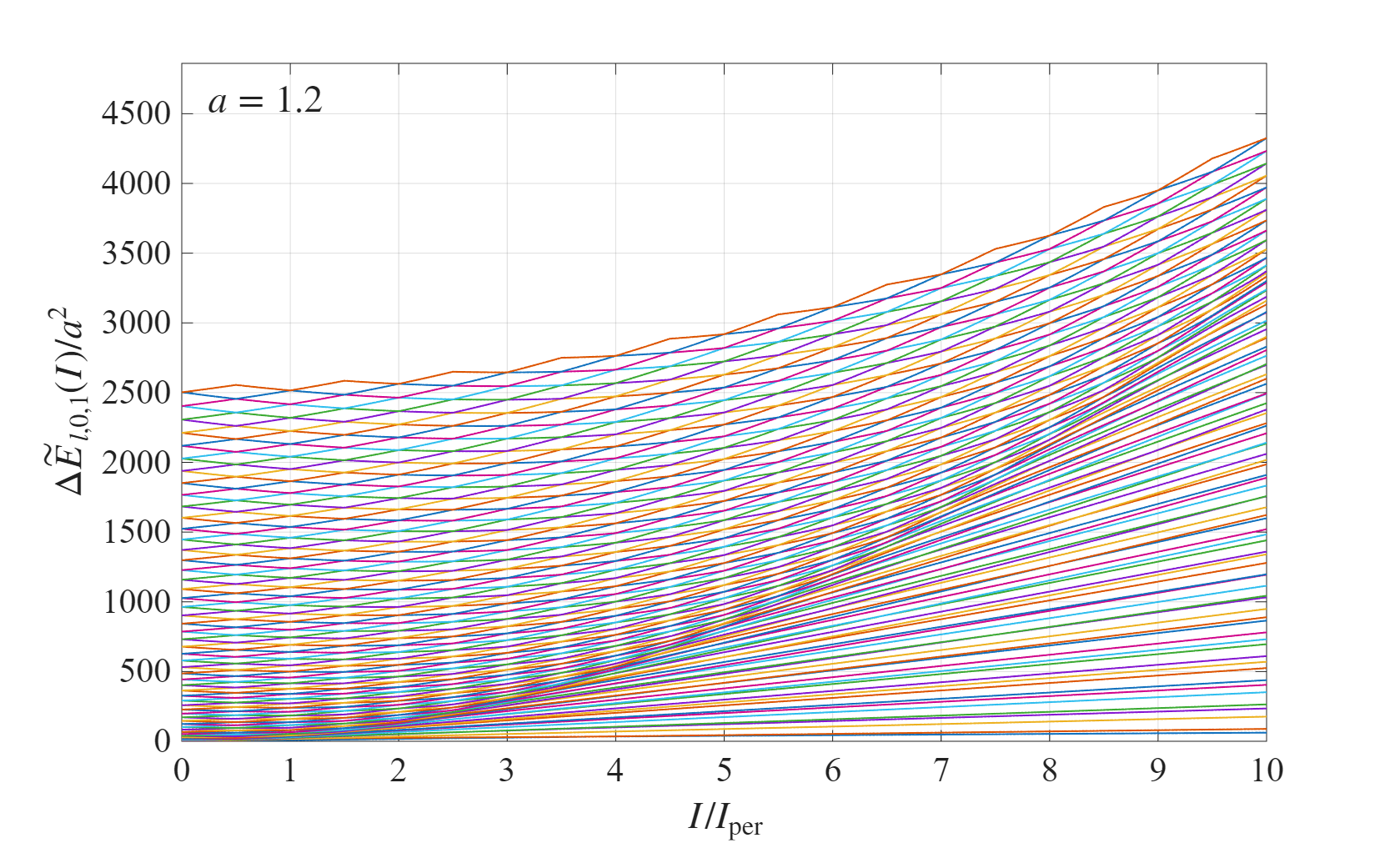}
	\includegraphics[width=0.48\linewidth]{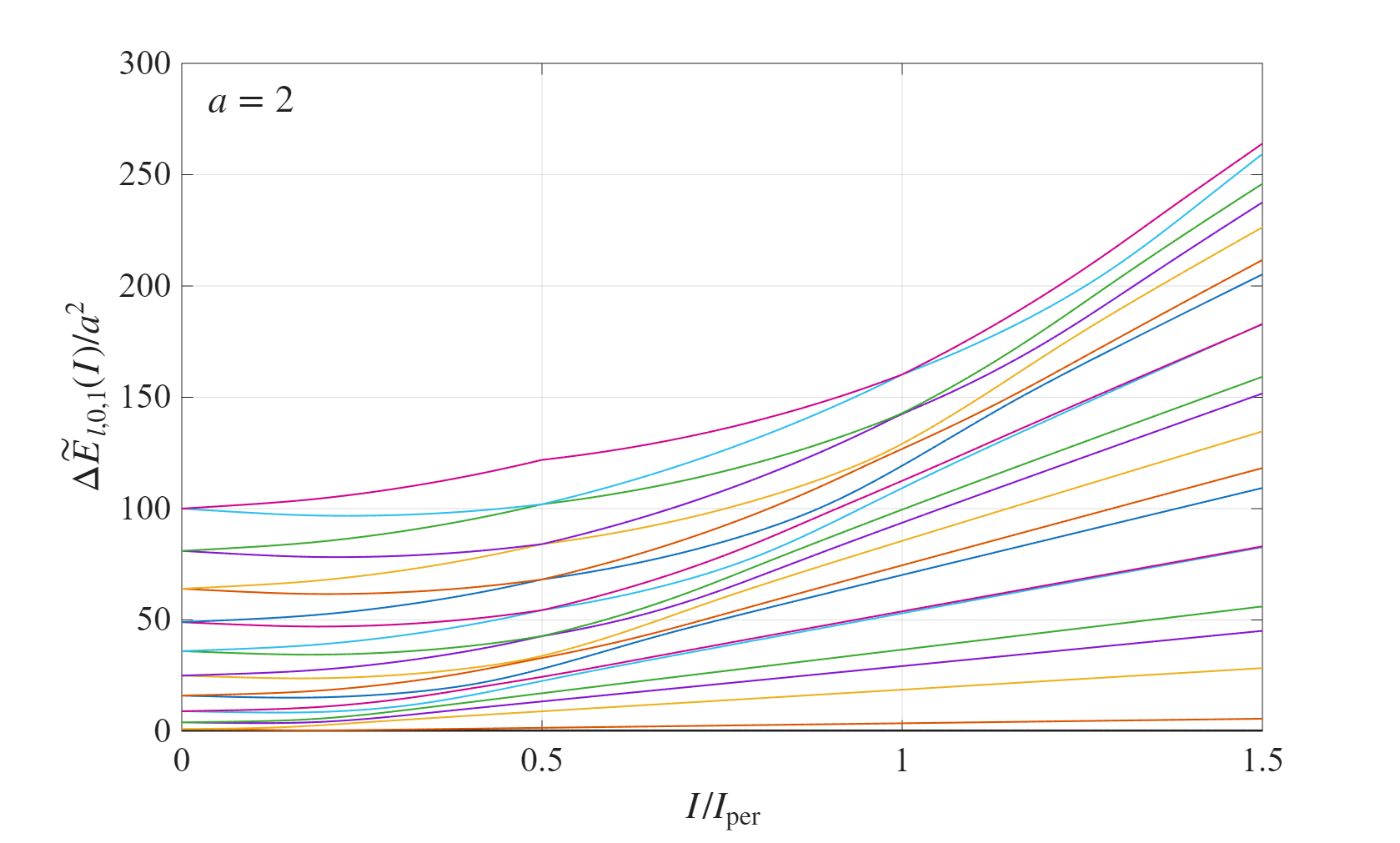}
	\includegraphics[width=0.48\linewidth]{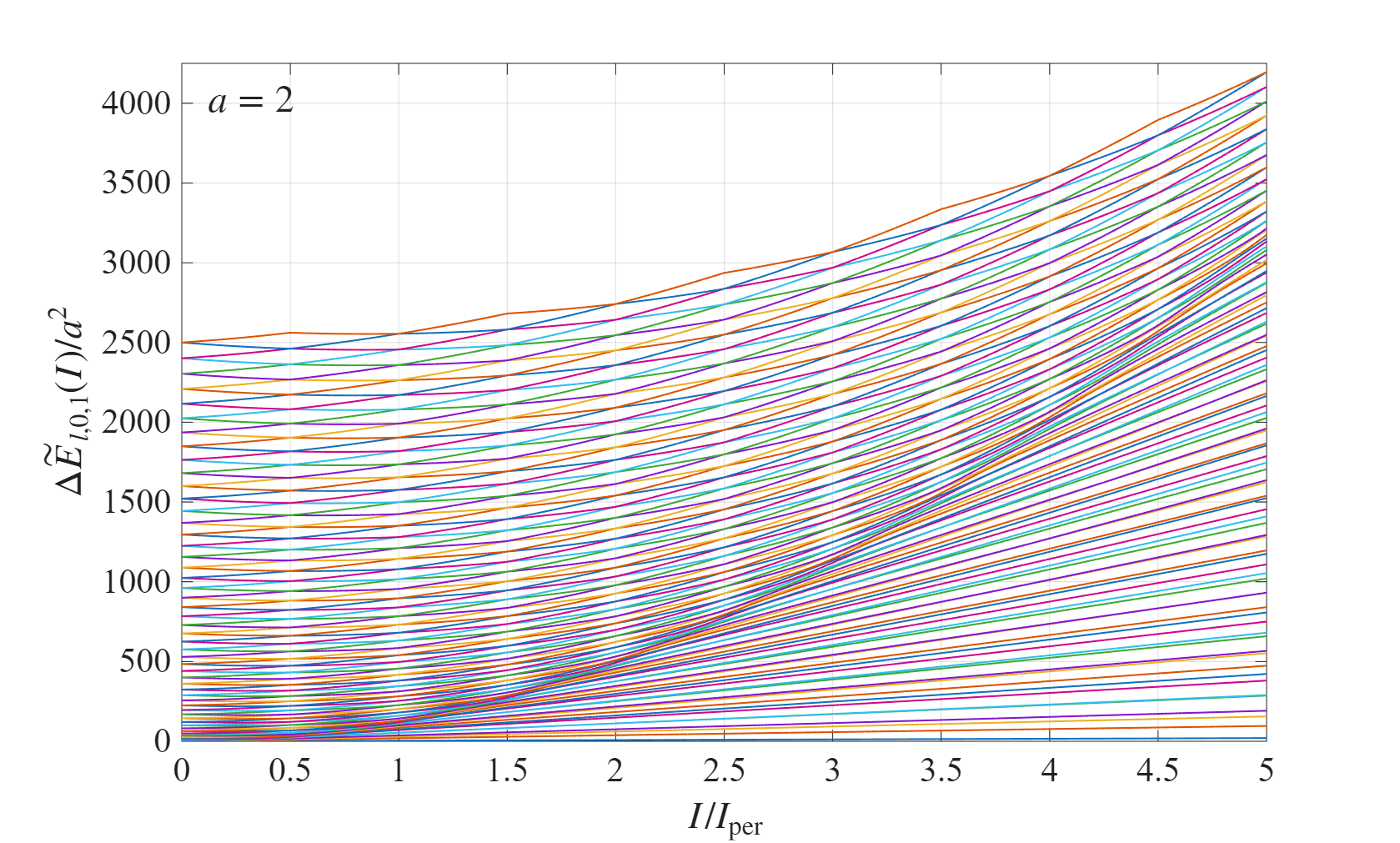}
	\includegraphics[width=0.48\linewidth]{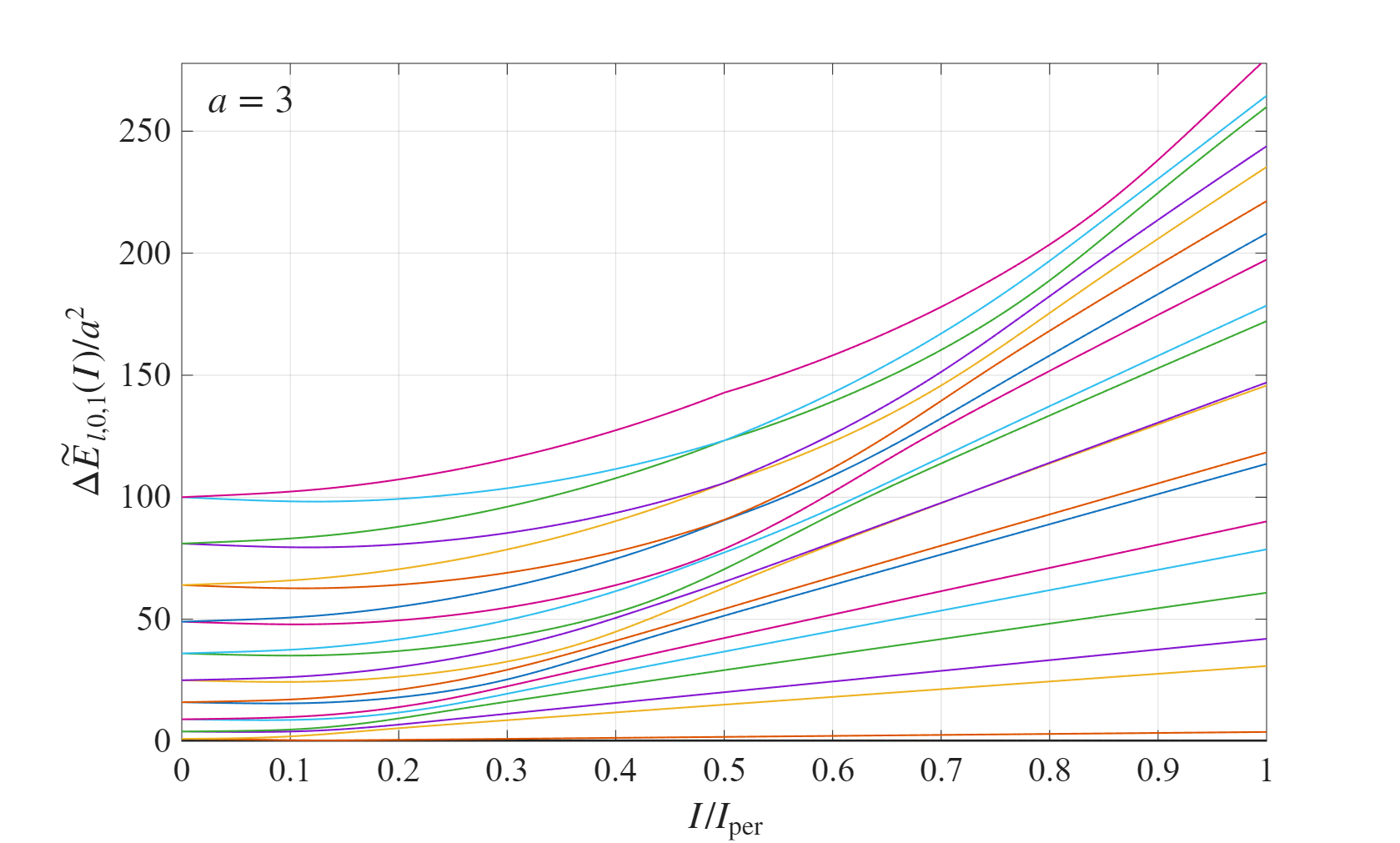}
	\includegraphics[width=0.48\linewidth]{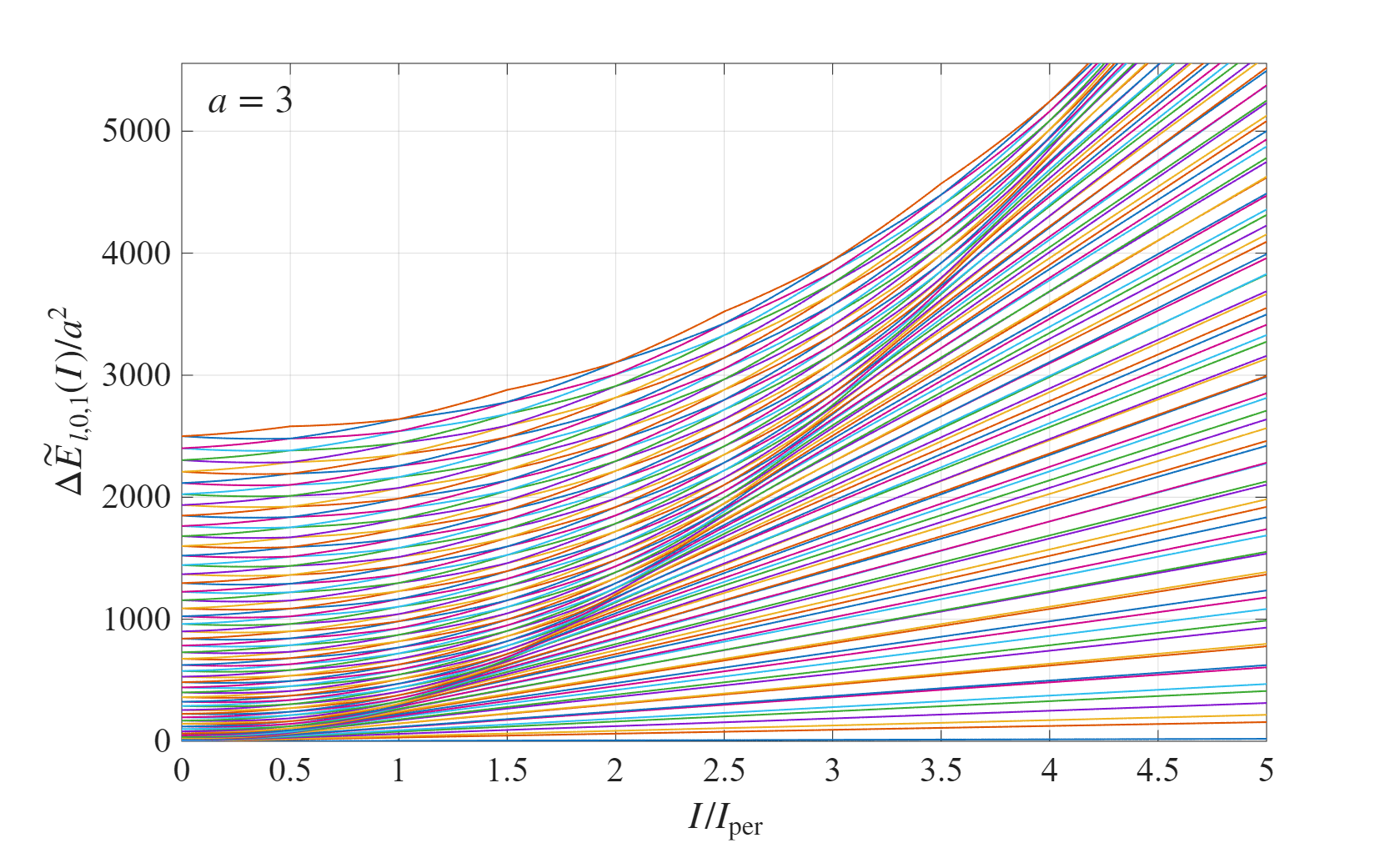}
	\caption{Excitation energies, scaled by $a^2$, $\Delta\tilde{E}_{l,0,1}(I)/a^2$ (for $m=0$ and $k=1$), as a function of scaled current $I/I_\text{per}$, for various values of $a$ (indicated on each panel).
	Left panels: levels $l = 1, 2, \dots, 20$. Right panels: levels $l = 1, 2, \dots, 100$. Apparent crossings of higher energy levels occur near $I/I_\text{per} = p/2$ (where $p$ is an integer)--the number of crossings increase with $l$. A legend identifying $l$ is shown in the upper-left panel for all left panels; a legend for the right panels is omitted due to lack of space.}
	\label{fig_Etot}
\end{figure}

\begin{figure}[t]
	\centering
	\includegraphics[width=0.48\linewidth]{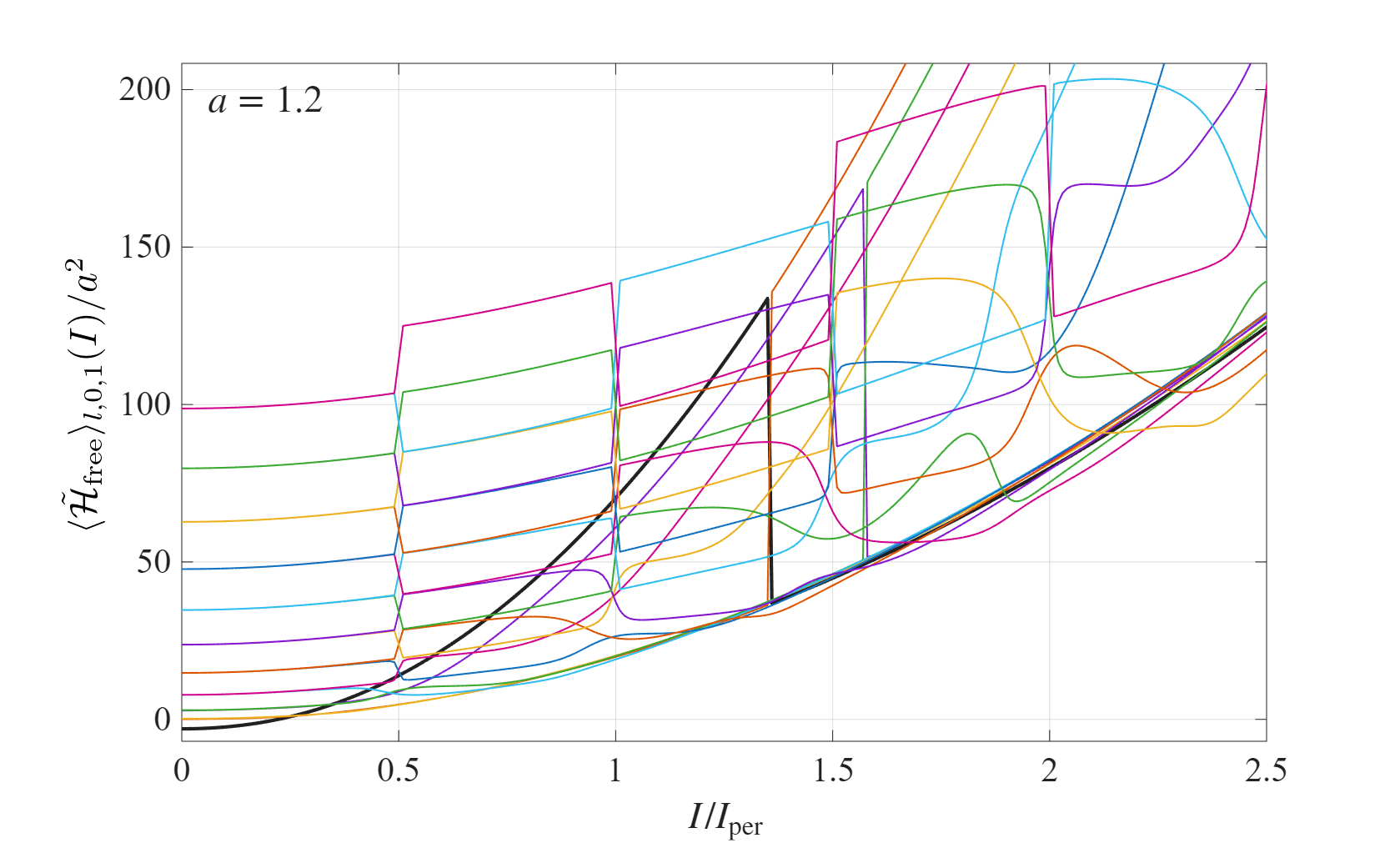}
	\includegraphics[width=0.48\linewidth]{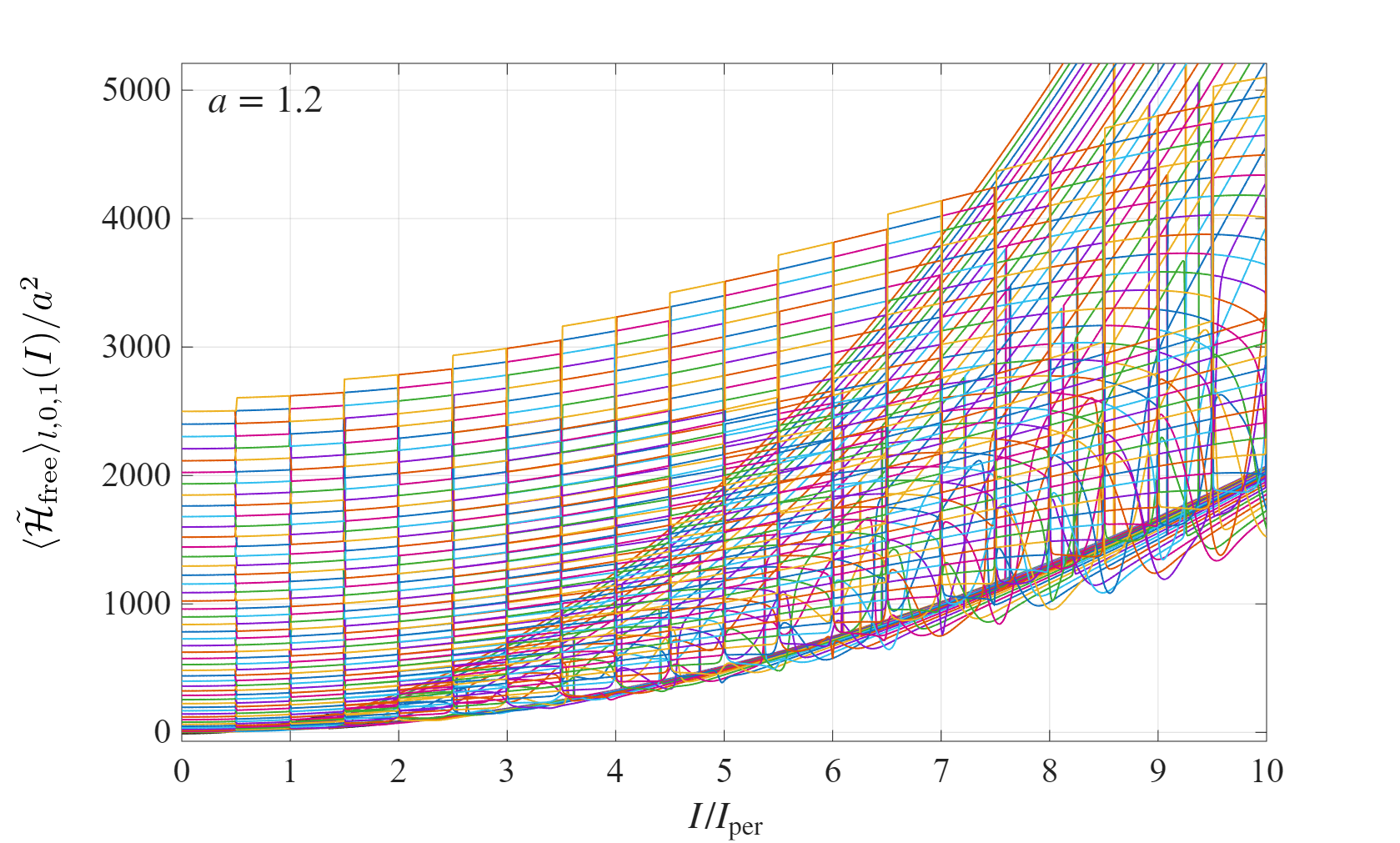}
	\includegraphics[width=0.48\linewidth]{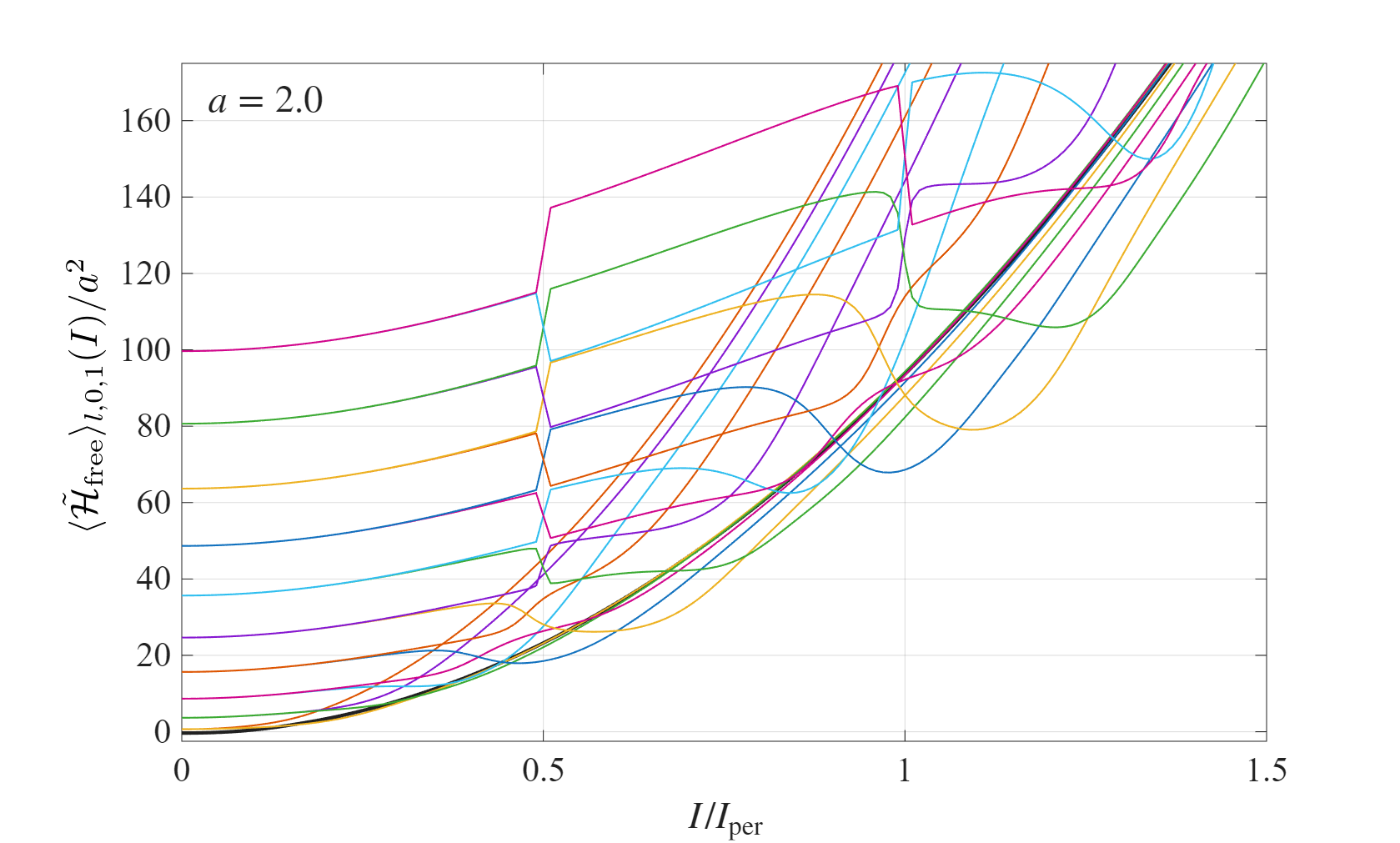}
	\includegraphics[width=0.48\linewidth]{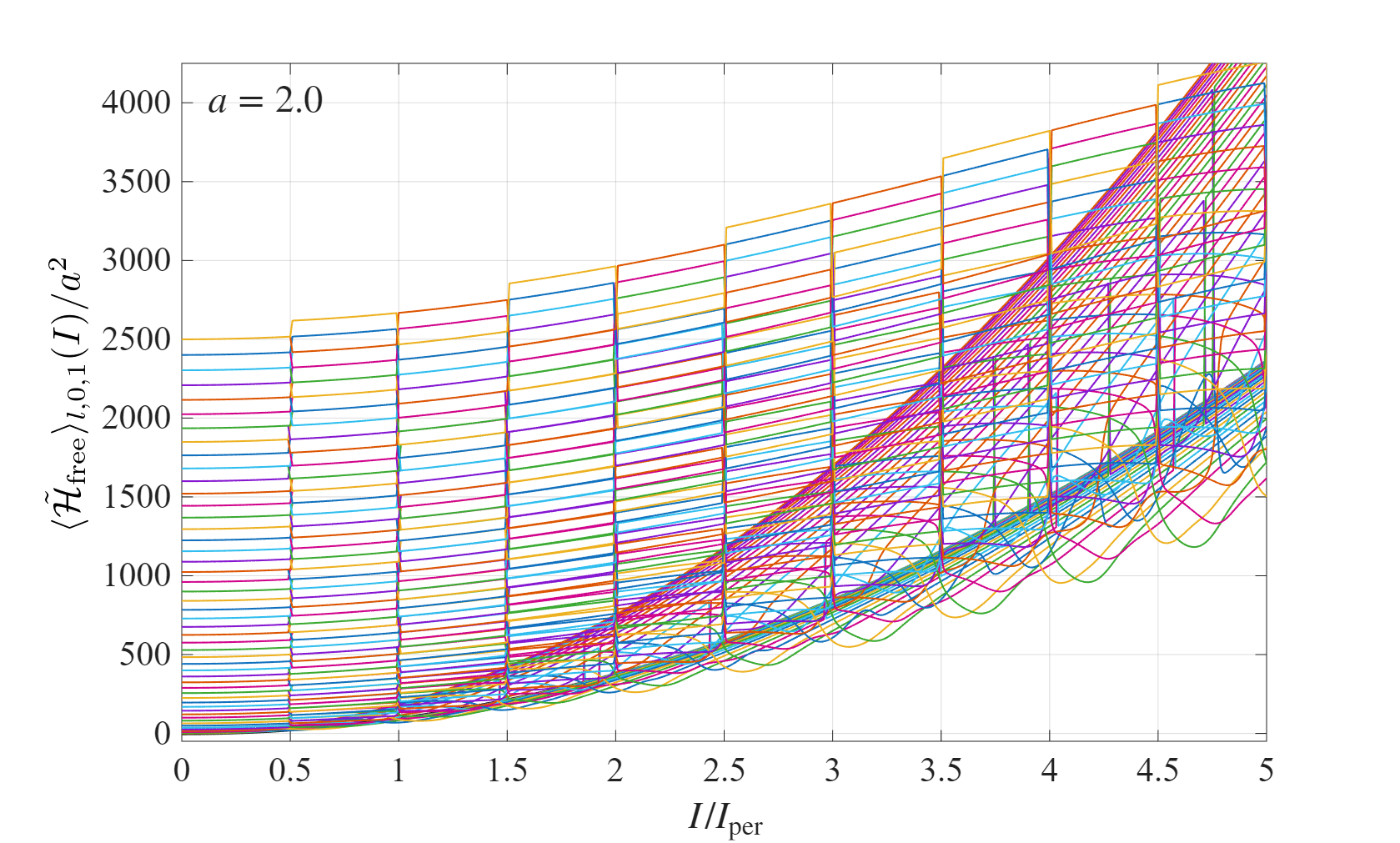}
	\includegraphics[width=0.48\linewidth]{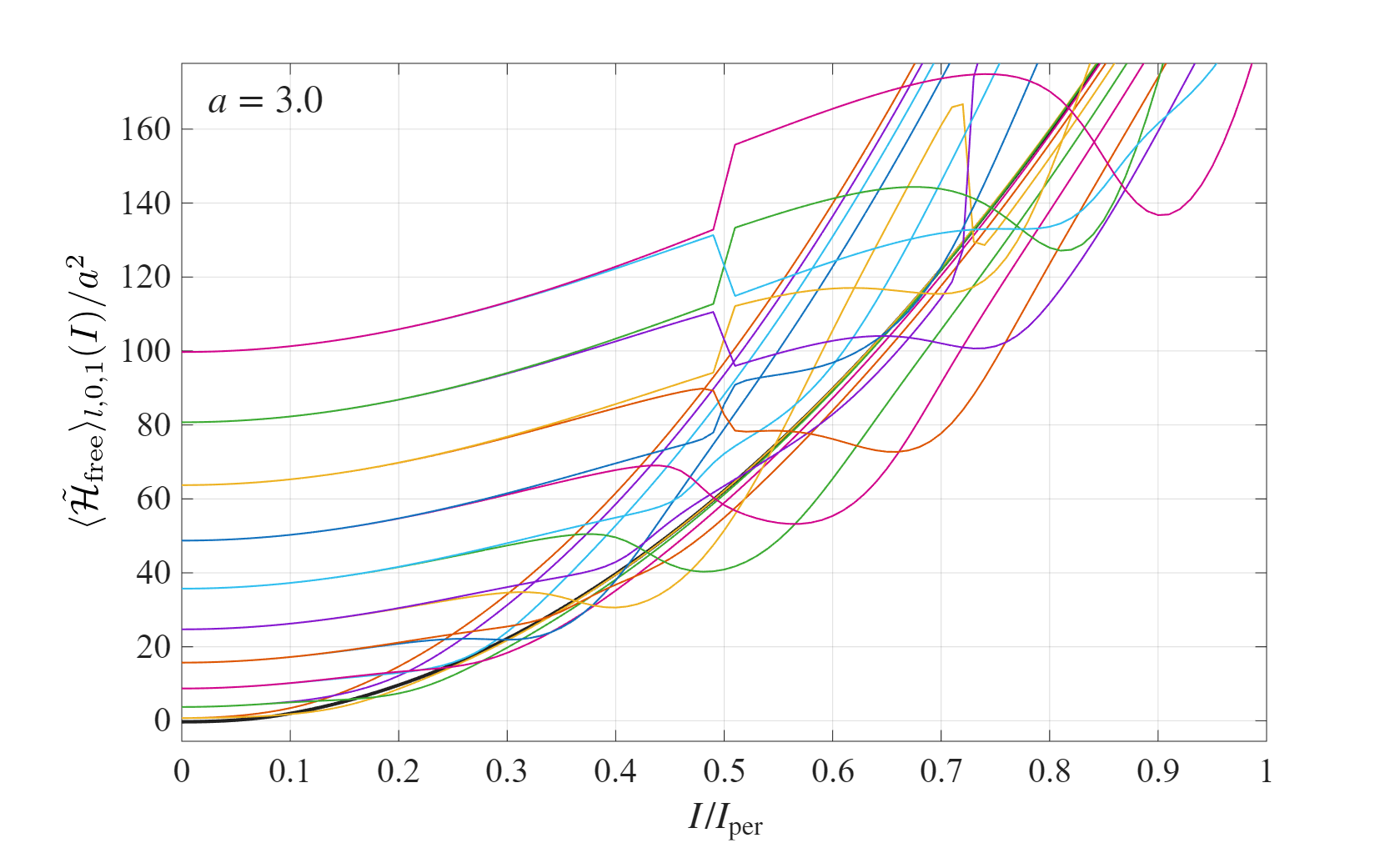}
	\includegraphics[width=0.48\linewidth]{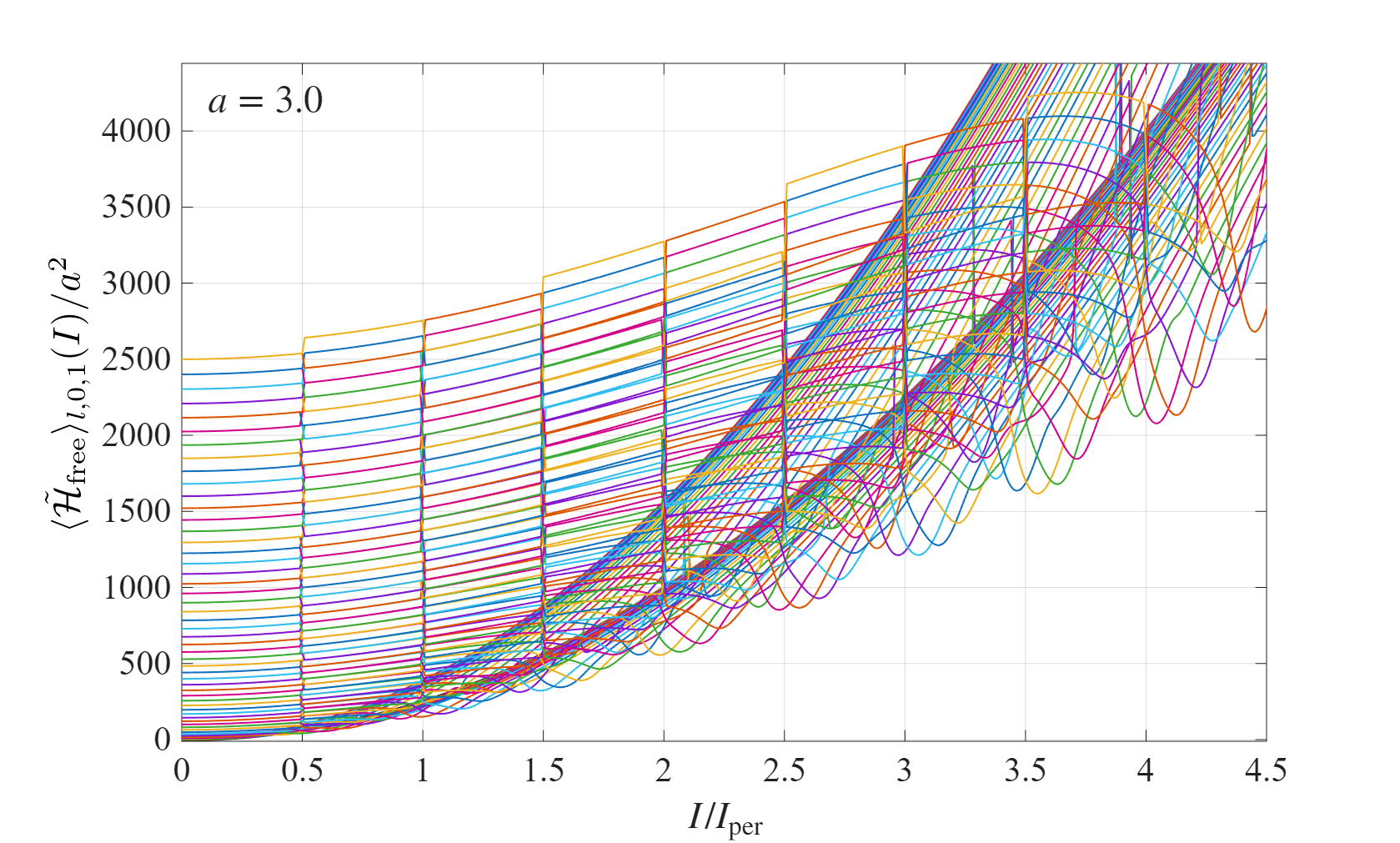}
	\caption{$\langle\tilde{\cH}_\text{free}\rangle_{l,0,1}(I)/a^2 \equiv \langle \tilde{\psi}_{l,0,1}(I) | \tilde{\cH}_\text{free} | \tilde{\psi}_{l,0,1}(I) \rangle/a^2$, versus scaled current $I/I_\text{per}$, for $m=0$, $k=1$, and $a=1.2, 2, 3$, as indicated in each panel. Left panels: $l = 1, \dots, 20$. Right panels: $l = 1, \dots, 100$.
	Abrupt changes of $\langle\tilde{\cH}_\text{free}\rangle_{l,0,1}(I)$ for higher levels occur near half-integer values of $I/I_\text{per}$. A legend for the energy level $l$ is shown in the upper-left panel for all left panels; the legends for the right panels are omitted.}
	\label{fig_Etot0}
\end{figure}

Next, we consider the full Hamiltonian.
We define $\Delta\tilde{E}_{l,m,1}(I) \equiv \tilde{E}_{l,m,1}(I) - \tilde{E}_{0,m,1}(I)$, where $\tilde{\psi}_{l,m,1}(I)$ and $\tilde{E}_{l,m,1}(I)$ are the eigenvectors and eigenvalues of $\tilde{\mathcal{H}}(I)$, respectively:
\begin{equation}
	\tilde{\mathcal{H}}(I) |\tilde{\psi}_{l,m,1}(I)\rangle = \tilde{E}_{l,m,1}(I) |\tilde{\psi}_{l,m,1}(I)\rangle.
\end{equation}
By definition, $\Delta\tilde{E}_{0,m,1}(I) \equiv 0$. The energy eigenvalues are ordered such that $\tilde{E}_{l,m,1}(I) \leq \tilde{E}_{l',m,1}(I)$ for $l < l'$ at any $I \geq 0$.  Figure~\ref{fig_Etot} plots $\Delta\tilde{E}_{l,0,1}(I)/a^2$ as a function of the scaled current $I/I_\text{per}$ for $l = 1, \ldots, 20$ (left panels) and $l = 1, \ldots, 100$ (right panels), with different panels corresponding to $a=1.2$, $2$, and $3$.

Figure~\ref{fig_Etot0} illustrates a sudden quench scenario.  The system is initially prepared in the $l$-th energy level ($m=0$, $k=1$) at $I=0$. The current is then quasistatically increased to a final value $I$, during which the system remains in the $l$-th energy eigenstate.  Subsequently, the current is instantaneously switched off (quenched) to $I=0$.  The wavefunction immediately after the quench is still $|\tilde{\psi}_{l,m,1}(I)\rangle$, but the Hamiltonian abruptly changes to $\tilde{\mathcal{H}}_\text{free} \equiv \tilde{\mathcal{H}}(I=0)$ (see Eq.~\ref{H_free_contribs}).
Therefore, $|\tilde{\psi}_{l,m,1}(I)\rangle$ is no longer an eigenstate, and the energy expectation value becomes $\langle\tilde{H}_\text{free}\rangle_{l,m,1}(I) \equiv \langle \tilde{\psi}_{l,m,1}(I) | \tilde{\mathcal{H}}_\text{free} | \tilde{\psi}_{l,m,1}(I) \rangle$.

\begin{figure}[t]
	\centering
	\includegraphics[width=0.48\linewidth]{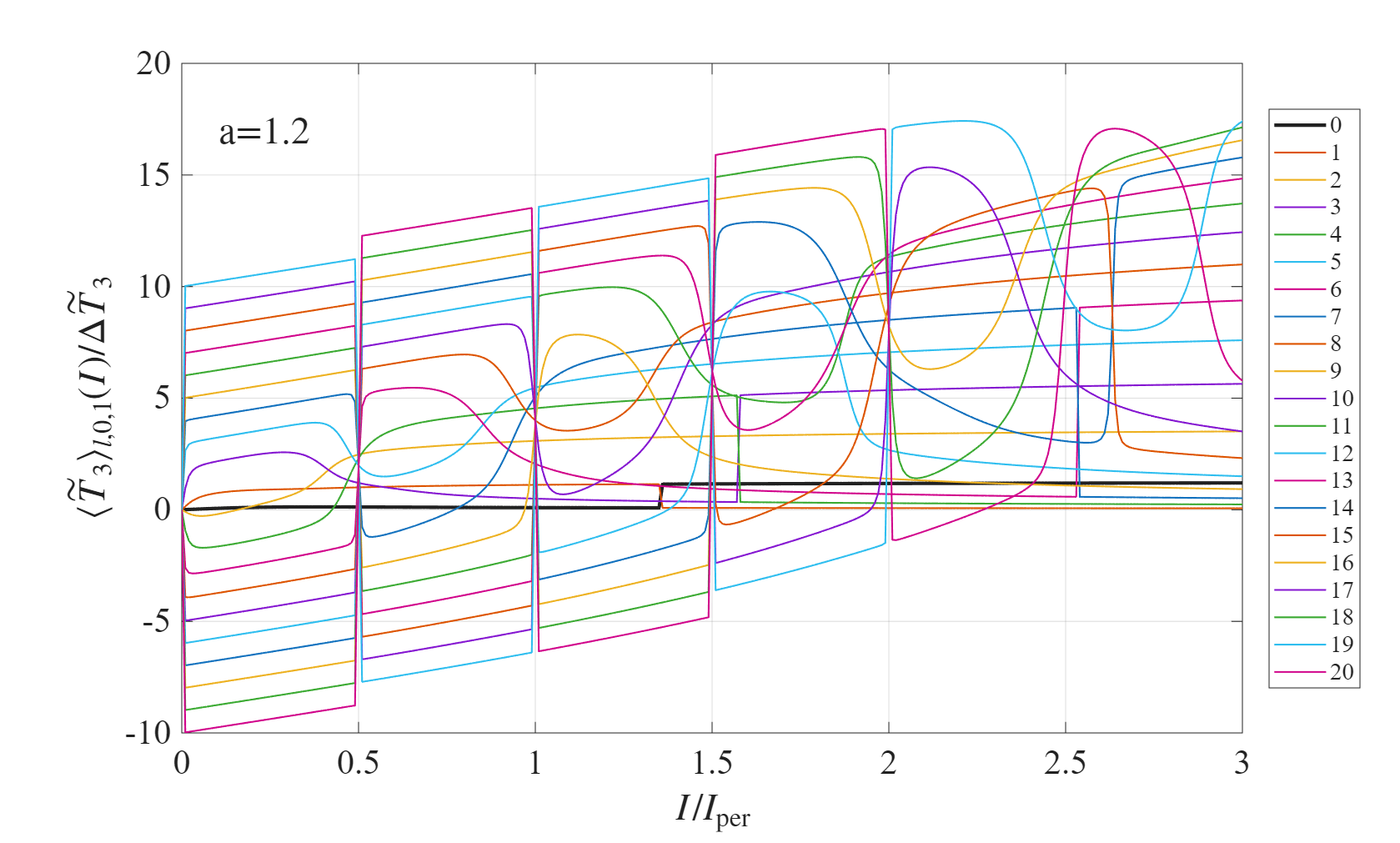}
	\includegraphics[width=0.48\linewidth]{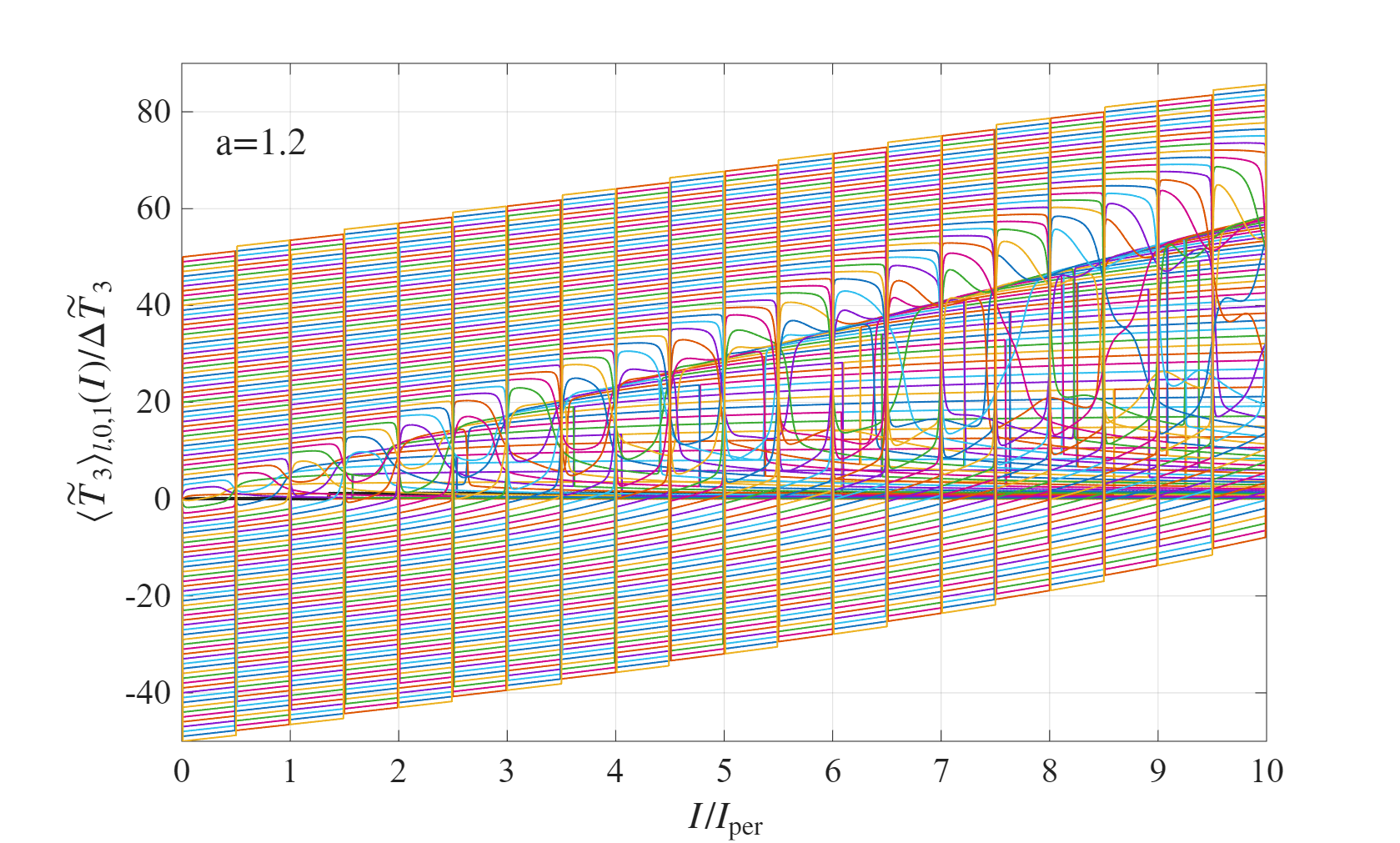}
	\includegraphics[width=0.48\linewidth]{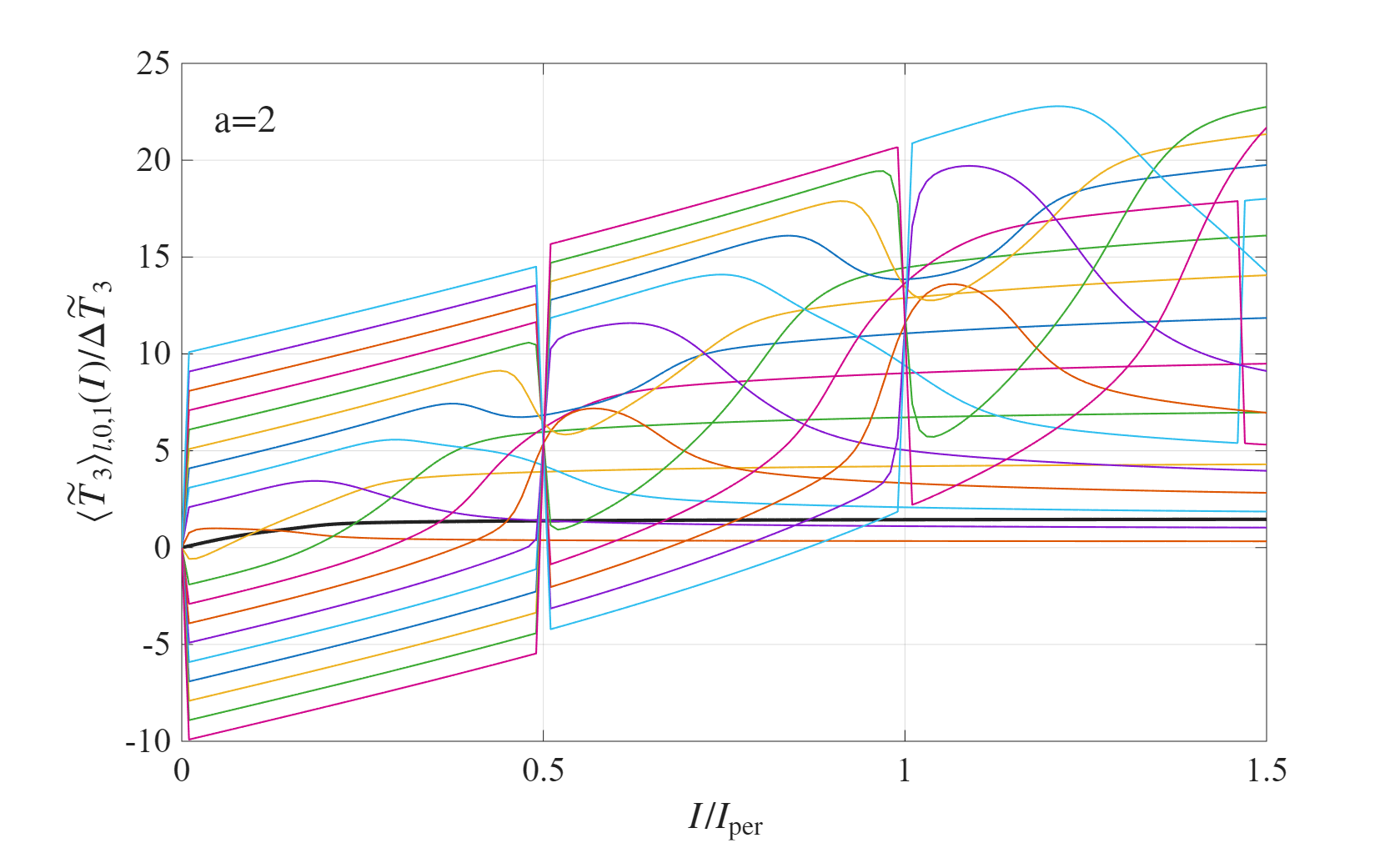}
	\includegraphics[width=0.48\linewidth]{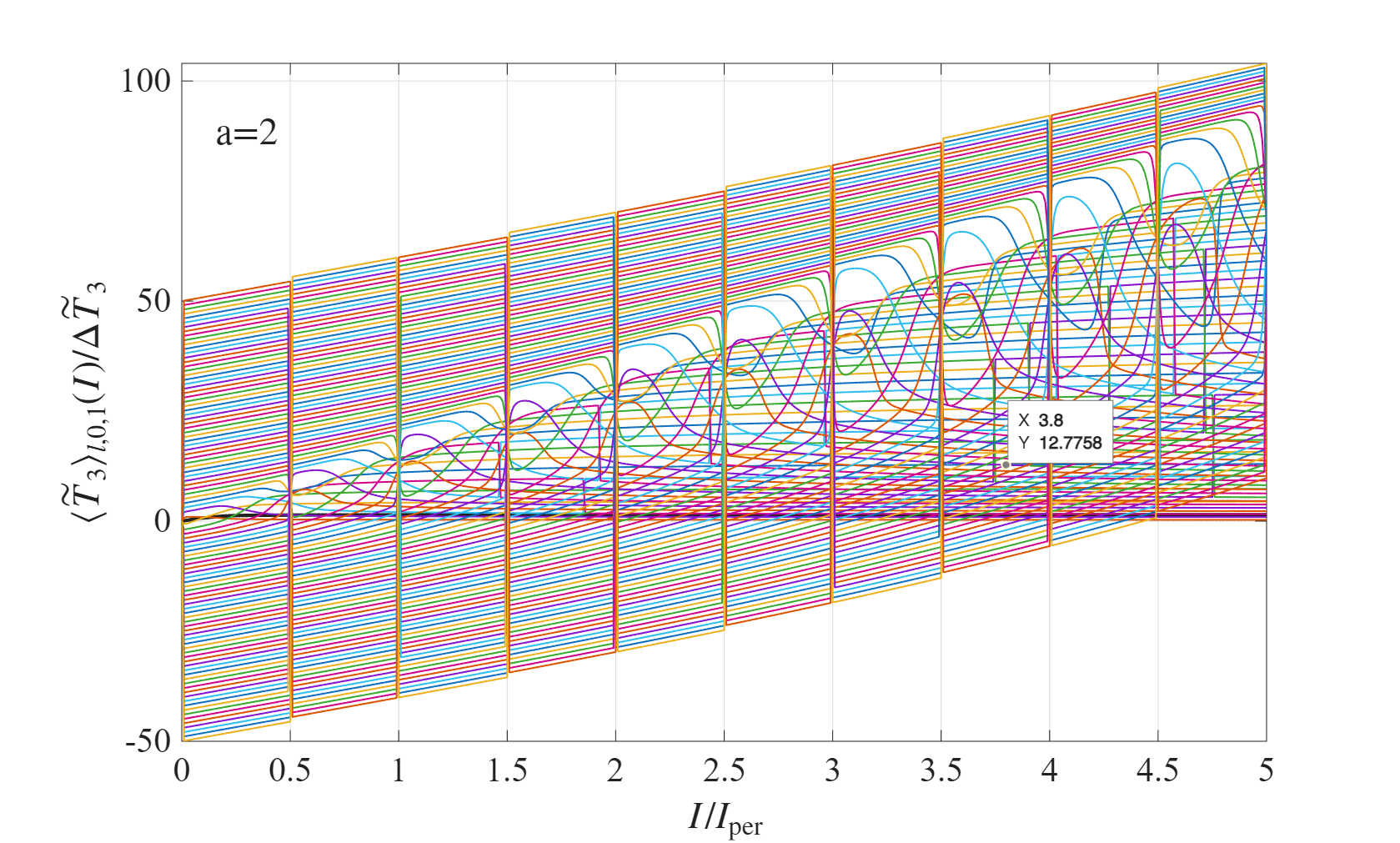}
	\includegraphics[width=0.48\linewidth]{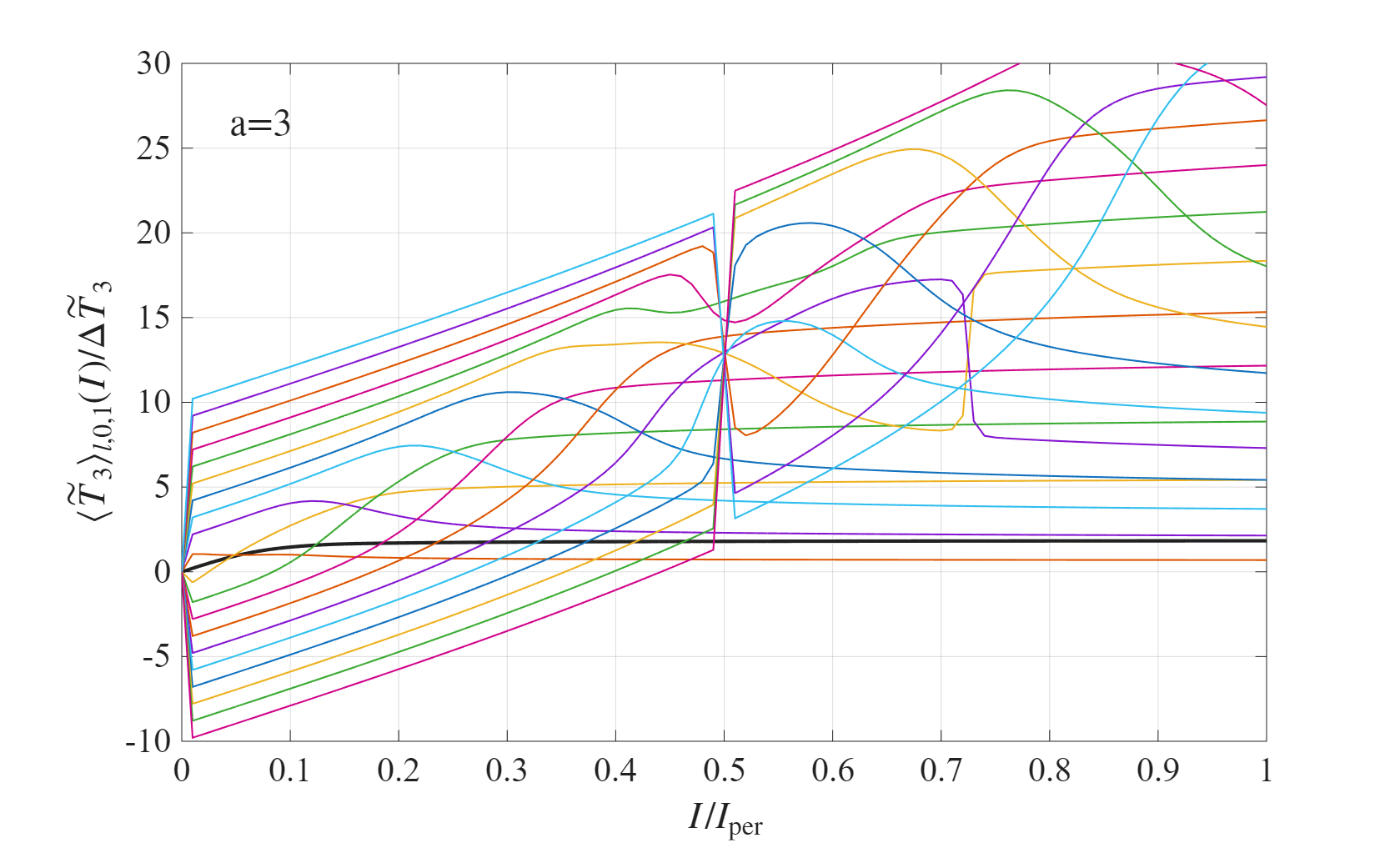}
	\includegraphics[width=0.48\linewidth]{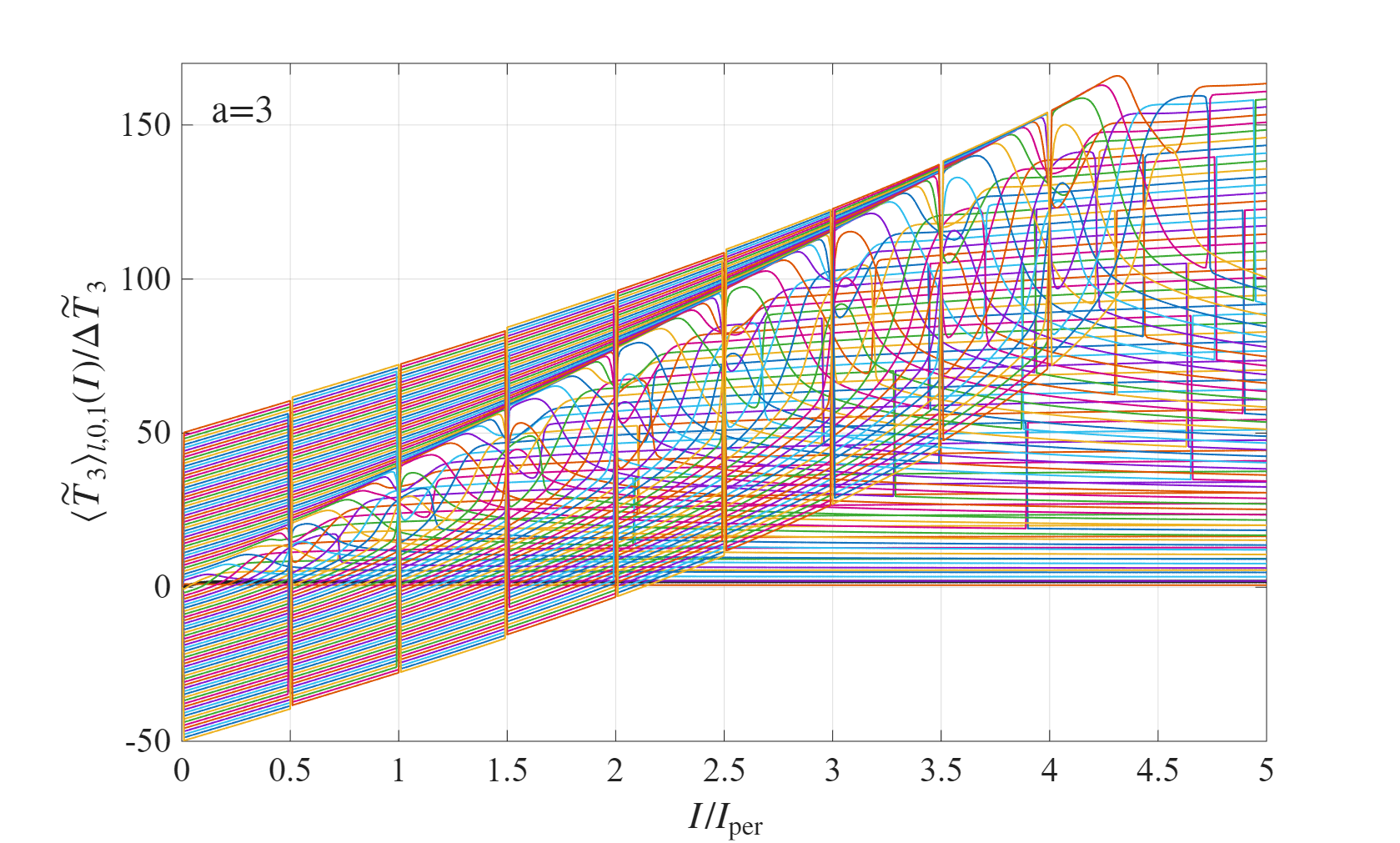}
	\caption{
	$\langle\tilde{T}_3\rangle_{l,0,1}(I)/\Delta \tilde{T}_3 \equiv \langle \tilde{\psi}_{l,0,1}(I) | \tilde{T}_3(I) | \tilde{\psi}_{l,0,1}(I) \rangle/\Delta \tilde{T}_3$, versus scaled current $I/I_\text{per}$, where $m=0$ and $k=1$. Left panels: $l = 0, 1, \dots, 20$. Right panels: $l = 0, 1, \dots, 100$. The black line indicates the ground state ($l=0$). Abrupt changes in $\langle\tilde{T}_3(I)\rangle_{l,0,1}/\Delta \tilde{T}_3$ for higher energy levels occur near half-integer values of $I/I_\text{per}$. Each panel corresponds to a different value of $a$, as indicated. A legend for $l$ is shown in the upper-left panel for all left panels; legends for the right panels are omitted.
	The thick black line corresponds to the ground state and $m=0$ in all the plots.
	}
	\label{fig_T3}
\end{figure}

\begin{figure}[t]
	\centering
	\includegraphics[width=0.48\linewidth]{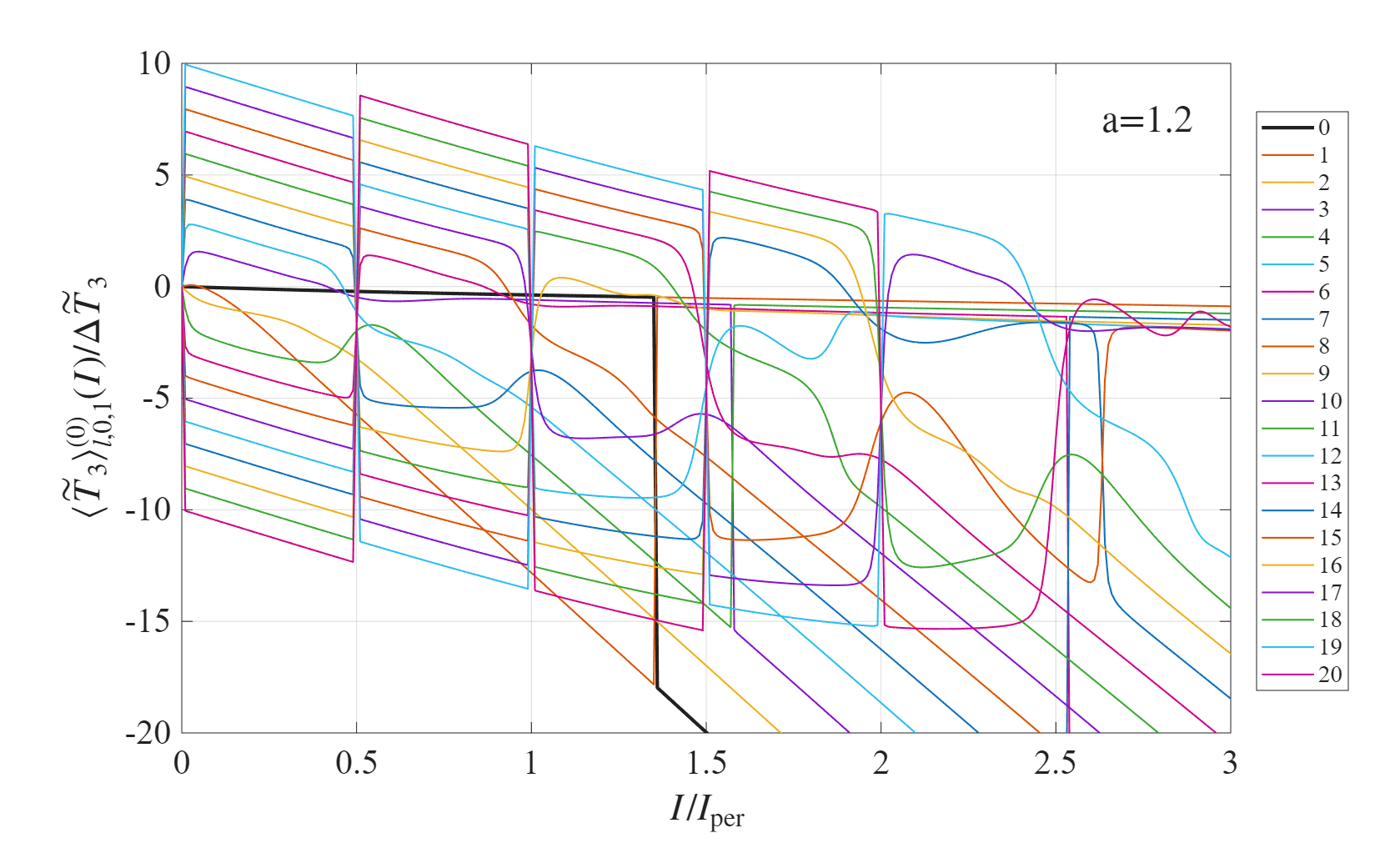}
	\includegraphics[width=0.48\linewidth]{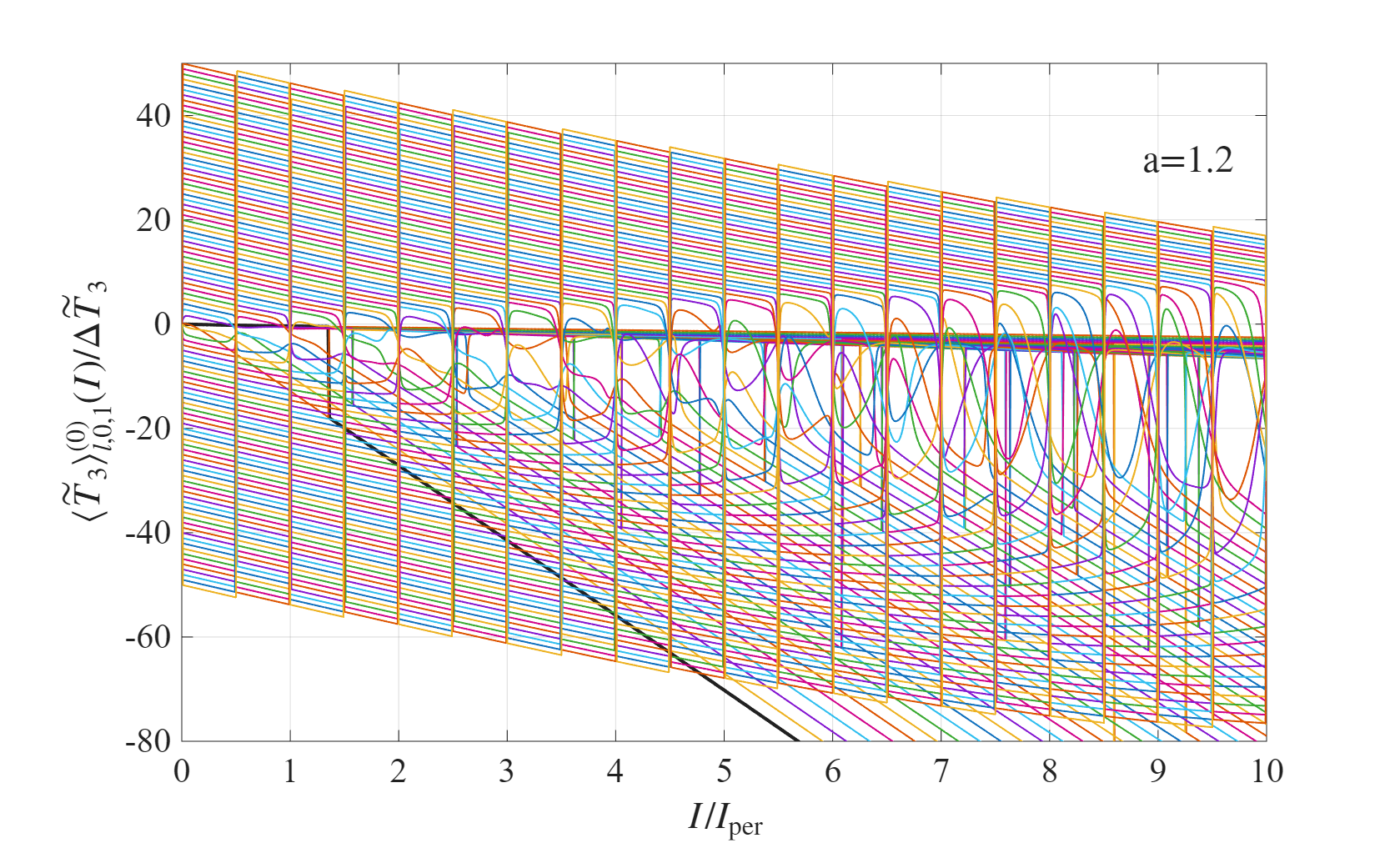}
	\includegraphics[width=0.48\linewidth]{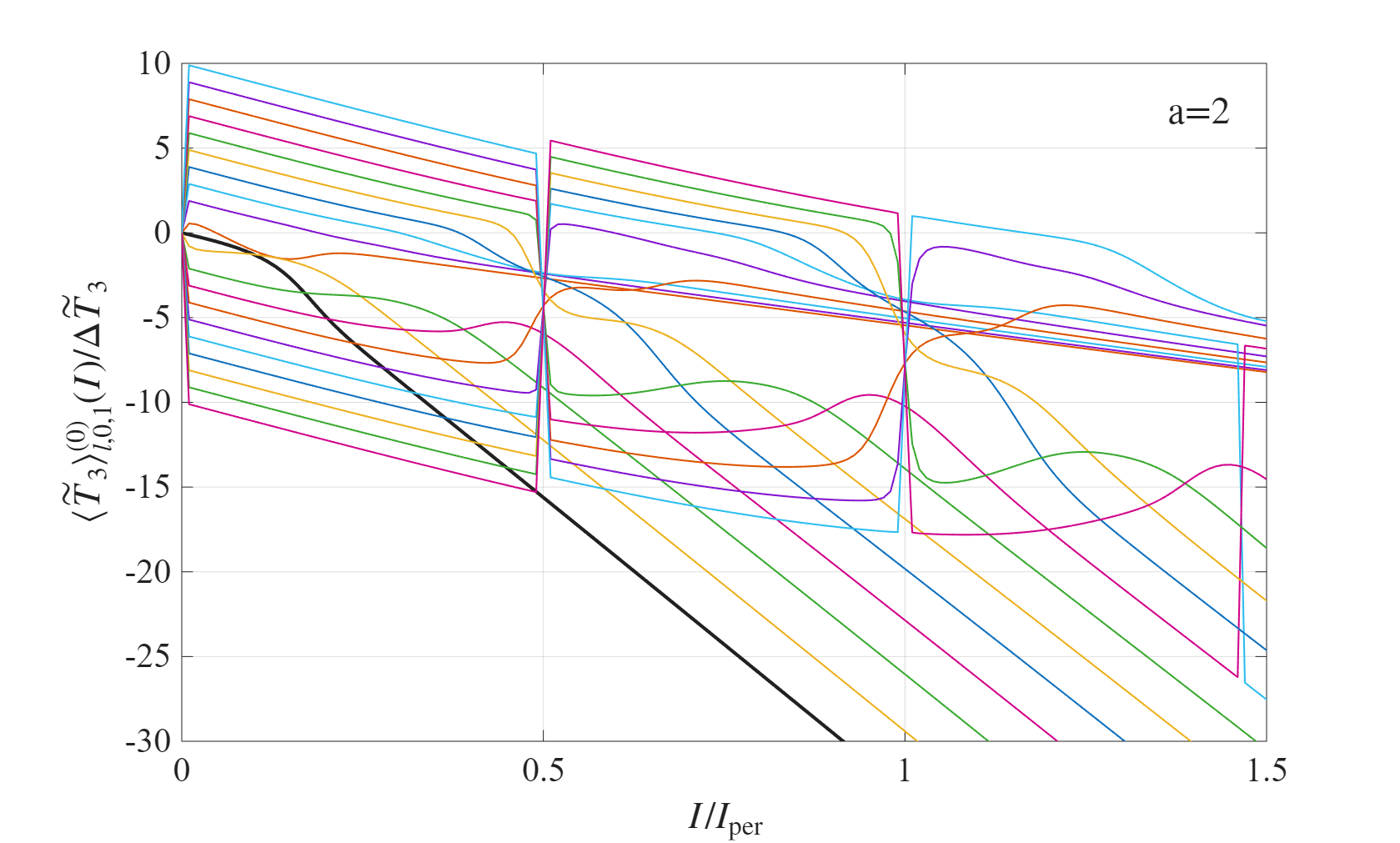}
	\includegraphics[width=0.48\linewidth]{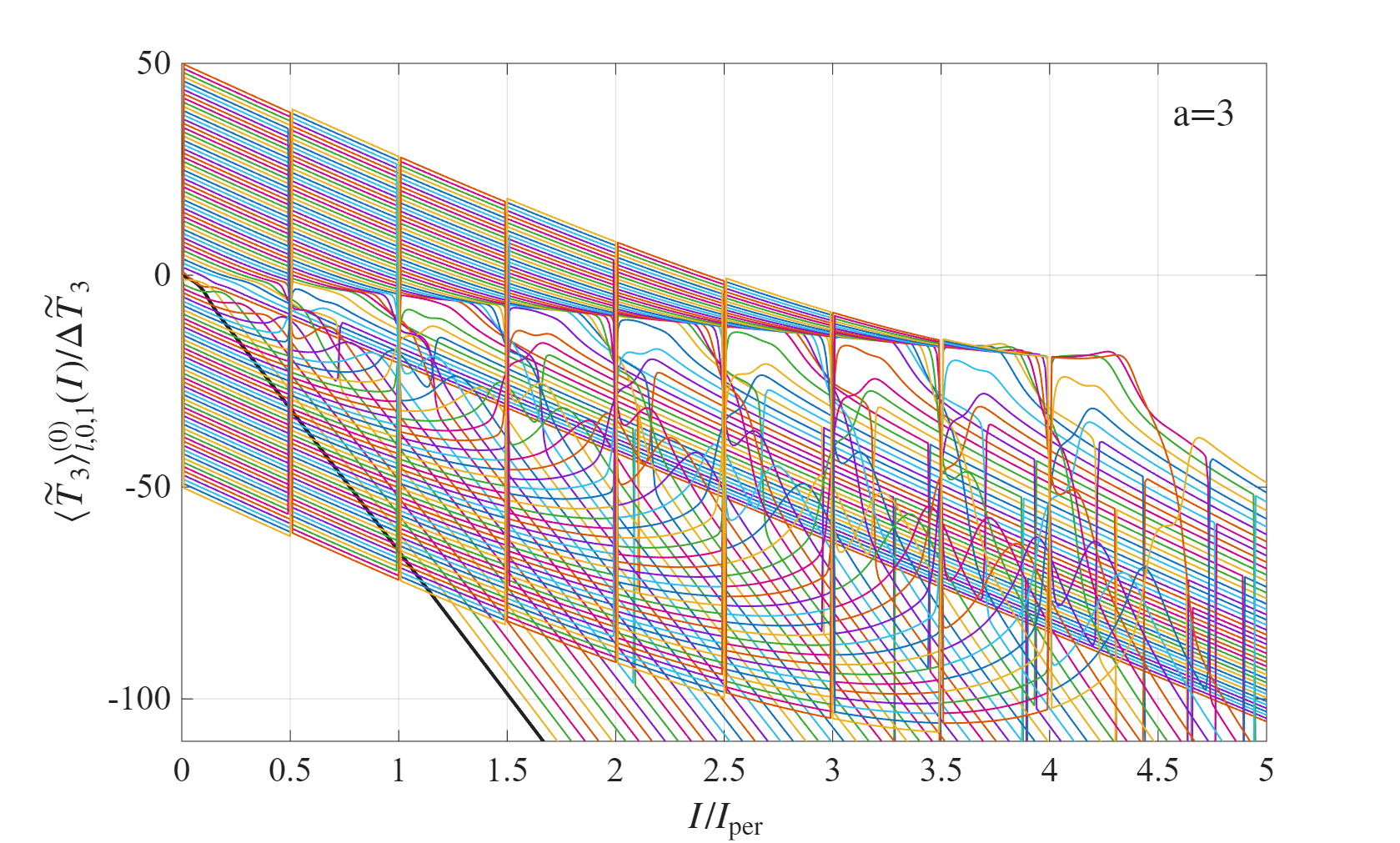}
	\includegraphics[width=0.48\linewidth]{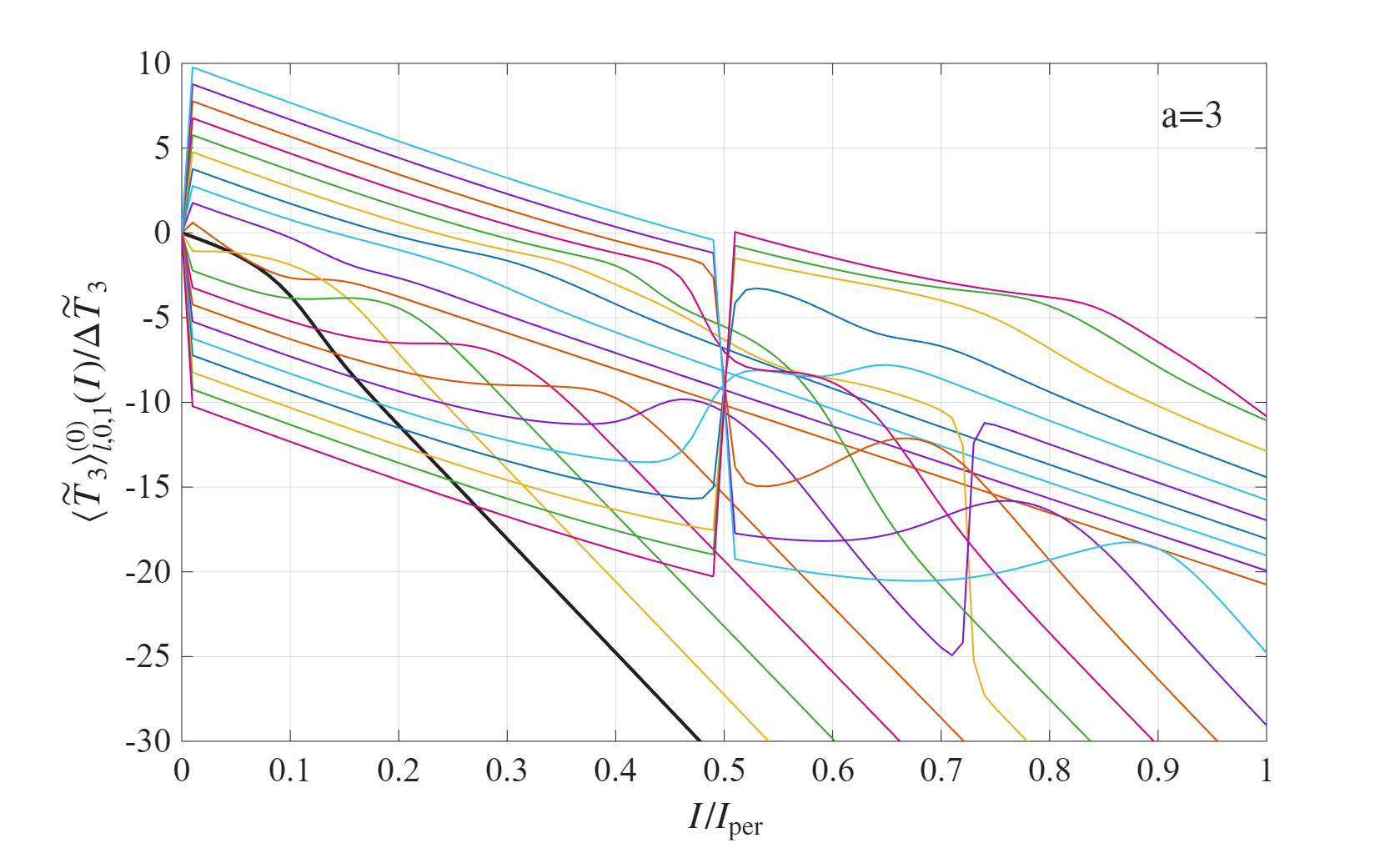}
	\includegraphics[width=0.48\linewidth]{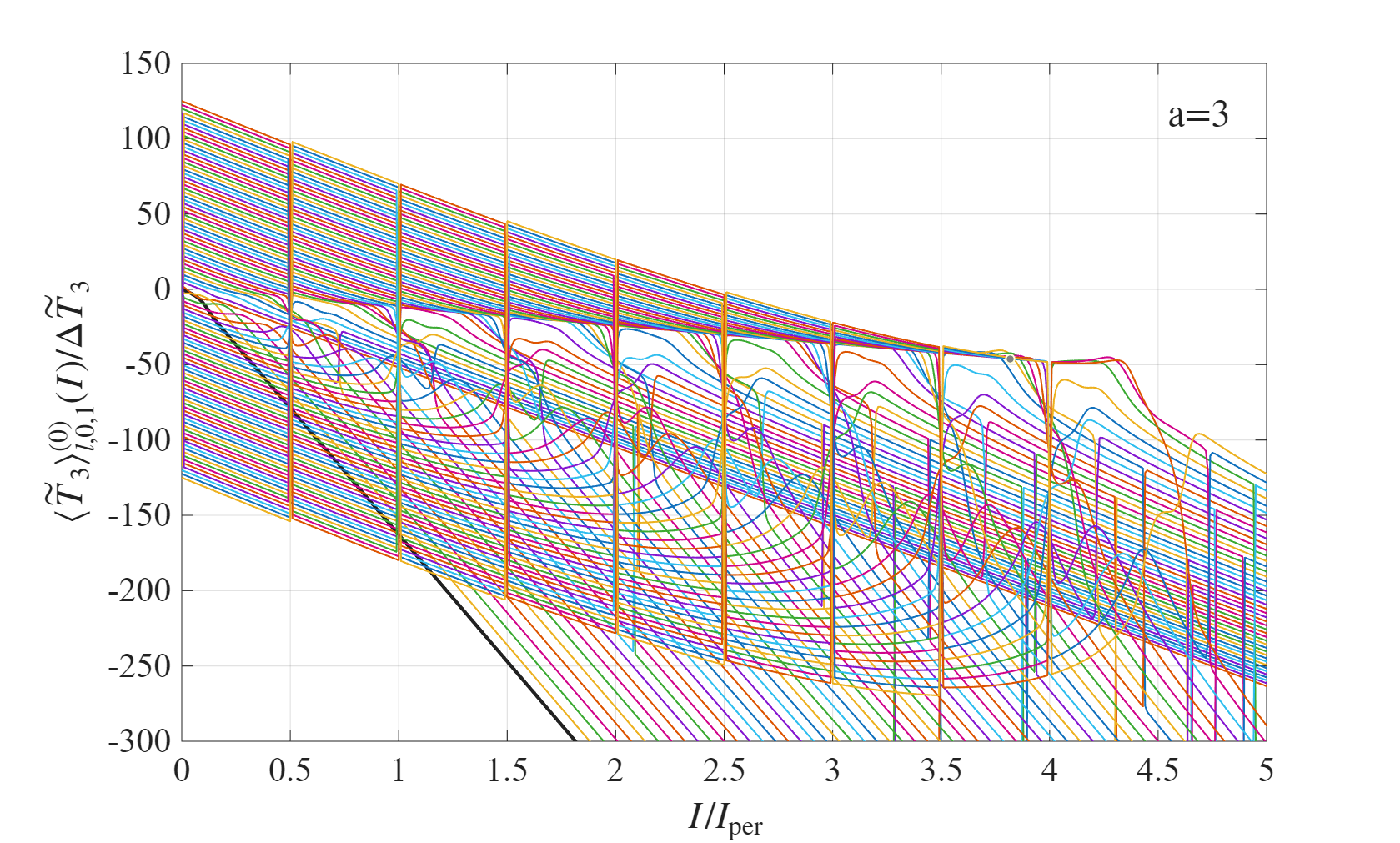}
	\caption{$\langle\tilde{T}_3\rangle^{(0)}_{l,0,1}(I)/\Delta\tilde{T}_3 \equiv \langle \tilde{\psi}_{l,0,1}(I) | \tilde{T}_3(0) | \tilde{\psi}_{l,0,1}(I) \rangle/\Delta\tilde{T}_3$, versus scaled current $I/I_\text{per}$, where $m=0$ and $k=1$. Left panels: $l = 0, 1, \dots, 20$. Right panels: $l = 0, 1, \dots, 100$. The black line indicates the ground state ($l=0$). Abrupt changes in $\langle\tilde{T}_3\rangle^{(0)}_{l,0,1}(I)/\Delta\tilde{T}_3$ for higher levels occur near half-integer values of $I/I_\text{per}$. Each panel corresponds to a different value of $a$, as indicated. A legend for $l$ is shown in the upper-left panel for all left panels; a legend for the right panels is omitted.
	The thick black line corresponds to the ground state and $m=0$ in all the plots.}
	\label{fig_T30}
\end{figure}

Figures~\ref{fig_T3} and \ref{fig_T30} show the expectation values of $\tilde{T}_3(I)$ and $\tilde{T}_3(0)$ (where $\tilde{T}_3(0)$ is defined by Eqs.~\ref{gen_T3_theta_xi_0} with $\mathbf{A}=0$), respectively, evaluated for the energy eigenvectors $|\tilde{\psi}_{l,0,1}(I)\rangle$, scaled by $\Delta\tilde{T}_3 = 2.5$~(\ref{T3_I_approx}), and plotted as functions of $I/I_\text{per}$.
Figure~\ref{fig_T3} depicts $\langle\tilde{T}_3\rangle_{l,0,1}(I)/\Delta\tilde{T}_3 \equiv \langle \tilde{\psi}_{l,0,1}(I) | \tilde{T}_3(I) | \tilde{\psi}_{l,0,1}(I) \rangle/\Delta\tilde{T}_3$, representing the expectation value during the quasistatic current increase.
Figure~\ref{fig_T30} depicts $\langle\tilde{T}_3\rangle^{(0)}_{l,0,1}(0)/\Delta\tilde{T}_3 \equiv \langle \tilde{\psi}_{l,0,1}(I) | \tilde{T}_3(0) | \tilde{\psi}_{l,0,1}(I) \rangle/\Delta\tilde{T}_3$, the expectation value after the abrupt current quench.  Thus, the plots in Figs.~\ref{fig_Etot} and~\ref{fig_Etot0} have a one-to-one correspondence with those in Figs.~\ref{fig_T3} and~\ref{fig_T30}, respectively.
Notably, the expectation value of $\tilde{T}_3$ after the quench, $\langle\tilde{T}_3\rangle^{(0)}_{l,0,1}(I\ne0)$, is generally different from its initial value, $\langle\tilde{T}_3\rangle_{l,0,1}(I=0)$.

To illustrate the effect of the finite-dimensional basis set on the computed eigenvectors, Table~\ref{tab_films} presents animations.  Each animation shows the evolution with $I$ of the probability amplitudes $|\langle \tilde{\psi}_{l,m,1}(I)|\tilde{\mathcal{F}}_{n_2,m,1}^{(\Lambda)} \rangle |^2$, where $n_2$ spans the interval $[-400,400]$.  Here, $\tilde{\psi}_{l,m,1}(I)$ denotes the eigenvector of $\tilde{\mathcal{H}}(I)$ for the $l$-th energy level, with fixed $m$ and $k=1$.  These animations are generated for specific $a$ and $l$ values, with $I$ ranging from $0$ to $10I_\text{per}$.
The animations demonstrate the adequacy of this basis for smaller $a$ and $I$ values, as used in Figs.~\ref{fig_Etot}, \ref{fig_Etot0}, \ref{fig_T3}, and \ref{fig_T30}.  This adequacy is evident from the vanishingly small components $|\langle \tilde{\psi}_{l,m,1}(I)|\tilde{\mathcal{F}}_{\pm 400,m,1}^{(\Lambda)} \rangle |^2$ at the boundaries ($n_2 = \pm 400$).  However, for larger $a$ or $I$, the basis becomes insufficient, potentially leading to numerical artifacts, as observed in the simulations from the lower part of Table~\ref{tab_films}.

Finally, the initial frames of the animations, and the left panels of Figs.~\ref{fig_T3} and \ref{fig_T30}, show that $\langle \tilde{T}_3 \rangle_{l,m,1}(I=0) = 0$.
This result, rigorously proven in Appendix~\ref{app_sec_sym_basis}, corrects numerical inaccuracies in Ref.~\cite{PhysicaA.598.127377.2022.Dolineanu}, where the diagonalization yielded non-zero values, $\langle \tilde{T}_3\rangle_{l,m,1}(I=0) \ne 0$, for large $l$.
These erroneous results are attributed to the quasi-double degeneracy of the corresponding energy levels, as we shall see in the next section.

\begin{table}[t]  
\centering          
\caption{
	Animations depicting the squared magnitudes of the inner products, $|\langle \tilde{\psi}_{l,m,1}(I) | \tilde{\mathcal{F}}_{n_2,m,1}^{(\Lambda)} \rangle|^2$ (vertical axis), between the energy eigenvectors $|\tilde{\psi}_{l,m,1}(I)\rangle$ of the reduced Hamiltonian and the basis functions $\tilde{\mathcal{F}}_{n_2,m,1}^{(\Lambda)}$. The horizontal axis represents the basis function index $n_2$ (ranging from -400 to 400), and the animations show the evolution as the adiabatic parameter $I$ increases from 0 to $10I_\text{per}$. Each animation corresponds to fixed values of the parameter $a$ and the energy level index $l$, with the azimuthal quantum number $m$ fixed at 0.  The numerical diagonalization of the Hamiltonian was performed using the full basis set $\{\tilde{\mathcal{F}}_{n_2,m,1}^{(\Lambda)}\}_{n_2=-400}^{400}$, although some animations display a smaller $n_2$ interval for enhanced clarity.
	} 
\label{tab_films}      
\begin{tabular}{|l|r|r|}  
	\hline             
	File name & a & $l$ \\  
	\hline             
	Eigenvector\_a1.2\_m0\_level1i\_per0-10\_max\_x100.gif & 1.2 & 1 \\
	Eigenvector\_a1.2\_m0\_level21i\_per0-10\_max\_x200.gif & 1.2 & 21 \\
	Eigenvector\_a1.2\_m0\_level101i\_per0-10\_max\_x200.gif & 1.2 & 101 \\
	Eigenvector\_a2\_m0\_level1i\_per0-10\_max\_x200.gif & 2 & 1 \\
	Eigenvector\_a2\_m0\_level21i\_per0-10\_max\_x200.gif & 2 & 21 \\
	Eigenvector\_a2\_m0\_level101i\_per0-10\_max\_x400.gif & 2 & 101 \\
	Eigenvector\_a3\_m0\_level1i\_per0-10\_max\_x400.gif & 3 & 1 \\
	Eigenvector\_a3\_m0\_level21i\_per0-10\_max\_x400.gif & 3 & 21 \\
	Eigenvector\_a3\_m0\_level101i\_per0-10\_max\_x400.gif & 3 & 101 \\
	Eigenvector\_a5\_m0\_level1i\_per0-10\_max\_x400.gif & 5 & 1 \\
	Eigenvector\_a5\_m0\_level21i\_per0-10\_max\_x400.gif & 5 & 21 \\
	Eigenvector\_a5\_m0\_level101i\_per0-10\_max\_x400.gif & 5 & 101 \\
	Eigenvector\_a10\_m0\_level1i\_per0-10\_max\_x400.gif & 10 & 1 \\
	Eigenvector\_a10\_m0\_level21i\_per0-10\_max\_x400.gif & 10 & 21 \\
	Eigenvector\_a10\_m0\_level101i\_per0-10\_max\_x400.gif & 10 & 101 \\
	Eigenvector\_a20\_m0\_level1i\_per0-10\_max\_x400.gif & 20 & 1 \\
	Eigenvector\_a20\_m0\_level21i\_per0-10\_max\_x400.gif & 20 & 21 \\
	Eigenvector\_a20\_m0\_level101i\_per0-10\_max\_x400.gif & 20 & 101 \\
	\hline             
\end{tabular}
\end{table}

\subsection{Approximate eigensolutions for $\cH^\text{(2D)}$ at arbitrary $I$} \label{subsec_appr_sol}

In Fig.~\ref{fig_H_2D} we notice an apparent degeneracy of the energy levels at $I = (\eta/2)I_\text{per}$ for any integer $\eta$.
For $I=\eta=0$, the Hamiltonian reduces to the free Hamiltonian, analyzed in Ref.~\cite{PhysicaA.598.127377.2022.Dolineanu, PhysicaA.617.127377.2022.Dolineanu}.
As we show in Appendix~\ref{app_sec_sym_basis}, in such a case it is best to solve the eigenvalues and eigenvectors problem for the Hamiltonian in the basis of symmetric and antisymmetric wavefunctions $\{\cF^{(\pm)}_{n_2,m,k}\}$~(\ref{defs_Fpm}).
For big enough values of $a$ or $n_2$ ($n_2\ge0$ in the definition of $\cF^{(\pm)}_{n_2,m,k}$), $\cF^{(\pm)}_{n_2,m,k}$ are good approximations for the Hamiltonian's eigenfunctions, with the approximate eigenvalues given in~(\ref{H_diag_Fpm}).
In this approximation, the ground state corresponds to $n_2=0$, with the eigenfunction $\cF^{(+)}_{0,m,k} \equiv \cF^{(\Lambda)}_{0,m,k}$ and eigenenergy~(\ref{H_diag_Fpm_0}).
For the excited states, corresponding to $n_2>0$, the eigenfunctions of the levels $2n_2-1$ and $2n_2$ are $\cF^{(-)}_{n_2,m,k}$ and $\cF^{(+)}_{n_2,m,k}$, respectively, corresponding to the eigenenergies given by Eqs.~(\ref{H_diag_Fpm}).

For general $I$, we use Eqs.~(\ref{H_transf_U}) and (\ref{T3_p_theta_U}) to transform $\tilde{H}^{\text{(2D)}}(I)$ into $\tilde{H}^{\text{(2D)}}_\text{free} \equiv \tilde{H}^{\text{(2D)}}(I=0)$ and $\tilde{T}_3^{(\theta)}(I)$ into $\tilde{T}_3^{(\theta)}(I=0)$, respectively.
In order to satisfy the boundary conditions of the transformed wavefunctions~(\ref{psi_cU_tot}), we transform also the basis~(\ref{basis_def}) into
\begin{equation}
	\tilde{\cF}_{n_2,m,k}^{(\Lambda, I)} (\theta, \phi, \alpha_\xi)
	\equiv \frac{\sin(\pi k \alpha_\xi/\alpha_{\xi_\text{max}}) \exp\left\{i\left[\left(n_2-\frac{\theta_\bA(I)}{2\pi}\right)\theta + m\phi\right]\right\}}{\pi \sqrt{2 \alpha_{\xi_\text{max}} (\alpha + \alpha_\xi) [1+(\alpha + \alpha_\xi)\cos\theta]}}
	\equiv \tilde{\cF}_{n_2^{(I)},m,k}^{(\Lambda)} (\theta, \phi, \alpha_\xi)
	, \label{basis_cU}
\end{equation}
which also forms a complete set of orthonormal functions; obviously, $\tilde{\cF}_{n_2^{(0)},m,k}^{(\Lambda)} \equiv \tilde{\cF}_{n_2,m,k}^{(\Lambda)}$~(\ref{basis_def}).
In the new basis~(\ref{basis_cU}), the matrix elements of $\tilde{H}^{\text{(2D)}}_\text{free}$ and $\tilde{T}_3^{(\theta)} (I = 0)$ are similar to~(\ref{H_elms_matr}) and (\ref{T30}), respectively, with the exception that $n_2$ should be replaced by the \textit{transformed quantum number}
\begin{equation}
	n_2^{(I)} \equiv n_2-\theta_\bA(I)/(2\pi) \equiv n_2 + I/I_\text{per} .
	\label{n2_cU}
\end{equation}
For big enough $n_2^{(I)}$ or $a$, the diagonal matrix elements dominate and the corresponding eigenvalues approximations are
%
\begin{equation}
	\tilde{E}_{n_2^{(I)}, m, 1}^\text{(2D)} \equiv
	a^2 \left[\left(n_2^{(I)}\right)^{2}-\frac{1}{4}
	+\frac{a}{\left(a^{2}-1\right)^{\frac{3}{2}}} \left(m^{2}-\frac{1}{4}\right)
	\right]
	, \label{E_n2_cU}
\end{equation}
as seen in Fig.~\ref{fig_H_2D}.

We notice that, in general, $\tilde{E}_{n_2^{(I)}, m, 1}^\text{(2D)}$~(\ref{E_n2_cU}) are not degenerate two-by-two, as are the diagonal matrix elements in~(\ref{H_elms_matr}) for $\pm n_2$, except in some special situations: to have double degeneracy in Eq.~(\ref{E_n2_cU}), there should be two distinct integers, $n_2$ and $n_2'$, such that
\begin{equation}
	n_2 + I/I_\text{per} = \pm \left[n_2' + I/I_\text{per}\right]
	. \label{deg_n2_1}
\end{equation}
For the plus sign on the right hand side of Eq.~(\ref{deg_n2_1}) we obtain the trivial, not interesting solution, $n_2=n_2'$.
The interesting solution of~(\ref{deg_n2_1}) corresponds to the minus sign on the right hand side, which gives
\begin{equation}
	\frac{\theta_\bA(I)}{2\pi} \equiv -\frac{I}{I_\text{per}} = \frac{n_2+n_2'}{2} ,
	\label{deg_n2_2}
\end{equation}
and corresponds to integer and half-integer values of $I/I_\text{per}$, as observed in Fig.~\ref{fig_H_2D}.

From Eqs.~(\ref{E_n2_cU}) and (\ref{deg_n2_2}) we may extract the energy level $l$ corresponding to each $n_2^{(I)}$,
\begin{equation}
l \equiv
\begin{cases}
	2 \left|\round{n_2^{(I)}}\right|, \vphantom{\frac{1}{\round{n_2^{(I)}}}} & \text{if $\left\{n_2^{(I)} > 0, n_2^{(I)} \ge \round{n_2^{(I)}}\right\}$
	or $\left\{n_2^{(I)} < 0, n_2^{(I)} < \round{ n_2^{(I)} }\right\}$} , \\
	2 \left| \round{n_2^{(I)}} \right| - 1 , & \text{if $\left\{n_2^{(I)} < 0, n_2^{(I)} \ge \round{n_2^{(I)}}\right\}$ or $\left\{n_2^{(I)} > 0, n_2^{(I)} < \round{n_2^{(I)}}\right\}$},
\end{cases}
\label{defs_l}
\end{equation}
where $\lfloor x \rceil$ is the integer part of $x$ (the closest integer to $x$, with the convention that $\round{0.5} \equiv 1$ and $\round{-0.5} \equiv 0$).

\subsection{The expectation values of $\tilde{T}_3^{(\theta)}$} \label{subsec_Disc_T3}

Similarly, the matrix elements of $\tilde{T}_3^{(\theta)}(I)$ may be read off Eqs.~(\ref{T30}) [for $\tilde{T}_3^{(\theta)} (I = 0) = \tilde{T}_{3,0}^{(\xi)} + \tilde{T}_{3,0}^{(\theta)}$], replacing $n_2$ by $n_2^{(I)}$.
Since in the Hamiltonian eigenfunctions the diagonal elements dominate, the expectation values of $\tilde{T}_3^{(\theta)} (I)$ may be approximated, in general, by
\begin{equation}
	\left\langle \tilde{T}_3^{(\theta)} (I) \right\rangle_{l,m,1}
	\approx
	- \frac{5n_2^{(I)}(l)}{2} \equiv - \Delta \tilde{T}_3 \left(n_2(l) + \frac{I}{I_\text{per}}\right)
	, \label{T3_I_approx}
\end{equation}
which is seen in Fig.~\ref{fig_T3_2D} for the upper energy levels.
Equation~(\ref{T3_I_approx}) exposes also the ``quanta'' of toroidal dipole $\Delta \tilde{T}_3 \equiv 5/2$ of our system.
Combining~(\ref{E_n2_cU}) with (\ref{T3_I_approx}), we obtain
\begin{equation}
	\tilde{E}_{l, m, 1}^\text{(2D)} \equiv
	a^2 \left[\frac{4}{25} \left\langle \tilde{T}_3^{(\theta)} (I) \right\rangle_{l,m,1}^{2}-\frac{1}{4}
	+\frac{a}{\left(a^{2}-1\right)^{\frac{3}{2}}} \left(m^{2}-\frac{1}{4}\right)
	\right]
	, \label{E_cU_T3}
\end{equation}
which is a direct way to calculate the toroidal dipole of the particle on the energy level $l$.

\subsubsection{Abrupt changes of $\langle \tilde{T}_3^{(\theta)} (I) \rangle_{l,m,1}$ around integer and half-integer values of $I/I_\text{per}$} \label{subsubsec_Dn_2I}

We show now that $\langle \tilde{T}_3^{(\theta)} (I) \rangle_{l,m,1}$~(\ref{T3_I_approx}) has abrupt changes around $I/I_\text{per} = \eta/2$, where $\eta$ is an integer.
For this, let us fix a sufficiently large excitation level $l$, such that Eq.~(\ref{E_n2_cU}) applies for our arguments, and a positive integer $\eta$; for clarity, let's assume that $l$ is even and $0 \le \eta < l/2$.
Then, we define the integers $n_2 \equiv l/2-\eta >0$ and $n_2' \equiv -l/2-\eta <0$, such that
\begin{equation}
	l = 2(n_2+\eta) \equiv 2 \left. n_2^{(I)}\right|_{n_2=l/2-\eta, I=\eta I_\text{per}}
	= 2(-n_2'-\eta) = 2|n_2'+\eta| \equiv
	2 \left| n_2^{(I)} \right|_{n_2=-(l/2+\eta), I=\eta I_\text{per}} .
	\label{jumps_l}
\end{equation}
If we choose $I$ so that $\eta \le I/I_\text{per} < \eta + 0.5$ (that is, $\round{I/I_\text{per}} = \eta$), then, according to Eqs.~(\ref{defs_l}) and the definition~(\ref{jumps_l}), $\left.n_2^{(I)}\right|_{n_2=l/2-\eta}$ corresponds to the level $l$ (notice that $l/2 \le \left.n_2^{(I)}\right|_{n_2=l/2-\eta} \equiv l/2-\eta+I/I_\text{per} < l/2+0.5$), whereas $\left.\left(n_2'\right)^{(I)}\right|_{n_2'= -l/2-\eta}$ corresponds to the level $l-1$ (notice that $-l/2+0.5 > \left.\left(n_2'\right)^{(I)}\right|_{n_2'= -l/2-\eta, I} \equiv -l/2-\eta + I/I_\text{per} \ge -l/2$).
On the other hand, if $\eta-0.5 \le I/I_\text{per} < \eta$ (again, $\round{I/I_\text{per}} = \eta$), then $\left.n_2^{(I)}\right|_{n_2=l/2-\eta}$ ($l/2-0.5 \le \left.n_2^{(I)}\right|_{n_2=l/2-\eta} \equiv l/2-\eta+I/I_\text{per} < l/2$) corresponds to the level $l-1$~(\ref{jumps_l}), whereas $\left.\left(n_2'\right)^{(I)}\right|_{n_2'= -l/2-\eta}$ corresponds to the level $l$ (since $-l/2 > \left.\left(n_2'\right)^{(I)}\right|_{n_2'= -l/2-\eta} \equiv -l/2-\eta + I/I_\text{per} \ge -l/2 - 0.5$).
Therefore, with increasing current, around $I = \eta I_\text{per}$, the level $l$, defined by~(\ref{jumps_l}), switches from the transformed quantum number $\left.\left(n_2'\right)^{(I)}\right|_{n_2'= -l/2-\eta, I = \eta I_\text{per}} \equiv -l/2$ to $\left. n_2^{(I)}\right|_{n_2= l/2-\eta, I=\eta I_\text{per}} = l/2$, whereas the level $l-1$ switches from the transformed quantum number $\left.n_2^{(I)}\right|_{n_2=l/2-\eta, I = \eta I_\text{per}} \equiv l/2$ to $\left.\left(n_2'\right)^{(I)}\right|_{n_2'= -l/2-\eta, I=\eta I_\text{per}} \equiv -l/2$.
This implies that the expectation value $\langle \tilde{T}_3^{(\theta)} (I) \rangle_{l,m,1}$ changes from $2.5 (l/2)$ to  $-2.5 (l/2)$, whereas $\langle \tilde{T}_3^{(\theta)} (I) \rangle_{l-1,m,1}$ changes from $-2.5 (l/2)$ to  $2.5 (l/2)$, as observed in Fig.~\ref{fig_T3_2D}.

Now we apply a similar argument for the interval $\eta+0.5 \le I/I_\text{per} < \eta+1$ (that is, $\round{I/I_\text{per}} = \eta+1$).
Then, according to~(\ref{defs_l}) and (\ref{jumps_l}), $\left.n_2^{(I)}\right|_{n_2=l/2-\eta-1}$ corresponds to the level $l-1$ (since $l/2-0.5 \le \left.n_2^{(I)}\right|_{n_2=l/2-\eta-1} \equiv l/2-\eta-1+I/I_\text{per} < l/2$), whereas $\left.\left(n_2'\right)^{(I)}\right|_{n_2'= -l/2-\eta-1}$ corresponds to the level $l$ (since $-l/2 > \left.\left(n_2'\right)^{(I)}\right|_{n_2'= -l/2-\eta-1} \equiv -l/2-\eta-1 + I/I_\text{per} \ge -l/2-0.5$).
Therefore, as the current increases, around $I = (\eta+0.5) I_\text{per}$, the level $l$ switches from the transformed quantum number $\left. n_2^{(I)}\right|_{n_2= l/2-\eta, I=(\eta+0.5) I_\text{per}} = l/2+0.5$ to $\left.\left(n_2'\right)^{(I)}\right|_{n_2'= -l/2-\eta-1, I=(\eta+0.5) I_\text{per}} = -l/2-0.5$, whereas the level $l-1$ switches from the transformed quantum number $\left.\left(n_2'\right)^{(I)}\right|_{n_2'= -l/2-\eta, I=(\eta+0.5) I_\text{per}} = -l/2+0.5$ to $\left.n_2^{(I)}\right|_{n_2=l/2-\eta-1, I = (\eta+0.5) I_\text{per}} \equiv l/2-0.5$.
This implies that $\langle \tilde{T}_3^{(\theta)} (I) \rangle_{l,m,1}$ changes from $-2.5 (l/2+0.5)$ to  $2.5 (l/2-0.5)$, whereas $\langle \tilde{T}_3^{(\theta)} (I) \rangle_{l-1,m,1}$ changes from $2.5 (l/2-0.5)$ to  $-2.5 (l/2-0.5)$, as also observed in Fig.~\ref{fig_T3_2D}.

These abrupt changes may also be observed in the simulations of Table~\ref{tab_films}.
In these simulations we see the change of the Hamiltonian eigenvectors as the current is gradually increased.
If, at low current, but high enough energy level or $a$, the vectors of the basis (\ref{basis_def}) or (\ref{basis_cU}) are good approximations for the energy eigenvectors, as I increases, more and more basis vectors start to significantly contribute to the eigenvectors.
This may seem to be in contradiction with our observations of Section~\ref{subsec_appr_sol}, where we approximate the \textit{transformed} eigenvectors~(\ref{psi_cU_tot}) by the vectors of the \textit{transformed} basis~(\ref{basis_cU}).
But, for example, $\tilde{\cF}_{n_2^{(I)},m,k}^{(\Lambda)} (\theta, \phi, \alpha_\xi)$~(\ref{basis_cU}) is an approximation for the eigenvector corresponding to the energy $\tilde{E}_{n_2^{(I)}, m, 1}^\text{(2D)}$~(\ref{E_n2_cU}), and this corresponds, in the original Hilbert space, to $\cU^{-1}\tilde{\cF}_{n_2^{(I)},m,k}^{(\Lambda)} (\theta, \phi, \alpha_\xi)$, which is not anymore one of the vectors in the basis~(\ref{basis_def}), so more and more basis vectors significantly contribute to the expansion of the energy eigenfunction, as $I$ increases.

\subsection{Special points: $I/I_\text{per}$ is integer or half-integer} \label{subsec_I_int}

When $I/I_\text{per} = \eta/2$, for some integer $\eta$, the set of transformed quantum numbers $\{n_2^{(I)}\}$ comprises all integers (the set $\mathbb{Z}$) if $I/I_{\text{per}}$ is an integer, or all half-integers if $I/I_{\text{per}}$ is a half-integer.
Consequently, the set $\{n_2^{(I)}\}$ is always symmetric about zero, meaning that if $n_2^{(I)}$ is an element, then so is $-n_2^{(I)}$. Additionally, $n_2^{(I)}=0$ is included in this set if $I/I_{\text{per}}$ is an integer.
Leveraging this symmetry property, we can construct symmetric and antisymmetric bases, as shown also in Appendix~\ref{app_sec_sym_basis}. These bases are denoted by $\{\tilde{\cF}^{(\pm)}_{n_2^{(I)},m,k}\}$, where $n_2^{(I)} \ge 0$.
If $I/I_{\text{per}}$ is an integer, then this basis $\{\tilde{\cF}^{(\pm)}_{n_2^{(I)},m,k}\}$ is identical to $\{\tilde{\cF}^{(\pm)}_{n_2,m,k}\}$ from Eq.~(\ref{defs_Fpm}) (although the subscript $n_2^{(I)}$ may be retained if it enhances textual clarity).
On the other hand, when $I/I_{\text{per}}$ is a half-integer, we define:
\begin{equation}
	\tilde{\cF}^{(\pm)}_{n_2^{(I)},m,k} \equiv \frac{1}{\sqrt{2}} \left[ \tilde{\cF}^{(\Lambda)}_{n_2^{(I)},m,k} \pm \tilde{\cF}^{(\Lambda)}_{-n_2^{(I)},m,k} \right] ,
	\quad \text{where $n_2^{(I)} = \frac{1}{2}, \frac{3}{2}, \frac{5}{2}, \ldots$}.
	\label{Fpm_h_int}
\end{equation}
%

The matrix elements of $\tilde{\cH}^\text{(2D)}$ in either basis $\{\tilde{\cF}^{(\pm)}_{n_2^{(I)},m,k}\}$ (corresponding to $I/I_{\text{per}}$ integer or half-integer) may be directly read off Eqs.~(\ref{tH_pm_matrix}).
We first observe that in both cases
\begin{subequations} \label{tH_pm_eta}
\begin{equation}
	0 = \left\langle \tilde{\cF}^{(+)}_{n_1^{(I)},m_1,k_1} \left| \tilde{\cH}^{(2D)}_\text{free} \right| \tilde{\cF}^{(-)}_{n_2^{(I)},m_2,k_2} \right\rangle
	= \left\langle \tilde{\cF}^{(-)}_{n_1^{(I)},m_1,k_1} \left| \tilde{\cH}^{(2D)}_\text{free} \right| \tilde{\cF}^{(+)}_{n_2^{(I)},m_2,k_2} \right\rangle ,
	\label{H_eta_mp}
\end{equation}
for any $n_1^{(I)}, n_2^{(I)} \ge 0$.
Second, if $I/I_\text{per}$ is an integer, we have
\begin{eqnarray}
	\left\langle \tilde{\cF}^{(+)}_{0,m,k} \left| \tilde{\cH}^{(2D)} \right| \tilde{\cF}^{(+)}_{0,m,k} \right\rangle
	&=& \left\langle \tilde{\cF}^{(\Lambda)}_{0,m,k} \left| \tilde{\cH}^{(2D)} \right| \tilde{\cF}^{(\Lambda)}_{0,m,k} \right\rangle
	= a^2 \left[ -\frac{1}{4}
	+ \frac{a}{\left(a^{2}-1\right)^{\frac{3}{2}}} \left(m^{2}-\frac{1}{4}\right) \right]
	, \label{H_diag_eta_0} \\
	\left\langle \tilde{\cF}^{(+)}_{0,m,k} \left| \tilde{\cH}^{(2D)} \right| \tilde{\cF}^{(+)}_{n_2^{(I)},m,k} \right\rangle
	&=& \frac{\sqrt{2} a^2 \left(n_2^{(I)} \sqrt{a^{2}-1}+a \right) \left(\sqrt{a^{2}-1}-a \right)^{n_2^{(I)}}}{\left(a^{2}-1\right)^{\frac{3}{2}}}
	\left(m^{2}-\frac{1}{4}\right)
	\nonumber \\
	&=& \left\langle \tilde{\cF}^{(+)}_{n_2^{(I)},m,k} \left| \tilde{\cH}^{(2D)} \right| \tilde{\cF}^{(+)}_{0,m,k} \right\rangle ,
	\label{H_diag_eta_0n2}
\end{eqnarray}
for any $n_2^{(I)} > 0$.
Finally, for $I/I_\text{per}$ integer or half-integer, and any $n_2^{(I)} , n_2^{(I)}-n > 0$, $n\ne 0$, we have
\begin{eqnarray}
	\left\langle \tilde{\cF}^{(\pm)}_{n_2^{(I)},m,k} \left| \tilde{\cH}^{(2D)} \right| \tilde{\cF}^{(\pm)}_{n_2^{(I)},m,k} \right\rangle
	&=& a^2 \left[ \left(n_2^{(I)}\right)^{2}-\frac{1}{4}
	+ \frac{a \pm \left({2n_2^{(I)}} \sqrt{a^{2}-1}+a \right) \left(\sqrt{a^{2}-1}-a \right)^{{2n_2^{(I)}}}}{\left(a^{2}-1\right)^{\frac{3}{2}}}
	\left(m^{2}-\frac{1}{4}\right) \right]
	, \label{H_diag_eta} \\
	\left\langle \tilde{\cF}^{(\pm)}_{n_2^{(I)}-n,m,k} \left| \tilde{\cH}^{(2D)} \right| \tilde{\cF}^{(\pm)}_{n_2^{(I)},m,k} \right\rangle
	&=&
	\frac{a^2}{\left(a^{2}-1\right)^{\frac{3}{2}}} \left[\left({|n|} \sqrt{a^{2}-1}+a \right) \left(\sqrt{a^{2}-1}-a \right)^{{|n|}}
	\right. \nonumber \\
	&& \left. \pm \left((2n_2^{(I)}-n) \sqrt{a^{2}-1}+a \right) \left(\sqrt{a^{2}-1}-a \right)^{{2n_2^{(I)}-n}}\right] \left(m^{2}-\frac{1}{4}\right)
	. \label{H_pp_mm_eta}
\end{eqnarray}
\end{subequations}
For the matrix elements of $\tilde{T}_3$ we immediately obtain in both cases
\begin{equation}
	\left\langle \tilde{\cF}^{(+)}_{n_2^{(I)}-n,m,k} \left| \tilde{T}_3(I=0) \right| \tilde{\cF}^{(+)}_{n_2^{(I)},m,k} \right\rangle
	= 	\left\langle \tilde{\cF}^{(-)}_{n_2^{(I)}-n,m,k} \left| \tilde{T}_3(I=0) \right| \tilde{\cF}^{(-)}_{n_2^{(I)},m,k} \right\rangle
	= 0 ,
	\label{T3_Fpm_I}
\end{equation}
as in the case with no current.

Equations~(\ref{tH_pm_eta}) and (\ref{T3_Fpm_I}) are analogous to Eqs.~(\ref{tH_pm_matrix}) and (\ref{T3_Fpm}) for the specific case where $I=0$.
The Hilbert space $H$ can be decomposed into two orthogonal subspaces, $H_I^{(+)}$ and $H_I^{(-)}$, spanned by the basis sets $\left\{\mathcal{F}^{(+)}_{n^{(I)},m,k}\right\}$ and $\left\{\mathcal{F}^{(-)}_{n^{(I)},m,k}\right\}$, respectively:  $H \equiv H_I^{(+)} \oplus H_I^{(-)}$.
Then, Eq.~(\ref{H_eta_mp}) implies that the eigenfunctions of the 2D Hamiltonian $\left.\hat{\mathcal{H}}^\text{(2D)}(I)\right|_{I/I_\text{per}=\eta/2}$ are confined entirely within either $H_I^{(+)}$ or $H_I^{(-)}$, whereas Eq.~(\ref{T3_Fpm_I}) implies that the expectation values of the operator $\tilde{T}_3^{(\theta)}$ on the Hamiltonian eigenvectors are identically zero, as it may be seen in Fig.~\ref{fig_T3_2D}.
Furthermore, the energy levels are better approximated by the expression~(\ref{H_diag_eta}), which give the energy splitting
\begin{eqnarray}
	\left. \Delta\tilde{E}_{n_2^{(I)}, m, 1}^\text{(2D)} \right|_{I/I_\text{per}=\eta/2}
	&\equiv& \left\langle \tilde{\cF}^{(\pm)}_{n_2^{(I)},m,k} \left| \tilde{\cH}^{(2D)} \right| \tilde{\cF}^{(\pm)}_{n_2^{(I)},m,k} \right\rangle
	- \left\langle \tilde{\cF}^{(\pm)}_{n_2^{(I)},m,k} \left| \tilde{\cH}^{(2D)} \right| \tilde{\cF}^{(\pm)}_{n_2^{(I)},m,k} \right\rangle
	\label{DE_Ieta2} \\
	&=&
	\frac{2a^2 \left({2n_2^{(I)}} \sqrt{a^{2}-1}+a \right) \left(\sqrt{a^{2}-1}-a \right)^{{2n_2^{(I)}}}}{\left(a^{2}-1\right)^{\frac{3}{2}}}
	\left(m^{2}-\frac{1}{4}\right)
	, \nonumber
\end{eqnarray}
rather than (\ref{E_n2_cU}), according to which $\tilde{E}_{n_2^{(I)}, m, 1}^\text{(2D)} = \tilde{E}_{-n_2^{(I)}, m, 1}^\text{(2D)}$. Consequently, these energy levels are not degenerate, although $\left. \Delta\tilde{E}_{n_2^{(I)}, m, 1}^\text{(2D)} \right|_{I/I_\text{per}=\eta/2}$ diminishes rapidly with increasing values of $a$ and $n_2^{(I)}$. 

\subsection{The full 3D problem} \label{subsec_3D}

The full effective Hamiltonian is given by Eq.~(\ref{def_Heff}), with the matrix elements of $\tilde{\cH}^{(\xi)}_{\bA,2}$ given by~(\ref{H_xi_A2}).
If the matrix elements of $\tilde{\cH}^{\text{(2D)}}$ are much bigger than those of $\tilde{\cH}^{(\xi)}_{\bA,2}$, then we can say that the system is quasi-2D.
If, moreover, the diagonal matrix elements of $\tilde{\cH}_\text{free}^\text{(2D)}$~(\ref{H_diag}) are dominant (as it happens when $a$ or the energy level $l$ are big enough, as explained before), we recover some of the properties described in the previous subsections of this section, like, for example, the quasi-double degeneracy of the high energy levels at $I/I_\text{per} = \eta/2$, as can be observed for the upper energy levels in Fig.~\ref{fig_Etot}.

\begin{figure}
	\centering
	\includegraphics[width=0.5\linewidth]{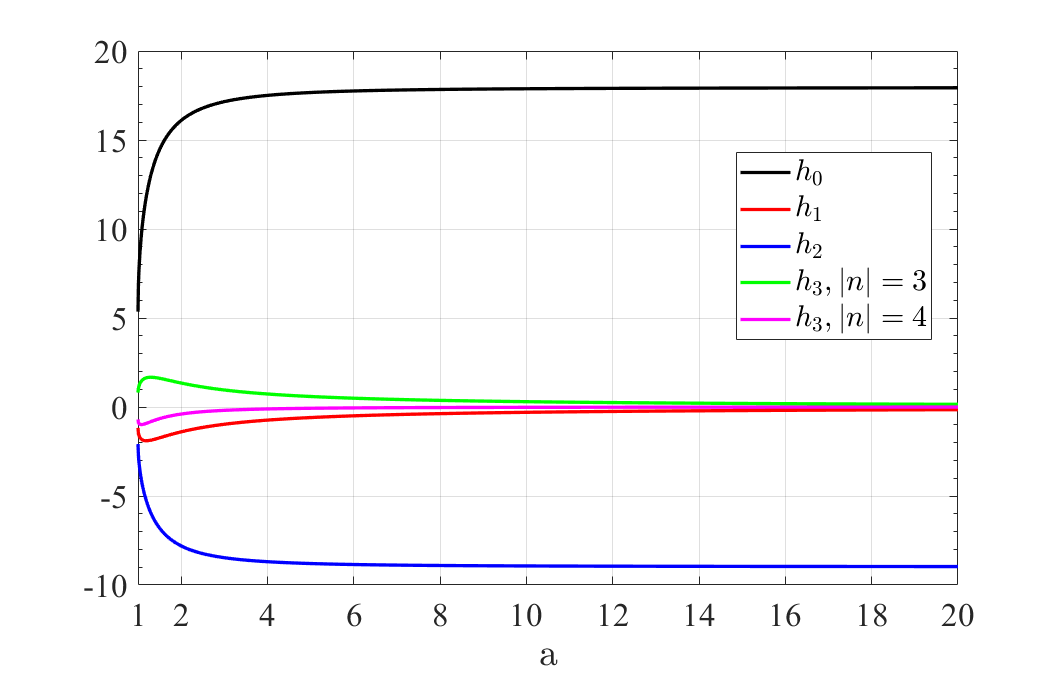}
	\caption{The functions $h_0\left(a; L/R\right)/a^2$, $h_1\left(a; L/R\right)/a^2$, $h_2\left(a; L/R\right)/a^2$, and $h_3\left(a; L/R, n\right)/a^2$, for $n=3,4$, and $L/R=10$.
	}
	\label{fig_h0h1h2h3a2}
\end{figure}

To better understand these properties, we compare the matrix elements~(\ref{H_xi_A2}) with the diagonal elements~(\ref{H_diag}).
Considering that the diagonal elements are proportional to $a^2$, for the matrix elements~(\ref{H_xi_A2}) we write
\begin{subequations} \label{H_xi_A2_comp}
\begin{equation}
	\left\langle \tilde{\cF}_{n_2-n,m,1}^{\Lambda} \left| \tilde{\cH}^{(\xi)}_{\bA,2} \right| \tilde{\cF}_{n_2,m,1}^{\Lambda} \right\rangle
	\equiv \frac{I^2}{I_\text{per}^2} a^2
	\left[ h_0\left(a;\frac{L}{R}\right) + h_1\left(a;\frac{L}{R}\right) + h_2\left(a;\frac{L}{R}\right) + h_3\left(a;\frac{L}{R}\right) \right]
	\label{H_xi_A2_h}
\end{equation}
where
\begin{eqnarray}
	h_0\left(a;\frac{L}{R}\right)
	&\equiv& \delta_{n,0} P(a)
	\left[ \log(\frac{2aL}{R}) \left\{ \log\left[\frac{8aL}{R \left(a+\sqrt{a^2-1}\right)^2}\right] - \left(a - \sqrt{a^2-1}\right)^{2} \right\}
	+ {\tilde{I}_0^{(\ln^2,-)}}(a)
	\right] , \nonumber \\
	&& 
	\label{C0} \\
	h_1\left(a;\frac{L}{R}\right)
	&\equiv& \delta_{|n|,1} P(a)
	\left[ - \log(\frac{2aL}{R}) \frac{2\left(-a +\sqrt{a^{2}-1}\right)^2 \left(a + 2\sqrt{a^{2}-1}\right)}{3}
	+ {\tilde{I}_n^{(\ln^2,-)}}(a)
	\right] , \label{C1} \\
	h_2\left(a;\frac{L}{R}\right)
	&\equiv& \delta_{|n|,2} P(a)
	\left[ - \log(\frac{2aL}{R})
	\frac{1}{2} \Biggl\{ \log\left[\frac{8aL}{R \left(a+\sqrt{a^2-1}\right)^2}\right]
	- \frac{\left(3a + 5\sqrt{a^{2}-1}\right) \left(a - \sqrt{a^{2}-1}\right)^{3}}{2}
	\Biggr\}
	\right. \nonumber \\
	&& \left.
	+ {\tilde{I}_n^{(\ln^2,-)}}(a) \right]
	, \label{C2} \\
	h_3\left(a;\frac{L}{R}, n\right)
	&\equiv& P(a)
	\Biggl[ - \log(\frac{2aL}{R}) \frac{4 \left[ \left(a^{2}-1\right) n^{2} + 2a|n|\sqrt{a^{2}-1} + 2 \right] \left(-a + \sqrt{a^2-1}\right)^{|n|}}{|n| (n^2-4)}
	+ {\tilde{I}_n^{(\ln^2,-)}}(a) \Biggr] , \
	\text{if $|n|>2$} , \nonumber \\
	&& \label{Cg2} \\
	\text{with} \quad P(a) &\equiv& \frac{1}{2 a^2 \left(a-\sqrt{a^2-1}\right)^2} ,
	\label{def_P}
\end{eqnarray}
\end{subequations}
are plotted in Fig.~\ref{fig_h0h1h2h3a2}.

Analyzing the prefactor $P(a)$, we observe that
\begin{equation}
	P(1) = \frac{1}{2} , \quad
	P(a) \stackrel{a\gg1}{\approx} 2-\frac{1}{a^{2}} , \quad \text{whereas} \quad
	P(a) < 2 , \quad
	\frac{\dd P(a)}{\dd a} = \frac{1}{\sqrt{a^{2}-1} \left(a -\sqrt{a^{2}-1}\right) a^{3}} > 0
	\quad \text{for $\forall a>1$.}
	\label{analysis_P}
\end{equation}
Further, we have the limiting cases of Eqs.~(\ref{H_xi_A2_comp}):
\begin{subequations} \label{h0_h1_h2_h3_lim1}
\begin{eqnarray}
	h_0\left(1;\frac{L}{R}\right)
	&=& \frac{I^2}{I_\text{per}^2} \frac{\delta_{n,0}}{2}
	\left[ \log(\frac{2L}{R}) \left\{ \log\left[\frac{8L}{R}\right] - 1 \right\}
	+ \tilde{I}_0^{(\ln^2,-)}(1) \right]
	, \label{h0_lim1} \\
	h_1\left(1;\frac{L}{R}\right)
	&=& \frac{I^2}{I_\text{per}^2} \frac{\delta_{|n|,1}}{2}
	\left[ - \log(\frac{2L}{R}) \frac{2}{3}
	+ \tilde{I}_1^{(\ln^2,-)}(1) \right] \vphantom{\Bigg(}
	, \label{h1_lim1} \\
	h_2\left(1;\frac{L}{R}\right)
	&=& \frac{I^2}{I_\text{per}^2} \frac{\delta_{|n|,2}}{2}
	\left[ - \log(\frac{2L}{R})
	\frac{1}{2} \left\{ \log\left[\frac{8L}{R}\right]
	- \frac{3}{2} \right\}
	+ \tilde{I}_2^{(\ln^2,-)}(1) \right]
	, \label{h2_lim1} \\
	h_3\left(1;\frac{L}{R}, n\right)
	&=& \frac{I^2}{I_\text{per}^2} \frac{1}{2}
	\left[ - \log(\frac{2L}{R}) \frac{8}{|n| (n^2-4)} \left(-1\right)^{|n|}
	+ {\tilde{I}_n^{(\ln^2,-)}}(1) \right] \vphantom{\Bigg(}
	, \nonumber \\
	&& \text{if $|n|>2$}
	, \label{h3_lim1}
\end{eqnarray}
\end{subequations}
with ${\tilde{I}_n^{(\ln^2,-)}}(1)$ plotted in Fig.~\ref{fig_inln2ma1},
and
\begin{subequations} \label{h0_h1_h2_h3_lim_inf}
\begin{eqnarray}
	h_0\left(a;\frac{L}{R}\right)
	&\stackrel{a\gg1}{\approx}& \delta_{n,0} \frac{I^2}{I_\text{per}^2}
	\left\{
	2 \ln \! \left(\frac{2 L}{R}\right)^{2}+\frac{1}{a^{2}} \left[ -\ln^2 \! \left(\frac{2 L}{R}\right) + \frac{1}{2} \ln \! \left(\frac{2 L}{R}\right) +\frac{1}{2}\right]
	\right\} , \label{h0_lim_inf} \\
	h_1\left(a;\frac{L}{R}\right)
	&\stackrel{a\gg1}{\approx}& \delta_{|n|,1} \frac{I^2}{I_\text{per}^2}
	\left[-\frac{\ln \! \left(\frac{2 L}{R}\right)+\ln \! \left(a \right)}{a}+\frac{2\ln \! \left(\frac{2 L}{R}\right) - 3 \ln a}{6 a^{3}} \right]
	, \label{h1_lim_inf} \\
	h_2\left(a;\frac{L}{R}\right)
	&\stackrel{a\gg1}{\approx}& - \delta_{|n|,2} \frac{I^2}{I_\text{per}^2}
	\left[\ln^2 \left(\frac{2 L}{R}\right) - 2\ln^2 \left(a \right)\right] \left(1-\frac{1}{2 a^{2}}\right)
	, \label{h2_lim_inf} \\
	h_3\left(a;\frac{L}{R}, n\right)
	&\stackrel{a\gg1}{\approx}& \frac{I^2}{I_\text{per}^2}
	\left[ - \log(\frac{2L}{R}) \frac{2 \left(4 |n| a^{2}+n^{2}-6 |n| +4\right)}{\left(|n| -2\right) |n|} \left(\frac{-1}{2a}\right)^{|n|}
	+ \tilde{I}_n^{(\ln^2,-)}(1) \right]
	, \quad \text{if $|n|>2$}
	, \label{h3_lim_inf}
\end{eqnarray}
\end{subequations}
where the expansions for $\tilde{I}_n^{(\ln^2,-)}(a)$ are given in~(\ref{In_ln2_p}) for $|n| \le 4$ and (\ref{asympt_exp_tIn_ln2}) for $|n| \ge 4$.
Obviously, in the limit $a\to\infty$ only the terms $h_0$ and $h_2$ survive, as can be seen also in Fig.~\ref{fig_h0h1h2h3a2}.

To make sure that $h_3\left(a;L/R, n\right)$ does not become big for certain values of $n$, from~(\ref{Cg2}) observe that
\begin{subequations} \label{h3_dep_n}
\begin{eqnarray}
	&& \left[ \left(a^{2}-1\right) n^{2} + 2a|n|\sqrt{a^{2}-1} + 2 \right] > 0 ,
	\label{h3_dep_n1} \\
	&& \frac{\partial}{\partial n} \left[ \frac{4 \left[ \left(a^{2}-1\right) n^{2} + 2a|n|\sqrt{a^{2}-1} + 2 \right]}{|n| (n^2-4)} \right]
	= \frac{-\left(a^{2}-1\right) n^{4}-4 \sqrt{a^{2}-1}\, a \,n^{3}-\left(2 a^{2}+1\right) n^{2}+8}{n^2 \left(n^{2}-4\right)^2} < 0
	, \label{h3_dep_n2} \\
	&& \text{and} \quad \frac{\partial \left(a - \sqrt{a^2-1}\right)^{|n|}}{\partial n} = \left(a -\sqrt{a^{2}-1}\right)^{|n|} \ln(a -\sqrt{a^{2}-1}) < 0
\end{eqnarray}
\end{subequations}
for any $|n|>2$ and $a>1$.
Combining Eqs.~(\ref{h3_dep_n}), we conclude that $|h_3(a;L/R, n)|$ is monotonically decreasing with $n$ in the region of interest, therefore its contributions to the eigenvalues and eigenfunctions problem is decreasing with $n$.

Now, let's compare the diagonal elements of $\tilde{\cH}_{\text{free}}^\text{(2D)}$ with the non-diagonal ones.
The ratio between the non-diagonal~(\ref{H_nondiag}) and diagonal~(\ref{H_diag}) elements is
\begin{subequations} \label{dominance}
\begin{equation}
	\cR^\text{(2D)}(a, n_2, n, m)
	\equiv
	\left| \frac{\langle \tilde{\cF}^{(\Lambda)}_{n_2-n,m, k} | \hat{\cH}^\text{(2D)}_\text{free} | \tilde{\cF}^{(\Lambda)}_{n_2,m, k} \rangle}{\langle \tilde{\cF}^{(\Lambda)}_{n_2,m, k} | \hat{\cH}^\text{(2D)}_\text{free} | \tilde{\cF}^{(\Lambda)}_{n_2,m, k} \rangle} \right|
	=
	\left| \frac{\frac{\left({|n|} \sqrt{a^{2}-1}+a \right) \left(\sqrt{a^{2}-1}-a \right)^{{|n|}}}{\left(a^{2}-1\right)^{\frac{3}{2}}} \left(m^{2}-\frac{1}{4}\right)}{\left(n_2^{2}-\frac{1}{4}\right)
	+\frac{a}{\left(a^{2}-1\right)^{\frac{3}{2}}} \left(m^{2}-\frac{1}{4}\right)} \right|
	\quad \text{for $n \ne 0$}.
	\label{domin_Hfree}
\end{equation}
We denote
\begin{equation}
	\cD_1 (a,n) \equiv \left({|n|} \sqrt{a^{2}-1}+a \right) \left(\sqrt{a^{2}-1}-a \right)^{{|n|}}
	\label{def_D1}
\end{equation}
and we have
\begin{equation}
	\frac{\partial \cD_1(a,n)}{\partial |n|}
	= \left(a -\sqrt{a^{2}-1}\right)^{|n|} \left[\left(|n| \sqrt{a^{2}-1}+a \right) \ln(a -\sqrt{a^{2}-1}) + \sqrt{a^{2}-1}\right]
	\quad \text{and} \quad
	\frac{\partial \cD_1(1,n)}{\partial |n|} = 0.
\end{equation}
But, since for $n \in \Z\setminus\{0\}$ and
\begin{equation}
\begin{cases}
	\vphantom{\bigg[}\Bigl[\left(|n| \sqrt{a^{2}-1}+a \right) \ln(a -\sqrt{a^{2}-1}) + \sqrt{a^{2}-1}\Bigr]_{a=1} = 0 , \\
	\frac{\partial\left[\left(|n| \sqrt{a^{2}-1}+a \right) \ln(a -\sqrt{a^{2}-1}) + \sqrt{a^{2}-1}\right]}{\partial a} = \frac{\left(n a +\sqrt{a^{2}-1}\right) \ln \! \left(a -\sqrt{a^{2}-1}\right)-n \sqrt{a^{2}-1}}{\sqrt{a^{2}-1}} \le 0,
	\quad \text{for $\forall a\ge 1$},
\end{cases}
\end{equation}
we conclude that the derivative $\partial \cD_1(a,n)/\partial |n|$ is negative for any $n \in \Z$, $a>1$, and $|n| \ge 1$, which implies that the non-diagonal terms~(\ref{H_nondiag}) decrease with $|n|$ for any $a$.
Furthermore, since
\begin{equation}
	\frac{\partial}{\partial a} \left[\frac{\left({|n|} \sqrt{a^{2}-1}+a \right)}{\left(a^{2}-1\right)^{\frac{3}{2}}}\right]
	=
	-\frac{2 a |n| \sqrt{a^{2}-1}+2 a^{2}+1}{\left(a^{2}-1\right)^{\frac{5}{2}}} < 0
	\quad \text{and} \quad
	\frac{\partial \left[\left(a -\sqrt{a^{2}-1}\right)^{|n|}\right]}{\partial a}
	=
	-\frac{|n| \left(a -\sqrt{a^{2}-1}\right)^{|n|}}{\sqrt{a^{2}-1}} < 0 ,
\end{equation}
\end{subequations}
than the non-diagonal terms~(\ref{H_nondiag}) decrease also with $a$ and we have the asymptotic expression
\begin{equation}
	\cR^\text{(2D)}(a, n_2, n, m) \stackrel{a\gg1}{\approx}
	\left(-\frac{1}{2 a}\right)^{n} \frac{n +1}{ a^{2} \left(n_{2}^{2}-\frac{1}{4}\right)} \left(m^{2}-\frac{1}{4}\right)
\end{equation}

In conclusion, the ratio $\cR^\text{(2D)}(a, n_2, n, m)$~(\ref{domin_Hfree}) decreases with $a$, $|n|$, and $|n_2|$, as long as the diagonal term~(\ref{H_diag}) is positive.
If $\cR^\text{(2D)}(a, n_2, n, m) \ll 1$ for any $n\ne0$, the non-diagonal terms may be neglected and the discussions of Sections~\ref{subsec_appr_sol}, \ref{subsec_Disc_T3}, \ref{subsec_I_int} apply to our problem.

To analyze the contribution of $\tilde{\cH}^{(\xi)}_{\bA,2}$ to the eigenvalues and eigenfunctions problem of the Hamiltonian, let's assume that $\cR^\text{(2D)}(a, n_2, n, m) \ll 1$ for any $n\ne0$, so that the eigenfunctions of $\tilde{\cH}^\text{(2D)}$ are well approximated by $\tilde{\cF}^{(\Lambda)}_{n_2,m, k}$ or $\tilde{\cF}^{(\pm)}_{n_2,m, k}$, as explained in Sections~\ref{subsec_appr_sol} and \ref{subsec_I_int}, respectively.
To estimate the contribution of the matrix elements of $\tilde{\cH}^{(\xi)}_{\bA,2}$, we define
\begin{subequations} \label{domin_Hxi}
\begin{equation}
	\cR^{(\xi)}(a, n_2, n, m)
	\equiv
	\frac{I_\text{per}^2}{I^2} \left| \frac{\langle \tilde{\cF}^{(\Lambda)}_{n_2\pm n,m, k} | \tilde{\cH}^{(\xi)}_{\bA,2} | \tilde{\cF}^{(\Lambda)}_{n_2,m, k} \rangle}{\langle \tilde{\cF}^{(\Lambda)}_{n_2,m, k} | \tilde{\cH}^\text{(2D)}_\text{free} | \tilde{\cF}^{(\Lambda)}_{n_2,m, k} \rangle}
	\right| , \label{domin_Hxi_n}
\end{equation}
so that
\begin{eqnarray}
	\cR^{(\xi)}(a, n_2, 0, m)
	&=&
	\left| \frac{h_0(a; L/R)}{n_2^{2}-\frac{1}{4}
	+\frac{a}{\left(a^{2}-1\right)^{\frac{3}{2}}} \left(m^{2}-\frac{1}{4}\right)}
	\right| , \label{domin_Hxi0} \\
	\cR^{(\xi)}(a, n_2, 1, m)
	&=&
	\left| \frac{h_1(a; L/R)}{n_2^{2}-\frac{1}{4}
	+\frac{a}{\left(a^{2}-1\right)^{\frac{3}{2}}} \left(m^{2}-\frac{1}{4}\right)} \right|
	, \label{domin_Hxi1} \\
	\cR^{(\xi)}(a, n_2, 2, m)
	&=&
	\left| \frac{h_2(a; L/R)}{n_2^{2}-\frac{1}{4}
	+\frac{a}{\left(a^{2}-1\right)^{\frac{3}{2}}} \left(m^{2}-\frac{1}{4}\right)} \right|
	, \label{domin_Hxi2} \\
	\cR^{(\xi)}(a, n_2, |n|>2, m)
	&=&
	\left| \frac{h_3(a; L/R, n)}{n_2^{2}-\frac{1}{4}
	+\frac{a}{\left(a^{2}-1\right)^{\frac{3}{2}}} \left(m^{2}-\frac{1}{4}\right)} \right|
	. \label{domin_Hxi3}
\end{eqnarray}
\end{subequations}
For small currents, $(I/I_\text{per})^2 \ll \left[\cR^{(\xi)}(a, n_2, n\ne0, m)\right]^{-1}$ (for the values of $a$ corresponding to Fig.~\ref{fig_h0h1h2h3a2}, it is enough to check that $(I/I_\text{per})^2 \ll \left[\cR^{(\xi)}(a, n_2, 2, m)\right]^{-1}$), the contribution from the diagonal terms are dominant and the eigenfunctions are well approximated by $\tilde{\cF}^{(\Lambda)}_{n_2,m, k}$ or $\tilde{\cF}^{(\pm)}_{n_2,m, k}$, as explained before.
In such a case, all the properties of the energy spectrum discussed in Sections~\ref{subsec_appr_sol} and \ref{subsec_I_int} are satisfied, except for the periodicity, since a term $(I/I_\text{per})^2 a^2 h_0(a; L/R)$ has to be added to all the energy eigenvalues:
\begin{eqnarray}
	\tilde{E}_{n_2^{(I)}, m, 1} &\equiv&
	a^2 \left[\left(n_2^{(I)}\right)^{2}-\frac{1}{4}
	+\frac{a}{\left(a^{2}-1\right)^{\frac{3}{2}}} \left(m^{2}-\frac{1}{4}\right)
	\right]
	+ \frac{I^2}{I_\text{per}^2}
	\left( \log(\frac{2aL}{R}) \left\{ \frac{\log\left[\frac{8aL}{R \left(a+\sqrt{a^2-1}\right)^2}\right]}{2 \left(a-\sqrt{a^2-1}\right)^2} - \frac{1}{2} \right\}
	\right. \nonumber \\
	&& \left. + \frac{{\tilde{I}_0^{(\ln^2,-)}}(a)}{2 \left(a-\sqrt{a^2-1}\right)^2}
	\right) . \label{E_n2_cU_3D}
\end{eqnarray}
This quadratic dependence of the energy on the current is observed in the upper energy levels of Fig.~\ref{fig_Etot}, for example.

As $I$ increases, when $(I/I_\text{per})^2$ becomes comparable to $\left[\cR^{(\xi)}(a, n_2, n, m)\right]^{-1}$ for at least one value $n\ne0$ (for the values of $a$ corresponding to Fig.~\ref{fig_h0h1h2h3a2}, we have the condition $(I/I_\text{per})^2 \approx \left[\cR^{(\xi)}(a, n_2, 2, m)\right]^{-1}$, since $|\cR^{(\xi)}(a, n_2, 2, m)| \ge |\cR^{(\xi)}(a, n_2, n, m)|$ for any $n\ne0$), the Hamiltonian matrix cannot be approximated anymore by its diagonal values and the eigenfunctions become significantly different from $\tilde{\cF}^{(\Lambda)}_{n_2,m, k}$ or $\tilde{\cF}^{(\pm)}_{n_2,m, k}$.
This can be seen in Figs.~\ref{fig_Etot} and \ref{fig_Etot0}, in the regions where the eigenvalues are not anymore quasi-degenerate with the periodicity $I/I_\text{per} = 1/2$.

The situation is similar for the expectation values of $\tilde{T}_3$, presented in Fig.~\ref{fig_T3}.
Combining Eqs.~(\ref{calc_T3xiA_fin}), (\ref{T2_cU}), and (\ref{T3_I_approx}) we obtain the approximation of the expectation values of the toroidal dipole at small currents (that is, when $(I/I_\text{per})^2 \ll \left[\cR^{(\xi)}(a, n_2, n, m)\right]^{-1}$, for any $n\ne0$):
\begin{eqnarray}
	\left\langle \tilde{T}_3 \right\rangle_{l,m,1}
	&\approx&
	- \frac{5n_2^{(I)}(l)}{2}
	+ \frac{I}{I_\text{per}}
	\Biggl\{
	\frac{(1 + 2a^2)}{2a (a - \sqrt{a^2-1})} \ln\left[\frac{4aL}{R \left(a+\sqrt{a^2-1}\right)}\right]
	+  \frac{1}{4a (a - \sqrt{a^2-1})} -  \frac{3}{2}
	\Biggr\} . \label{T3_I_approx_tot}
\end{eqnarray}
The linear dependence of $\langle \tilde{T}_3^{(\theta)} (I) \rangle_{l,m,1}$ on $I$ and its abrupt changes around the current values $I = (\eta/2)I_\text{per}$, where $\eta$ is an integer, may be observed in Fig.~\ref{fig_T3}, in regions where $(I/I_\text{per})^2 < \left[\cR^{(\xi)}(a, n_2, n, m)\right]^{-1}$.
Nevertheless, at transitions between different energy levels, say from the level $l_1$ to the level $l_2$, the quantization observed in Eq.~(\ref{T3_I_approx}) remains valid, namely
\begin{equation}
	\left\langle \tilde{T}_3 \right\rangle_{l_2,m,1} - \left\langle \tilde{T}_3 \right\rangle_{l_1,m,1} = \frac{5}{2} [n_2(l_1) - n_2(l_2)]
	\label{Delta_T3_l1l2}
\end{equation}
is an integer multiple of the quanta $2.5$, where $n_2(l_1), n_2(l_2)$ are given by Eq.~(\ref{defs_l}).
When $(I/I_\text{per})^2$ becomes comparable to $\left[\cR^{(\xi)}(a, n_2, n, m)\right]^{-1}$ for some values of $n$, the current dependence of $\langle \tilde{T}_3^{(\theta)} (I) \rangle_{l,m,1}$ changes qualitatively.

A more interesting consequence of Eqs.~(\ref{E_n2_cU_3D}) and (\ref{T3_I_approx_tot}) is that the transition energy has the very simple form
\begin{eqnarray}
	\tilde{E}_{n_2^{(I)}(l_2), m, 1} - \tilde{E}_{n_2^{(I)}(l_1), m, 1}
	&=& a^2 \left[n_2^2(l_2) - n_2^2(l_1)\right]
	- \frac{I}{I_\text{per}} \frac{4a^2}{5} \left[\left\langle \tilde{T}_3 \right\rangle_{l_2,m,1} - \left\langle \tilde{T}_3 \right\rangle_{l_1,m,1}\right]
	\nonumber \\
	&=& a^2 \left\{\left[n_2^2(l_2) - n_2^2(l_1)\right]
	+ 2 \frac{I}{I_\text{per}} \left[n_2(l_2) - n_2(l_1)\right] \right\}
	. \label{trans_en}
\end{eqnarray}
According to~(\ref{trans_en}), the transition energy is always linear in $I/I_\text{per}$ and the slope is proportional to the jump in the toroidal dipole between the two states.
Moreover, this slope is integer multiple of $2a^2$.
This provides a direct and simple method to observe and measure the toroidal dipole expectation values on the energy eigenstates.

\section{Conclusions} \label{sec_conclusions}

In this work, we have provided a complete theoretical framework for the quantum toroidal dipole (TD) in a nanostructure and proposed the first method for its direct, quantitative measurement. This resolves a foundational challenge that has persisted since the toroidal multipole family was conceptualized over 60 years ago.

Our approach is centered on linking the TD to a direct spectroscopic observable. By analyzing a quantum particle on a toroidal manifold, we demonstrated that an external current $I$, acting as a conjugate field, shifts the system's energy eigenvalues. The central result of this paper is the proof that this energy shift is directly proportional to the expectation value of the TD operator. We find that the transition energy $\Delta E$ between two eigenstates is linear in the applied current, following the simple relation $\Delta E \approx C_1 + C_2 \cdot I$, where the slope $C_2$ is directly proportional to the change in the toroidal dipole moment between the states, $\Delta \langle \hat{T}_3 \rangle$.

This finding provides a clear and unambiguous "smoking gun" for experiment. We have shown that the toroidal dipole, in a quantum-mechanical system, is quantized. The slope of the transition energy versus current is not continuous but is determined by the discrete quantum numbers of the eigenstates. Measuring this slope provides a direct, quantitative value for the TD, in contrast to all previous indirect methods, which rely on far-field scattering and simulation.

This work bridges the gap between the classical, macroscopic toroidal dipoles studied in metamaterials and the fundamental quantum nature of matter. It builds directly upon our previous work \cite{PhysicaA.598.127377.2022.Dolineanu, PhysicaScripta.98.015223.2023.Anghel}, which established the toroidal dipole operator as a self-adjoint (Hermitian) quantum observable. Having established it as an observable in theory, we now provide the practical blueprint to observe it in an experiment.

The principles laid out here are experimentally feasible in a variety of modern systems, including semiconductor quantum rings, bent carbon nanotubes, or engineered optical lattices. The ability to measure and control the quantum TD opens significant new avenues for research, from engineering quantum metamaterials with novel electromagnetic properties to probing fundamental physics, such as parity-violating anapole moments in molecules. By providing the tools to measure what was once hidden, this work completes the quantum multipole toolkit.

\section{Acknowledgments}

Discussions with Profs. V. O. Nesterenko and S. Sen are gratefully acknowledged.
We acknowledge the financial support by the Ministry of Education, UEFISCDI
projects PN~23210101 and PN~23210204.
We would like to thank Google's AI assistant, Gemini, for its help in refining the language and structure of the manuscript, improving the cover letter, and formatting LaTeX code.

\appendix

\section{Analytical expression for $I_n(a)$} \label{app_sec_AnSol}

To prove Eq.~(\ref{In_Inln}) we first write
\begin{eqnarray}
	\frac{1}{2}(I_{n+1}+I_{n-1})+aI_n = \frac{1}{2\pi}\int_0^{2\pi} \frac{e^{in\theta}\frac{1}{2}(e^{i\theta}+e^{-i\theta})+e^{in\theta} a}{a+\cos(\theta)} \dd\theta =\frac{1}{2\pi} \int_0^{2\pi} e^{in\theta} \dd\theta = \delta_{n} .
	\label{req11}
\end{eqnarray}
If we assume that $I_n(a) \equiv C(a) \lambda^n(a)$, then, for $n>0$ Eq.~(\ref{req11}) gives us the recurrence relation
\begin{eqnarray}
	\lambda^2+2a\lambda+1=0 \label{ch_eq_In}
\end{eqnarray}
with the general solutions
\[
I_n(a) = C(a) \left(-a\pm\sqrt{a^2-1}\right)^n .
\]
Observing that
\begin{equation}
	- a - \sqrt{a^2-1} = \frac{1}{ - a + \sqrt{a^2-1}} ,
	\quad 	I_n = I_n^* = I_{-n} ,
	\quad {\rm and} \quad
	I_0(a) > 0 ,
	\label{identities}
\end{equation}
we obtain the final solution
\begin{equation}
	I_n(a) = \frac{\left(-a+\sqrt{a^2-1}\right)^{|n|}}{\sqrt{a^2-1}}
	, \label{In_final}
\end{equation}
which is the first equation of~(\ref{In_Inln}).
The second equation of~(\ref{In_Inln}) follows simply from the first one.

\section{Asymptotic relations for $I_n^{(\ln^2)}(a)$ and $\tilde{I}_n^{(\ln^2,\pm)}(a)$} \label{app_sec_asympt_Iln2}

\begin{figure}
	\centering
	\includegraphics[width=0.5\linewidth]{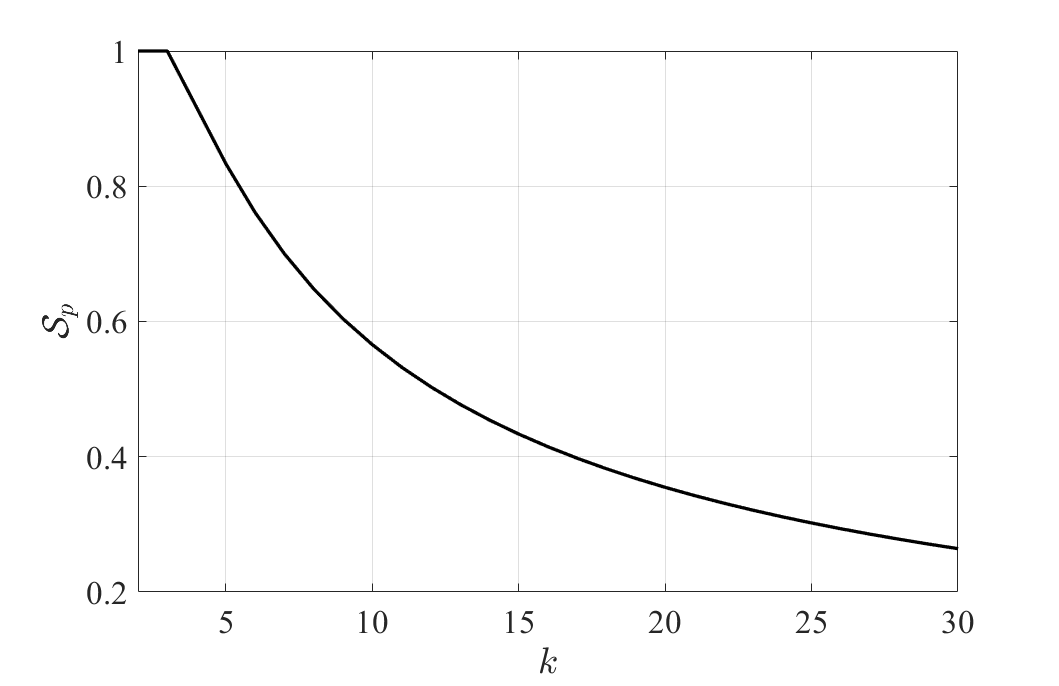}
	\caption{The partial summation $\cS_\text{part}(k)$ vs $k$. We observe a rapid decrease with $k$.}
	\label{fig_sumlkml}
\end{figure}

Since $|\cos\theta/a| < 1$, we may write
\begin{eqnarray}
	I_n^{(\ln^2)}(a) &\equiv& \frac{1}{2\pi} \int_{0}^{2\pi} e^{in\theta} \ln^2(a + \cos\theta) \dd \theta
	=
	\ln(a) \left[2\ln(\frac{a+\sqrt{a^2-1}}{2}) - \ln(a) \right]_{n=0}
	- 2\ln(a) \left.\frac{\left(-a + \sqrt{a^2-1}\right)^{|n|}}{|n|}\right|_{n\ne0}
	\nonumber \\
	&& + \sum_{k=2}^{\infty} \left\{
	\left(\frac{-1}{2a}\right)^k \left[\sum_{l=1}^{k-1} \frac{1}{l(k-l)}\right] \left[\sum_{l=0}^{k} \binom{k}{l} \delta_{|n|,2l-k}\right]
	\right\}
	. \label{In_ln2_serie}
\end{eqnarray}
In Fig.~\ref{fig_sumlkml} we plot the partial summation
\begin{subequations} \label{def_Sp}
\begin{equation}
	\cS_\text{part}(k) \stackrel{k\ge2}{\equiv} \sum_{l=1}^{k-1} \frac{1}{l(k-l)}
	= \frac{2}{k} \sum_{l=1}^{k-1} \frac{1}{l}
	\equiv \frac{2}{k} H_{k-1}
	, \quad \text{with $\cS_\text{part}(2) = \cS_\text{part}(3) = 1$ and $\cS_\text{part}(n>3) < 1$}
	, \label{def_Sp1}
\end{equation}
where $H_{k-1}$ is the harmonic number.
Using this observation, we may write an asymptotic expression for $\cS_\text{part}(k)$:
\begin{equation}
	\cS_\text{part}(k)
	\stackrel{k\gg1}{\approx}
	\frac{2}{k} \left[\ln(k) + \gamma + \cO\left(\frac{1}{k}\right)\right]
	, \label{def_Sp2}
\end{equation}
\end{subequations}
where $\gamma \approx 0.57721$ is the Euler's constant.

Using~(\ref{def_Sp}) we analyze the summation
\begin{subequations}\label{S_n}
\begin{equation}
	\cS(n) \equiv \sum_{k=2}^{\infty} \left\{
	\left(\frac{-1}{2a}\right)^k \cS_\text{part}(k) \left[\sum_{l=0}^{k} \binom{k}{l} \delta_{|n|,2l-k}\right]
	\right\}
	\equiv \sum_{k=2}^{\infty} \cS_1(n,k) \label{def_cS}
\end{equation}
from Eq.~(\ref{In_ln2_serie}) and write its expressions for even and odd $n$:
\begin{eqnarray}
	\cS(n) 
	&\equiv&
	\sum_{t=\max(s,1)}^{\infty} \left\{ \left(\frac{1}{2a}\right)^{2t} \frac{(2t)!}{(t+s)! (t-s)!} \cS_\text{part}(2t)
	\right\} , \quad \text{for $|n|\equiv 2s$}, \label{S_2r} \\
	\cS(n) 
	&\equiv&
	- \sum_{t=\max(s,1)}^{\infty} \left\{
	\left(\frac{1}{2a}\right)^{2t+1} \frac{(2t+1)!}{(s+t+1)!(t-s)!} \cS_\text{part}(2t+1)
	\right\} , \quad \text{for $|n|\equiv 2s+1$} , \label{S_2rp1}
\end{eqnarray}
\end{subequations}
where $s\ge0$ is an integer.
The convergence of the summations in Eqs.~(\ref{S_n}) can be readily demonstrated by analyzing the asymptotic behavior of $\cS_1(n,k)$.
For $k, k-n\gg1$ we may use the Stirling approximation for the factorials, treating $n$ as a continuous variable in~(\ref{def_cS}), whereas for $n/k\ll1$ we may use the Taylor expansion.
Then, the logarithm of the absolute value of a term of the summation $\cS(n)$ may be approximated as
\begin{subequations} \label{ansympt_exp_T1_T2}
\begin{eqnarray}
	&& \log\left[\left| \cS_1(n,k) \right|\right] =
	\log\left\{
	\left(\frac{1}{2a}\right)^k \cS_\text{part}(k)
	\frac{k!}{\left(\frac{k+n}{2}\right)! \left(\frac{k-n}{2}\right)!}
	\right\}
	\approx
	\log[\cS_\text{part}(k)] - k\log(a) - \frac{n^2}{k}
	, \label{ansympt_exp_T1}
\end{eqnarray}
which leads to
\begin{equation}
	\left| \cS_1(n,k) \right| \approx \cS_\text{part}(k) e^{-n^2/k} a^{-k}
	. \label{ansympt_exp_S1}
\end{equation}
Therefore, the series~(\ref{S_n}) decrease more rapidly than a geometric progression with common ratio $a^{-1} < 1$, leading to the conclusion that these series are absolutely convergent.
On the other hand, from any of Eqs.~(\ref{S_n}) it follows that $k\ge |n|$, therefore, for any $n$ the lowest order term is
\begin{equation}
	\cS_1(n,|n|) =
	\left(\frac{-1}{2a}\right)^{|n|} \cS_\text{part}(|n|)
	. \label{lowest_S1}
\end{equation}

The other nonzero term from Eq.~(\ref{In_ln2_serie}) has the asymptotic expansion:
\begin{equation}
	- 2\ln(a) \left.\frac{\left(-a + \sqrt{a^2-1}\right)^{|n|}}{|n|}\right|_{n\ne0}
	\stackrel{a\gg1}{\approx}
	- \left(\frac{-1}{2a}\right)^{|n|} 2\ln(a) \left(\frac{1}{|n|}+\frac{1}{4 a^{2}} + \cO\left(\frac{1}{na^4}\right)\right)
	. \label{asympt_exp_T2}
\end{equation}
\end{subequations}
Therefore, from~(\ref{lowest_S1}) and (\ref{asympt_exp_T2}) we see that the dominant term in the expansion of $I_n^{(\ln^2)}(a)$ with respect to $a$ is
\begin{equation}
	\left(\frac{-1}{2a}\right)^{|n|} \left[\cS_\text{part}(|n|) - \frac{2\ln(a)}{|n|}\right]
	, \quad \text{for any $|n| \ge 2$} , \label{asympt_exp_In_ln2}
\end{equation}
which explains the rapid decrease of $\left|I_n^{(\ln^2)}(a)\right|$ with increasing values of both $n$ and $a$.

The results above may be readily applied to obtain the asymptotic behavior of $\tilde{I}_n^{(\ln^2)}(a)$.
For $|n|\le 4$, the dominant terms are given in Eq.~(\ref{In_ln2_p}), whereas for $|n|\ge4$, we use the identity $\tilde{I}_n^{(\ln^2,\pm)}(a) \equiv I_n^{(\ln^2)}(a) \pm \left[I_n^{(\ln^2)}(a) + I_n^{(\ln^2)}(a)\right]/2$ and Eq.~(\ref{asympt_exp_In_ln2}):
\begin{equation}
	\tilde{I}_n^{(\ln^2,\pm)}(a) \stackrel{|n|\ge4}{\approx} \pm \frac{1}{2} \left(\frac{-1}{2a}\right)^{|n|-2} \left[\cS_\text{part}(|n|-2) - \frac{2\ln(a)}{|n|-2}\right] , \label{asympt_exp_tIn_ln2}
\end{equation}
where the result for $|n|=4$ is identical to the one in Eq.~(\ref{In_ln2_p}).

\section{Symmetric and antisymmetric basis vectors} \label{app_sec_sym_basis}

Exploiting the symmetry of the diagonal matrix elements to the change of $n_2$ into $-n_2$~(\ref{H_diag}) and the fast decrease of non-diagonal elements with $|n|$~(\ref{H_nondiag}), we introduce the symmetric ($+$) and antisymmetric ($-$) basis vectors
\begin{subequations} \label{defs_Fpm}
\begin{equation}
	\tilde{\cF}^{(+)}_{0,m,k} \equiv \tilde{\cF}^{(\Lambda)}_{0,m,k}
	\quad {\rm and} \quad
	\tilde{\cF}^{(\pm)}_{n_2,m,k} \equiv \frac{1}{\sqrt{2}} \left[ \tilde{\cF}^{(\Lambda)}_{n_2,m,k} \pm \tilde{\cF}^{(\Lambda)}_{-n_2,m,k} \right] ,
	\label{def_Fpm}
\end{equation}
for any $m\in \Z$ (integer) and $n_2,k \in \N^*$ (positive integers).
The functions $\tilde{\cF}^{(\pm)}_{n_1,m,k}$ and $\tilde{\cF}^{(\pm)}_{n_2,m,k}$ satisfy the conditions
\begin{equation}
	\langle \tilde{\cF}^{(+)}_{n_1,m_1,k_1} | \tilde{\cF}^{(+)}_{n_2,m_2,k_2} \rangle = \langle \tilde{\cF}^{(-)}_{n_1,m_1,k_1} | \tilde{\cF}^{(-)}_{n_2,m_2,k_2} \rangle = \delta_{n_1,n_2} \delta_{m_1,m_2} \delta_{k_1,k_2},
	\quad
	\langle \tilde{\cF}^{(+)}_{n_1,m_1,k_1} | \tilde{\cF}^{(-)}_{n_2,m_2,k_2} \rangle = 0.
	\label{norm_Fpm}
\end{equation}
\end{subequations}

In the basis~(\ref{defs_Fpm}) we have ($n_2 > 0$ and $n<n_2$)
\begin{subequations} \label{tH_pm_matrix}
\begin{eqnarray}
	0 &=& \langle \tilde{\cF}^{(+)}_{n_1,m_1,k_1} | \tilde{\cH}^{(2D)}_\text{free} | \tilde{\cF}^{(-)}_{n_2,m_2,k_2} \rangle
	= \langle \tilde{\cF}^{(-)}_{n_1,m_1,k_1} | \tilde{\cH}^{(2D)}_\text{free} | \tilde{\cF}^{(+)}_{n_2,m_2,k_2} \rangle ,
	\label{H_mm_Fpm_mp} \\
	\langle \tilde{\cF}^{(+)}_{0,m,k} | \tilde{\cH}^{(2D)}_\text{free} | \tilde{\cF}^{(+)}_{0,m,k} \rangle
	&=& \langle \tilde{\cF}^{(\Lambda)}_{0,m,k} | \tilde{\cH}^{(2D)}_\text{free} | \tilde{\cF}^{(\Lambda)}_{0,m,k} \rangle
	= a^2 \left[ -\frac{1}{4}
	+ \frac{a}{\left(a^{2}-1\right)^{\frac{3}{2}}} \left(m^{2}-\frac{1}{4}\right) \right]
	, \label{H_diag_Fpm_0} \\
	\langle \tilde{\cF}^{(+)}_{0,m,k} | \tilde{\cH}^{(2D)}_\text{free} | \tilde{\cF}^{(+)}_{n_2,m,k} \rangle
	&=& \sqrt{2} \frac{a^2 \left(n_2 \sqrt{a^{2}-1}+a \right) \left(\sqrt{a^{2}-1}-a \right)^{n_2}}{\left(a^{2}-1\right)^{\frac{3}{2}}}
	\left(m^{2}-\frac{1}{4}\right)
	= \langle \tilde{\cF}^{(+)}_{n_2,m,k} | \tilde{\cH}^{(2D)}_\text{free} | \tilde{\cF}^{(+)}_{0,m,k} \rangle ,
	\label{H_diag_Fpm_0n2} \\
	\langle \cF^{(\pm)}_{n_2,m,k} | \tilde{\cH}^{(2D)}_\text{free} | \cF^{(\pm)}_{n_2,m,k} \rangle
	&=& a^2 \left[ n_2^{2}-\frac{1}{4}
	+ \frac{a \pm \left({2n_2} \sqrt{a^{2}-1}+a \right) \left(\sqrt{a^{2}-1}-a \right)^{{2n_2}}}{\left(a^{2}-1\right)^{\frac{3}{2}}}
	\left(m^{2}-\frac{1}{4}\right) \right]
	, \label{H_diag_Fpm} \\
	\langle \cF^{(\pm)}_{n_2-n,m,k} | \tilde{\cH}^{(2D)}_\text{free} | \cF^{(\pm)}_{n_2,m,k} \rangle
	&=&
	\frac{a^2 \left[\left({|n|} \sqrt{a^{2}-1}+a \right) \left(\sqrt{a^{2}-1}-a \right)^{{|n|}} \pm \left((2n_2-n) \sqrt{a^{2}-1}+a \right) \left(\sqrt{a^{2}-1}-a \right)^{{2n_2-n}}\right]}{\left(a^{2}-1\right)^{\frac{3}{2}}}
	\nonumber \\
	&& \times \left(m^{2}-\frac{1}{4}\right)
	, \quad \text{for $n \ne 0$.} \label{H_pp_mm_Fpm}
\end{eqnarray}
\end{subequations}
Equations~(\ref{H_diag_Fpm}) and (\ref{H_pp_mm_Fpm}) may be grouped in a single line, for any $n_2>0$ and $n<n_2$:
\begin{eqnarray}
	&& \langle \tilde{\cF}^{(\pm)}_{n_2-n,m, 1} | \tilde{\cH}^\text{(2D)}_\text{free} | \tilde{\cF}^{(\pm)}_{n_2,m, 1} \rangle
	= a^2 \left[\left(n_2^{2}-\frac{1}{4}\right) \delta_{n,0}
	\right. \nonumber \\
	&& \left. + \frac{\left({|n|} \sqrt{a^{2}-1}+a \right) \left(\sqrt{a^{2}-1}-a \right)^{{|n |}} \pm \left((2n_2-n) \sqrt{a^{2}-1}+a \right) \left(\sqrt{a^{2}-1}-a \right)^{{2n_2 - n}}}{\left(a^{2}-1\right)^{\frac{3}{2}}} \left(m_1^{2}-\frac{1}{4}\right)
	\right] ,
	\label{H_elms_matr_sym}
\end{eqnarray}
Further, using~(\ref{sym_T3}) we obtain
\begin{equation} \label{T3_Fpm}
	\langle \tilde{\cF}^{(+)}_{n_1,m_1,k_1} | \tilde{T}_3^{(\theta)}(\bA=0) | \tilde{\cF}^{(+)}_{n_2,m_2,k_2} \rangle
	= \langle \tilde{\cF}^{(-)}_{n_1,m_1,k_1} | \tilde{T}_3^{(\theta)}(\bA=0) | \tilde{\cF}^{(-)}_{n_2,m_2,k_2} \rangle
	= 0 .
\end{equation}
%

The basis~(\ref{defs_Fpm}) provides a more suitable description of the system at $I=0$ for the following reasons. First, Eq.~(\ref{H_mm_Fpm_mp}) suggests a natural decomposition of the Hilbert space $H$ into two orthogonal subspaces: $H \equiv H^{(+)} \oplus H^{(-)}$. Here, $H^{(+)}$ and $H^{(-)}$ are spanned by the basis sets $\left\{\mathcal{F}^{(+)}_{n,m,k}\right\}$ and $\left\{\mathcal{F}^{(-)}_{n,m,k}\right\}$, respectively. Consequently, the eigenfunctions of $\hat{\mathcal{H}}^{(2D)}_\text{free}$ reside entirely within either $H^{(+)}$ or $H^{(-)}$; they are never linear combinations of vectors from both subspaces. Second, this separation, combined with Equation~(\ref{T3_Fpm}), indicates that the expectation values of the operator $\tilde{T}_3^{(\theta)}(\mathbf{A}=0)$ evaluated on the Hamiltonian eigenvectors are identically zero.
Therefore, diagonalizing the Hamiltonian in the bases~(\ref{defs_Fpm}) at $I=0$ ensures that this property is strictly satisfied, even in the case when $a \gg 1$ or $n_2\gg 1$ and proper numerical diagonalization is difficult due to the quasi double degeneracy of the energy levels~\cite{PhysicaA.598.127377.2022.Dolineanu, PhysicaA.617.127377.2022.Dolineanu}.

\end{document}